\documentclass[10pt,a4paper]{article}

\usepackage{amsmath,amssymb,amsfonts}
\usepackage{graphicx}
\usepackage[hidelinks]{hyperref}
\usepackage{cite}
\usepackage{bm}
\usepackage{mathrsfs}
\usepackage{xcolor}

\usepackage{tikz}
\usetikzlibrary{decorations.pathmorphing}

\usepackage{geometry}
\usepackage[mathscr]{euscript}
\usepackage{setspace}
\usepackage{authblk}

\numberwithin{equation}{section}

\definecolor{linkblue}{rgb}{0.1,0.3,.7}
\definecolor{forestgreen}{rgb}{0.13, 0.55, 0.13}
\definecolor{lava}{rgb}{0.81, 0.06, 0.13}
\definecolor{dblue}{rgb}{0.55,0.25,3.0}

\newcommand{\beq}{\begin{equation}}
\newcommand{\eeq}{\end{equation}}
\newcommand{\bea}{\begin{eqnarray}}
\newcommand{\eea}{\end{eqnarray}}
\newcommand{\beal}{\begin{equation}\begin{aligned}}
\newcommand{\eeal}{\end{aligned}\end{equation}}
\newcommand{\bei}{\begin{itemize}}
\newcommand{\eei}{\end{itemize}}

\newcommand{\sgn}[1]{\text{sgn}\!\left(#1\right)}
\newcommand{\ff}{{_2F_1}}
\newcommand{\xx}{{\bf x}}
\newcommand{\kk}{{\bf k}}
\newcommand{\cc}{\mathcal{C}}
\renewcommand{\l}{\left(}
\renewcommand{\r}{\right)}
\renewcommand{\L}{\left[}
\newcommand{\R}{\right]}
\newcommand{\la}{\left\langle}
\newcommand{\ra}{\right\rangle}
\newcommand{\ii}{\mathrm{i}}

\newcommand{\scP}{\mathscr{P}}
\newcommand{\scD}{\mathscr{D}}

\newcommand{\mathf}{\mathcal{F}}
\newcommand{\mathk}{\mathcal{K}}

\newcommand{\hpm}{h_{\pm}}

\newcommand{\hpl}{h_+}
\newcommand{\hmi}{h_-}
\newcommand{\chp}{\chi^+}
\newcommand{\chm}{\chi^-}
\newcommand{\chpm}{\chi^{\pm}}

\newcommand{\dpl}{\Delta_+}
\newcommand{\dmi}{\Delta_-}
\newcommand{\dpm}{\Delta_{\pm}}
\newcommand{\tilF}{\tilde{F}}
\newcommand{\tilG}{\tilde{G}}
\newcommand{\vare}{\varepsilon}

\newcommand{\iso}{$ \text{SL}(2,\mathbb{R})_+ \times \text{SL}(2,\mathbb{R})_- \,$}
\newcommand{\ee}{\text{even}}
\newcommand{\oo}{\text{odd}}

\title{\bf Spectrum of a 2D QFT analogue of the SYK model}

\author{
Avik Chakraborty\thanks{avikchakraborty589@gmail.com},\; Pratik K. Das\thanks{dask.pratik@gmail.com}\; and  Manavendra Mahato\thanks{manav@iiti.ac.in}
\\ 
{\small Department of Physics, Indian Institute of Technology Indore, \\ Khandwa road, Indore 453552, India}
}
\date{}

\begin{document}
\maketitle

\begin{abstract}
 We study the spectrum of the two-dimensional generalisation of the SYK model proposed by Turiaci and Verlinde, in which chiral Majorana fermions with a topological kinetic term interact through random $2q$-fermion couplings. We compute the four-point function at order $1/N$ in the conformal limit. In addition to the conformal Casimir, the ladder kernels commute with the operator that exchanges a single lightcone coordinate. As a result, the four-point function splits into an even and an odd sector, which the kernels do not mix. In each sector we construct a complete eigenbasis of the Casimir, find the eigenvalues of the kernels in closed form and sum the ladder diagrams. The poles in the ladder sum lie on a spectral curve in the plane of the two conformal weights. A conformal solution in the IR limit breaks the symmetries of the model under chiral reparametrisations and local Lorentz transformations, and the kernel has eigenvalue one on the corresponding modes. The reparametrisation modes, with conformal weights $(2,0)$ and $(0,2)$, are non-normalisable, whereas the Lorentz mode of the odd sector, with weights $(1,1)$, is normalisable. We determine the region where the sum of ladders can be represented in a contour form analogous to that of the one-dimensional SYK model.
\end{abstract}
\vfill
\pagebreak

\tableofcontents

\section{Introduction}

 The Sachdev-Ye-Kitaev (SYK) model \cite{Sachdev_1992,kitaevtalks} is a system of $N$ Majorana fermions $\psi^i$ that is tractable in the large-$N$ limit. It is governed by the action
\bea
\label{eq:SYK}
 S_{\text{SYK}} = \int  dt \,\l  \sum_{i}  \frac{\ii}{2} \psi^i \partial_t \psi^i\, - \, {\ii}^{\frac q 2}\! \sum_{i_1 < \dots < i_q}  J_{i_1 \dots i_q} \psi^{i_1} \dots \psi^{i_q}\r ,
\eea
where the couplings $J_{i_1 \dots i_q}$ are chosen randomly and independently from a Gaussian distribution with zero mean and $\la J_{i_1 \dots i_q}^2 \ra = (q-1)!J^2/N^{q-1}$. In the large-$N$ limit, the perturbative expansion is dominated by melonic diagrams, which makes the model tractable. At low temperature and strong coupling or in the low energy (IR) limit, the SYK model acquires a reparametrisation symmetry, which is broken to SL(2,$\mathbb{R}$) by a conformal solution of the two-point function and the corresponding Goldstone mode is governed by the Schwarzian action \cite{kitaevtalks,Maldacena_2016}. The effective action of the Goldstone mode also describes the boundary theory of dilaton gravity in AdS$_2$ \cite{Almheiri:2014cka,Jensen:2016pah,maldacena2016conf,Engels_y_2016}, which makes the SYK model useful as a toy model in the study of holography. In the strong coupling regime, the out-of-time-ordered (OTO) thermal correlator grows exponentially and saturates the chaos bound \cite{kitaevfirsttalk,Maldacena:2015waa}.

The SYK model inspires the study of many interesting generalisations of the model including proposals involving higher dimensional continuum theories \cite{Witten:2016iux,Berkooz_2017_thirring,Turiaci_2017,Lian_2019,Altland_2025}, higher dimensional lattices \cite{Gu_2017,Berkooz_2017,PhysRevD.96.106008,PhysRevB.95.155131}, models with complex fermions \cite{Sachdev_2015,PhysRevB.95.155131}, global symmetry \cite{Gross:2016kjj}, supersymmetry \cite{Fu:2016vas}, and extra quadratic fermions \cite{Banerjee:2016ncu}. In two spacetime dimensions (2D), free fermions with a canonical kinetic term acquire mass dimension $1/2$, which renders the $q$-fermion interaction irrelevant for $q>4$. This complicates any effort to get SYK-like physics in two or more dimensions. This work involves the nonchiral continuum model \cite{Turiaci_2017} with spatially homogeneous SYK interaction, where the usual kinetic term was replaced by a topological ultraviolet (UV) term that makes the fermionic fields dimensionless and preserves the essential SYK features. The proposed 2D generalisation is described by chiral Majorana fields $\psi_\pm^i$ and is governed by the action
\beal
 \label{eq:action}
  & \quad \quad S = S_{\text{UV}} + S_{\text{IR}} , \\ 
  S_{\text{UV}} &= \sum_i \int d^2 \xx \, \epsilon^{\mu\nu}  \psi^i_+ \partial_\mu \psi^i_+\, \psi^i_- \partial_\nu \psi^i_- , \\
  S_{\text{IR}} = \sum_{i_1 < \dots < i_q} &\sum_{j_1 < \dots < j_q} \int d^2 \xx \, J_{i_1 \dots j_q} \psi^{i_1}_- \dots \psi^{i_q}_- \psi^{j_q}_+ \dots \psi^{j_1}_+ ,
\eeal
where $J_{i_1 \dots j_q}$ are Gaussian random couplings with zero mean and
\beq
 \la \l J_{i_1 \dots j_q}\r ^2 \ra = \frac{J^2 (q-1)! q!}{N^{2q-1}}.
\eeq
 The interaction contains $2q$ Majorana fermions, which under conjugation gives
\beq
 \l\psi^{i_1}_-\cdots\psi^{i_q}_-\psi^{j_q}_+\cdots\psi^{j_1}_+\r^\dagger=(-1)^q\,\psi^{i_1}_-\cdots\psi^{i_q}_-\psi^{j_q}_+\cdots\psi^{j_1}_+ .
\eeq
 Therefore, $S_{\text{IR}}$ is Hermitian for real couplings only if $q$ is even, and we take $q$ to be even throughout. In the 1D model a product of $q$ Majorana fermions picks up $(-1)^{q(q-1)/2} = (-1)^{q/2}$ under conjugation, which the factor $\ii^{q/2}$ in Eq. \eqref{eq:SYK} compensates. The corresponding factor for $2q$ fermion interaction in $S_\text{IR}$ is real. No phase is therefore required in $S_{\text{IR}}$ for even $q$. The fermions $\psi^i_\pm$ are analogous to the chiral components of a Dirac fermion but they follow the unconventional kinetic term in $S_{\text{UV}}$. The UV action describes $N$ topological conformal field theories (TCFTs). In the IR limit, it exhibits emergent conformal symmetry and its dynamics are governed by Goldstone modes associated with broken 2D reparametrisation symmetry. Reference \cite{Turiaci_2017} showed that their effective action can also be obtained from AdS$_3$ gravity, which suggests the existence of a holographic dual.

 In the 1D SYK model, the operators that appear in the operator product expansion (OPE) of two fermions and their OPE coefficients are obtained from the four-point function at order $1/N$ \cite{Maldacena_2016,kitaevtalks,Kitaev_2017}. The ladder kernel commutes with the Casimir of SL$(2,\mathbb{R})$, and its eigenvalue $k_c(h)$ on the eigenfunctions of the Casimir is known in closed form. Then the sum over ladders can be written as a contour integral over $h$, and closing the contour gives the dimensions of the operators as the solutions of $k_c(h)=1$ and their OPE coefficients as the residues. The state $h=2$, for which $k_c(2)=1$ exactly, corresponds to the reparametrisation mode. Its enhanced contribution away from the conformal limit is governed by the Schwarzian action.
 
 For the 2D model, the two-point function, the effective action of the reparametrisations and its relation to AdS$_3$ gravity have been discussed \cite{Turiaci_2017}, but the spectrum of the bilinears has not been determined. The aim of this work is to determine the spectrum and the operator content of the 2D model at order $1/N$ in the conformal limit.

Three features of the 2D model make this more than a direct generalisation. First, the model is defined in Lorentzian signature with independent lightcone coordinates $x^\pm$, and the conformal group is \iso. The conformal spin $j$ of the fermions is not quantised, so the Euclidean SL$(2,\mathbb{C})$ analysis of 2D SYK-like models \cite{Murugan_2017} does not apply. Second, the eigenvalues of the Casimir are degenerate, and an additional operator that commutes with the kernels is needed to obtain a basis of eigenfunctions. Third, the kernels mix the two chiralities of the fermions, so that their eigenvalues form a $2\times2$ matrix that acts on chirality vectors.

Our main results are as follows. We show that the conformal kernels commute with the exchange of a single lightcone coordinate, which acts on the cross ratios as the reflection $\chi^\pm\to\chi^\pm/(\chi^\pm-1)$ of one of them. Therefore, the four-point function splits into an even and an odd sector, which the kernels do not mix. Both sectors are sourced by the zero-rung ladder. In each sector we construct a complete basis of eigenfunctions of the Casimir. For each lightcone it consists of a continuum $h=\frac{1}{2}+\ii s$ with real $s > 0$ and of discrete states, $h=2n$ in the even sector and $h=2n-1$ in the odd sector with integer $n \geq 1$. The even sector reproduces the eigenfunctions of the 1D model. The two sectors are generated by the two conformally covariant three-point functions of two fermions and an operator of weights $(\hpl,\hmi)$, and we obtain the eigenvalues of the kernels in both sectors in closed form, as products of chiral functions of the two weights. The projections of the zero-rung ladder on the eigenfunctions are given by the same chiral functions. This allows us to sum the ladder diagrams and to write the result as a double contour integral.

Closing one of the two contours expands the four-point function in conformal blocks of one lightcone. Unlike in the 1D model, the poles lie on a spectral curve, $\det(\mathbb{I}-\mathk^\sigma(\hpl,\hmi))=0$, which we study numerically. For a weight $\hmi$ on the principal series, the weights $\hpl$ fill continuous bands that end at the weights of the generalised free field. We determine the region of the $(q,j)$ plane in which the contour representation holds (Fig.~\ref{fig:validity_region}). The kernel has eigenvalue one at the conformal weights $(2,0)$ and $(0,2)$ of the chiral reparametrisations and at weights $(1,1)$ of the local Lorentz transformations. Unlike in the 1D model, the reparametrisation modes are not normalisable and do not belong to the eigenbasis. The state $(1,1)$ of the odd sector is normalisable. We identify it with the odd part of the local Lorentz transformations of the conformal solution and compute its contribution in the conformal limit. Finally, we discuss the reparametrisation modes, their effective action and their relation to the stress tensor, and what they imply for a 3D bulk dual.

The paper is organised as follows. In Sec. \ref{sec:model} we review the model, its topological UV limit and the conformal solution of the Dyson-Schwinger equations. Sec. \ref{sec:four_point_fn} is devoted to the four-point function. We set up the conformal kernels and the two reflection sectors in Sec. \ref{subsec:conf_symmetry}, construct the eigenfunctions of the Casimir in Sec. \ref{subsec:casimir_eigenfunctions} and compute the eigenvalues of the kernels in Sec. \ref{subsec:eigenvalues}. We sum the ladders in Sec. \ref{subsec:summ_ladders}, discuss the operator content in Sec. \ref{subsec:operators} and treat the Lorentz mode in Sec. \ref{subsec:lorentz_mode}. In Sec. \ref{sec:repara_modes} we discuss the reparametrisation modes and a possible bulk dual, and we conclude in Sec. \ref{sec:conclusion}. The appendices contain the conformal kernels (Appendices \ref{app:conf_kernel} and \ref{app:kernel_cross_ratio}), the eigenbasis (Appendix \ref{app:eigenbasis_of_Casimir}), the eigenvalues (Appendix \ref{app:eigenvalues}) and the zero-rung ladder (Appendix \ref{app:zero-rung_ladder}).

\paragraph{Notations and conventions.} We work in 2D Minkowski space with the lightcone coordinates $x^\pm=x\pm t$, and write $\xx=(x^+,x^-)$ and $\xx_{ij}=\xx_i-\xx_j$. The chiral Majorana fermions are $\psi^i_\pm$, $i=1,\dots,N$. We call the label $a=\pm$ their chirality and write $\bar a=-a$. The scaling dimension is $\Delta=1/q$, the spin is $j$, and $\dpm=\Delta\pm j$, so that $\psi_a$ has the weights $\l\frac{\dpl}{2},\frac{\dmi}{2}\r$ on the lightcones $\l x^a,x^{\bar a}\r$. The weights of the operators that appear in the OPE are denoted $(\hpl,\hmi)$, and $\chpm$ are the cross ratios on the two lightcones. The superscript $\lbrace \sigma=\ee,\oo \rbrace $ labels the two reflection sectors, with $\tau=+1$ for $\sigma=\ee$ and $\tau=-1$ for $\sigma=\oo$. $K_{ab}$ denotes the ladder kernels and $\mathk^\sigma_{ab}(\hpl,\hmi)$ their eigenvalues. Results for the 1D model are quoted in the conventions of reference \cite{Maldacena_2016}.


\section{A 2D QFT analogue of SYK model}\label{sec:model}


 In this section, we briefly revisit the symmetries and large-$N$ limit of the 2D generalisation of SYK model proposed in \cite{Turiaci_2017}. We summarise the relevant results required for a detailed study of the four-point function in Sec. \ref{sec:four_point_fn}.

 \subsection{Topological CFT in the UV limit}\label{subsec:tcft}

  The UV action $S_{\text{UV}}$ in Eq. \eqref{eq:action} decouples into $N$ topological theories, each consisting of a pair of chiral Majorana fermions. We suppress the flavour indices and focus on a single pair of fermions in this subsection. Introducing Hubbard-Stratonovich fields $e^{\pm}_{\mu}$ into each UV sector, the action can be rewritten as
  \beq
  \label{eq:topological_action}
   S = \int d^2 \xx \; \epsilon^{\mu \nu} \l e^{a}_{\mu} \psi_a \partial_\nu \psi_a - \frac{1}{2} \epsilon_{ab} e^{a}_{\mu}e^{b}_{\nu} \r ,
  \eeq
  where $a=\pm$. It is straightforward to see that imposing the equations of motion for $e^{\pm}_{\mu}$ into the action in Eq. \eqref{eq:topological_action} leads to the UV action in eq. \eqref{eq:action}. The fields $e^{\pm}_{\mu}$ can be treated as Cartan zweibeins parametrising a dynamical 2D metric with local Lorentz frame. The UV theory is evidently reparametrisation and local Lorentz invariant,
  \bea
  \label{eq:UV_reparametrisation}
   e^{\pm}_{\mu}(\xx) \to \frac{\partial \tilde{x}^\nu}{\partial x^{\mu}} e^{\pm}_{\nu}(\tilde{\xx}(\xx)), &\quad & \psi_\pm(\xx) \to \psi_\pm(\tilde{\xx}(\xx)), \\ 
   e^{\pm}_{\mu}(\xx) \to \Lambda^{\mp 2}(\xx)e^{\pm}_{\mu}(\xx), &\quad & \psi_\pm(\xx) \to \Lambda^{\pm 1}(\xx) \psi_\pm(\xx),
  \eea
  where the fermions transform as scalars under reparametrisations and with $\Lambda^{\pm 1}$ under the local Lorentz tranformations.

  The UV theory is independent of the choice of a metric. It describes $N$ copies of 2D topological Ising CFT with gauged Virasoro symmetry, called the topological Ising CFT \cite{Turiaci_2017}. It is the simplest example of a topological rational CFT, and its properties are derived in reference \cite{Turiaci_2017}. The gauging projects out the Virasoro descendants and leaves three gauge invariant local operators: the identity operator $\mathbb{I}$, the spin field $\sigma$ and the energy operator\footnote{The definition of $\epsilon$ here differs from reference \cite{Turiaci_2017} by the factor of $\ii$, which is necessary to make the energy operator Hermitian.}
  \beq
   \epsilon(\xx) = \ii \, \psi_+(\xx)\psi_-(\xx).
  \eeq
  All three operators have scaling dimension zero. Their Euclidean correlation functions are independent of the positions and equal the number of chiral conformal blocks of the corresponding correlation function of the Ising CFT. The operators form a commutative and associative algebra isomorphic to the Ising fusion rules,
  \beq
  \label{eq:ising_fusion}
   \epsilon \times \epsilon = \mathbb{I}, \quad \epsilon \times \sigma = \sigma, \quad \sigma \times \sigma = \mathbb{I} + \epsilon .
  \eeq
  
  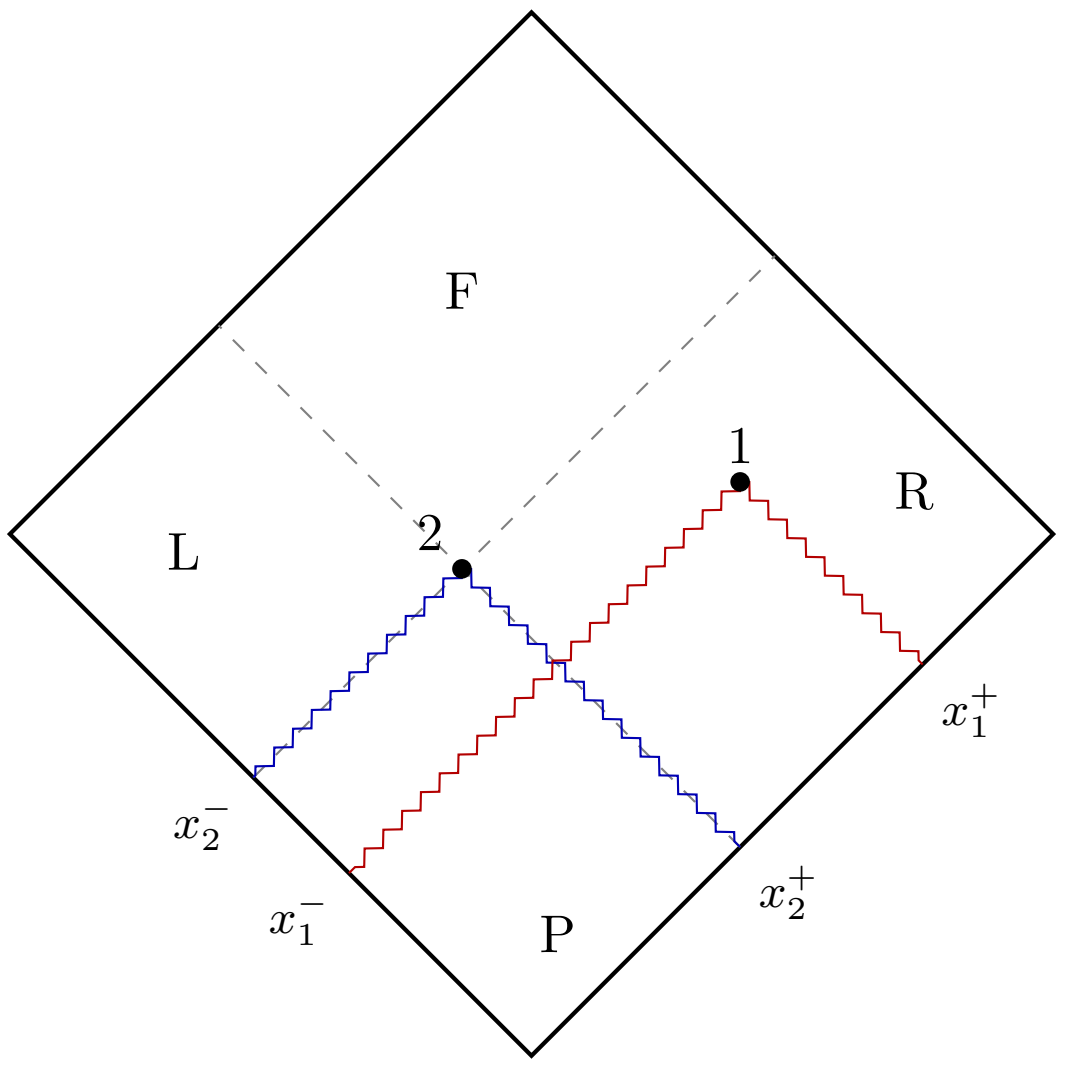
\begin{figure}[t]
  \centering
  \begin{tikzpicture}[scale=1.15]
   \draw[thick] (0,3)--(3,0)--(0,-3)--(-3,0)--cycle;
   \draw[dashed,gray] (-1.6,-1.4)--(1.4,1.6);
   \draw[dashed,gray] (1.2,-1.8)--(-1.8,1.2);
   \draw[decorate,decoration={zigzag,segment length=5pt,amplitude=1.2pt},blue!70!black] (-0.4,-0.2)--(1.2,-1.8);
   \draw[decorate,decoration={zigzag,segment length=5pt,amplitude=1.2pt},blue!70!black] (-0.4,-0.2)--(-1.6,-1.4);
   \draw[decorate,decoration={zigzag,segment length=5pt,amplitude=1.2pt},red!70!black] (1.2,0.3)--(2.25,-0.75);
   \draw[decorate,decoration={zigzag,segment length=5pt,amplitude=1.2pt},red!70!black] (1.2,0.3)--(-1.05,-1.95);
   \fill (-0.4,-0.2) circle (1.6pt) node[above left] {$2$};
   \fill (1.2,0.3) circle (1.6pt) node[above] {$1$};
   \node at (-0.4,1.4) {F}; \node at (0.15,-2.3) {P}; \node at (-2.0,-0.1) {L}; \node at (2.2,0.25) {R};
   \node[below right] at (1.2,-1.8) {\small $x_2^+$}; \node[below right] at (2.25,-0.75) {\small $x_1^+$};
   \node[below left] at (-1.6,-1.4) {\small $x_2^-$}; \node[below left] at (-1.05,-1.95) {\small $x_1^-$};
  \end{tikzpicture}
  \caption{Local operators of the TCFT are attached to two light-like Wilson lines that end on past null infinity, shown in the conformal diagram of 2D Minkowski space. The end-points $x_i^+$ lie on the right and the end-points $x_i^-$ on the left component of past null infinity. The dashed lines are the lightcone of operator 2, which divides the plane into the regions P, F, L and R. In the configuration shown, operator 1 is in region R of operator 2, and the end-points have the same ordering along $x^+$ and $x^-$.}
  \label{fig:wilson_lines}
  \end{figure}

  In Minkowski space, the correlation functions acquire a dependence on the ordering of the operators. We use the lightcone coordinates $x^\pm = x \pm t$. Each local operator is attached to two light-like Wilson lines that extend to past null infinity (Fig.~\ref{fig:wilson_lines}), and the asymptotic end-points $x_i^\pm$ of the Wilson lines label the position of the operator. Since the bulk has no metric, only the ordering of the end-points along $x^+$ and along $x^-$ is meaningful. Every pair of operators is therefore separated in one of four topologically distinct ways: past (P), future (F), left (L) or right (R). The pair is spacelike separated (L, R) if $x_{12}^+x_{12}^->0$, and timelike separated (P, F) if $x_{12}^+x_{12}^-<0$. Moving a pair from spacelike to timelike separation crosses exactly one lightcone, and the chiral conformal blocks of that chirality acquire the braiding phase of the two chiral components. For $\epsilon$ this phase is the sign $-1$ of Fermi statistics, and the time-ordered two-point function is
  \beq
  \label{eq:tcft_two_point}
   \la \mathbb{T} \; \epsilon(\xx_1)\epsilon(\xx_2) \ra_{\text{TCFT}} = \sgn{x_{12}^+}\sgn{x_{12}^-},
  \eeq
  where $\xx_{12} = \xx_1-\xx_2$. It equals $+1$ for spacelike separation, where it agrees with the Euclidean correlation function, and $-1$ for timelike separation. In the $n$-point function every timelike pair contributes a factor $-1$,
  \beq
  \label{eq:tcft_n_point}
   \la \mathbb{T} \; \epsilon(\xx_1) \dots \epsilon(\xx_n) \ra_{\text{TCFT}} = \prod_{i<j}\sgn{x_{ij}^+}\sgn{x_{ij}^-} = \text{Pf}\l \sgn{x_{ij}^+} \r \text{Pf}\l \sgn{x_{ij}^-} \r ,
  \eeq
  for even $n$, while it vanishes for odd $n$. The non-chiral $n$-point function factorises into two chiral factors, which defines the $n$-point functions of the chiral Majorana fermions,
  \beq
  \label{eq:tcft_chiral}
   \la \mathbb{T}\, \psi_\pm(\xx_1) \dots \psi_\pm(\xx_n) \ra_{\text{TCFT}} = \text{Pf}\l \sgn{x_{ij}^{\pm}} \r,
  \eeq
  where $\text{Pf}$ represents the Pfaffian. The fields $\psi_\pm(x^\pm)$ can thus be referred to as topological \textit{2D chiral Majorana-Weyl fermions}. In terms of them, the time ordering acts as $\mathbb{T} = \mathbb{T}_ + \mathbb{T}_-$, where $\mathbb{T}_\pm$ orders the operators $\psi_\pm$ by their lightcone coordinate $x^\pm$, with larger $x^\pm$ to the left, in the same way as time ordering acts on the 1D Majorana fermion, $\la\mathbb{T}\,\psi(\tau_1)\psi(\tau_2)\ra=\sgn{\tau_{12}}$.

  The UV theory of the model is the tensor product of $N$ topological Ising CFTs, followed by a chiral projection analogous to the GSO projection, which admits the bilinears $\psi_+^i \psi_-^j$ with $i\neq j$ as local operators \cite{Turiaci_2017}. Locality with respect to these bilinears removes the spin fields $\sigma_i$ of the individual copies, and the correlation functions of the remaining operators follow from Eq. \eqref{eq:tcft_chiral} and Fermi statistics.

 \subsection{Dyson-Schwinger equation and the conformal limit}

 In this subsection, we analyse the model in the large-$N$ limit at the level of the two-point function and the self-energy. We assume that the model does not undergo a spin-glass transition, so that the disorder average can be taken directly on the generating functional. At leading order in $1/N$, this annealed average leads to the same saddle-point equations as the replica-diagonal quenched average assumed in \cite{Turiaci_2017}. The model is defined in Lorenzian time. Therefore, the disorder-averaged generating functional is
 \beq
  \la Z \ra_J = \int \mathcal{D} J_{i_1 \dots j_q} \mathcal{D} \psi^i_\pm \mathcal{D} e^{\pm}_\mu e^{\ii S}, \qquad \mathcal{D} J_{i_1 \dots j_q} = \prod_{\substack{i_1<\dots<i_q\\ j_1<\dots<j_q}}dJ_{i_1\dots j_q}\exp\l-\frac{J_{i_1\dots j_q}^2N^{2q-1}}{2J^2(q-1)! \, q!}\r.
 \eeq
 We shall use the UV action in the form of Eq. \eqref{eq:topological_action} with the frame fields replaced by their flavour average, since only this collective field participates in the large-$N$ dynamics \cite{Turiaci_2017}\footnote{More precisely, $S_{\text{UV}} = \int d^2 x \; \epsilon^{\mu \nu} \l \sum_{i,a} e^{a}_{\mu} \psi_a^i \partial_\nu \psi_a^i - N e^{+}_{\mu}e^{-}_{\nu} \r , $ where $e^a_{\mu} = \frac{1}{N} \sum_i e^a_{i,\mu} $.}. Performing the Gaussian integral over the random couplings we obtain
 \beq
  \la e^{\ii S_{\text{IR}}} \ra_J = \exp\L-\frac{J^2}{2qN^{2q-1}}\int d^2x_1d^2x_2\l\sum_{i}\psi^i_+(\xx_1)\psi^i_+(\xx_2)\r^q\l\sum_{i}\psi^i_-(\xx_1)\psi^i_-(\xx_2)\r^q\R .
 \eeq
 We introduce the dynamical mean fields $G_\pm(\xx_1,\xx_2)=\frac1N\sum_i\psi^i_\pm(\xx_1)\psi^i_\pm(\xx_2)$ into the path integral by inserting the identity 
 \beq
  1=\int\mathcal DG_\pm\mathcal D\Sigma_\pm\exp\L-\frac N2\int d^2x_1d^2x_2\,\Sigma_\pm(\xx_1,\xx_2)\l G_\pm(\xx_1,\xx_2)-\frac1N\sum_{i=1}^N\psi^i_\pm(\xx_1)\psi^i_\pm(\xx_2)\r\R ,
 \eeq
 where the self-energies $\Sigma_\pm$ act as Lagrange multipliers enforcing the constraints. The fermions now appear quadratically. Integrating them out, the averaged generating functional can be written in terms of the bilinears, $\la Z\ra_J=\int\mathcal DG_\pm\mathcal D\Sigma_\pm\mathcal De^\pm_\mu\,e^{-NI_{\rm eff}}$, with the effective action
 \beal
  I_{\rm eff}=&-\sum_{a=\pm}\log\mathrm{Pf}\l2\ii\,\epsilon^{\mu\nu}e^a_\mu\partial_\nu+\Sigma_a\r +\frac12\sum_{a=\pm}\int d^2x_1d^2x_2\,\Sigma_a(\xx_1,\xx_2)G_a(\xx_1,\xx_2)\nonumber\\
  &+\frac{J^2}{2q}\int d^2x_1d^2x_2\,G_+(\xx_1,\xx_2)^q\,G_-(\xx_1,\xx_2)^q+\ii\int d^2x\,\epsilon^{\mu\nu}e^+_\mu e^-_\nu .
 \eeal
 where $\text{Pf}$ denotes the Pfaffian. In the limit $N\to\infty$, the dynamics is governed by the saddle point of $I_{\rm eff}$. The DS equations are obtained by varying $I_{\rm eff}$ with respect to $\Sigma_\pm$, $G_\pm$ and $e^\pm_\mu$,
 \beq
  \label{eq:ds1}
  \Sigma_{\pm}(\xx_{12}) = -J^2 G_{\pm}(\xx_{12})^{q-1} G_{\mp}(\xx_{12})^{q} ,
 \eeq
 \beq
 \label{eq:ds2}
  \l 2\ii \epsilon^{\mu \nu} e^{\pm}_{\mu} \partial_\nu G_{\pm} + \Sigma_{\pm} * G_{\pm} \r (\xx_{12}) = -\delta^2 (\xx_{12}) ,
 \eeq
 \beq
 \label{eq:ds3}
  e^{\pm}_{\mu} (\xx_1) = \left. \frac{\partial G_{\mp}(\xx_{12})}{\partial \xx_2^{\mu}} \right|_{\xx_2 \to \xx_1} ,
 \eeq
 where $G_\pm$ and $\Sigma_\pm$ now denote the on-shell values of the bilinears, $*$ denotes the convolution $(\Sigma*G)(\xx_1,\xx_2)=\int d^2x_3\,\Sigma(\xx_1,\xx_3)G(\xx_3,\xx_2)$, and translational invariance has been used to write the bilinears as functions of $\xx_{12}=\xx_1-\xx_2$. The self-energy in Eq. \eqref{eq:ds1} differs by an overall sign from the one in reference \cite{Turiaci_2017}\footnote{The effective action of \cite{Turiaci_2017} has the same structure as the Euclidean dynamical mean-field action of the 1D SYK model \cite{Maldacena_2016}, in which the path-integral weight is $e^{-S_E}$ with Euclidean action $S_E$. In 1D, the factor $\ii^{q/2}$ that the Hamiltonian requires for Hermiticity contributes $(-1)^{q/2}$ to the disorder average. This cancels the sign $(-1)^{q(q-1)/2}=(-1)^{q/2}$ from rearranging $q$ fermions, and the result is $\Sigma=+J^2G^{q-1}$. In the 2D model, both of these signs are absent for even $q$. A Euclidean formulation is not available here without giving up the real and independent lightcone coordinates $x^\pm$, on which the scaling ansatz of the two-point function and its Fourier transforms depend. The factor $2\ii$ in the UV term of Eq. \eqref{eq:ds2} also comes from the real-time weight and the normalisation of Eq. \eqref{eq:topological_action}. This term is dropped in the IR analysis, and its normalisation plays no role there.}. We show below that the sign in Eq. \eqref{eq:ds1} is also the one for which the conformal solution exists with a real propagator and a positive variance $J^2$.

 The SYK coupling $J$ has mass dimension two, since the fermions are dimensionless in the UV. At finite temperature $\beta^{-1}$, the dimensionless coupling is therefore $\beta^2 J$, and the strong-coupling regime $\beta^2 J \gg 1$ describes the IR limit of the model. The UV term in Eq. \eqref{eq:ds2} can be neglected in this limit and the equation reduces to
 \beq
 \label{eq:ds2_IR}
  \l G_{\pm} * \Sigma_{\pm} \r (\xx_{12}) = -\delta^2 (\xx_{12}).
 \eeq
 Assuming translational invariance of the model, this translates to the momentum space equation
 \beq
 \label{eq:ds_IR_momentum space}
  G_{\pm}(\kk) \Sigma_{\pm}(\kk) = -1 .
 \eeq

 The IR equations \eqref{eq:ds1} and \eqref{eq:ds2_IR} are covariant under chiral reparametrisations that preserve the lightcone ordering. Under these reparametrisations, $x^\pm \to \phi_\pm (x^\pm)$ with $\phi'_\pm >0$, the bilinears transform as
 \beal
 \label{eq:bilinear_transformations}
  G_\pm(\xx_1,\xx_2)&\to\L \phi_+'(x_1^+)\phi_+'(x_2^+)\R^{\frac{\dpm}{2}}\L \phi_-'(x_1^-)\phi_-'(x_2^-)\R^{\frac{\Delta_\mp}{2}} G_\pm \l \phi(\xx_1),\phi(\xx_2) \r ,\\
  \Sigma_\pm(\xx_1,\xx_2)&\to\L \phi_+'(x_1^+)\phi_+'(x_2^+)\R^{1-\frac{\dpm}{2}}\L \phi_-'(x_1^-)\phi_-'(x_2^-)\R^{1-\frac{\Delta_\mp}{2}} \Sigma_\pm \l \phi(\xx_1),\phi(\xx_2) \r,
 \eeal
 with $\dpm=\Delta \pm j$, where $\Delta$ and $j$ represent the conformal dimension and the conformal spin, respectively. Covariance of Eq. \eqref{eq:ds1} requires $\Delta = 1/q$ whereas the spin $j$ is left undetermined. The UV term is not invariant under the transformations in Eq. \eqref{eq:bilinear_transformations}, since the fermions transform as scalars in the UV (Eq. \eqref{eq:UV_reparametrisation}), and it explicitly breaks the chiral reparametrisation symmetry.
 
 The emergent chiral reparametrisation invariance in the IR limit allows us to take a scaling ansatz for the two-point function of the form
 \beq
  \label{eq:two-point_fn}
  G_{\pm}(\xx) = \mathtt{b} \frac{\sgn{x^\pm}}{|x^+|^{\dpm}|x^-|^{\Delta_{\mp}}} ,
 \eeq
 where $\mathtt{b}$ is a real constant. The sign functions implement Fermi statistics and match the ordering structure of the UV two-point functions in Eq. \eqref{eq:tcft_chiral}. The ansatz is invariant only under \iso, so it spontaneously breaks the chiral reparametrisation symmetry. Substituting the conformal solution $G_{\pm}$ in the DS equation \eqref{eq:ds1}, the self-energy in the IR limit is obtained as,
 \beq
 \label{eq:self_energy}
  \Sigma_{\pm}(\xx) = -J^2 \mathtt{b}^{2q-1} \frac{\sgn{x^\pm}}{|x^+|^{2 - \dpm}|x^-|^{2- \Delta_{\mp}}}.
 \eeq
 The conformal ansatz breaks the diffeomorphism invariance of the UV theory and specifies the choice of lightcone directions. Substituting the Fourier transforms of the scaling ansatz \eqref{eq:two-point_fn} and the self-energy \eqref{eq:self_energy} into the IR DS equation \eqref{eq:ds_IR_momentum space} expresses the scaling dimension and the constant $\mathtt{b}$ in terms of SYK parameters,
 \beq
 \label{eq:alpha0}
  \Delta = \frac{1}{q}, \quad \text{and} \quad J^2 \mathtt{b}^{2q} = \frac{(1- \dpl)(1- \dmi)}{4\pi^2 \cot{\frac{\pi \dpl}{2}} \tan{\frac{\pi \dmi}{2}}} \equiv \alpha_0^2.
 \eeq
 This does not determine the conformal spin $j$, which will be treated as a free parameter.

 Since $J^2$ is the variance of real Gaussian couplings and $\mathtt{b}$ is real, a consistent solution requires $\alpha_0^2>0$. For $\Delta=1/q\le 1/2$, the right-hand side of Eq. \eqref{eq:alpha0} is positive precisely for $|j|<\Delta$. In this window both chiral weights $\dpm/2$ of the fermions are positive, so the two-point functions decay along both lightcone directions. The chiral limit $j\to\Delta$ lies at the edge of this window. There $\tan\frac{\pi\dmi}{2}\to0$, and the limit has to be taken together with $J\to\infty$ at fixed $\mathtt{b}$, as observed in reference \cite{Turiaci_2017}\footnote{The minus sign in Eq. \eqref{eq:ds1} is essential for this result. If we had used $\Sigma_\pm=+J^2G_\pm^{q-1}G_\mp^q$ as in reference \cite{Turiaci_2017}, the same computation would give $J^2\mathtt{b}^{2q}=-\alpha_0^2$. That requires either a complex $\mathtt{b}$ for $|j|<\Delta$, or a spin in the range $\Delta<|j|<1-\Delta$, where one of the chiral weights is negative. Therefore, the positive value of $J^2 \mathtt{b}^{2q}$ quoted in reference \cite{Turiaci_2017} is consistent only with the real-time sign of the self-energy. The same sign enters the kernel of the ladder diagrams in Sec. \ref{sec:four_point_fn}. There it ensures that the reparametrisation modes $(h_+,h_-)=(2,0)$ and $(0,2)$ correspond to eigenvalue one of the kernel.}.


\section{Four-point function}\label{sec:four_point_fn}

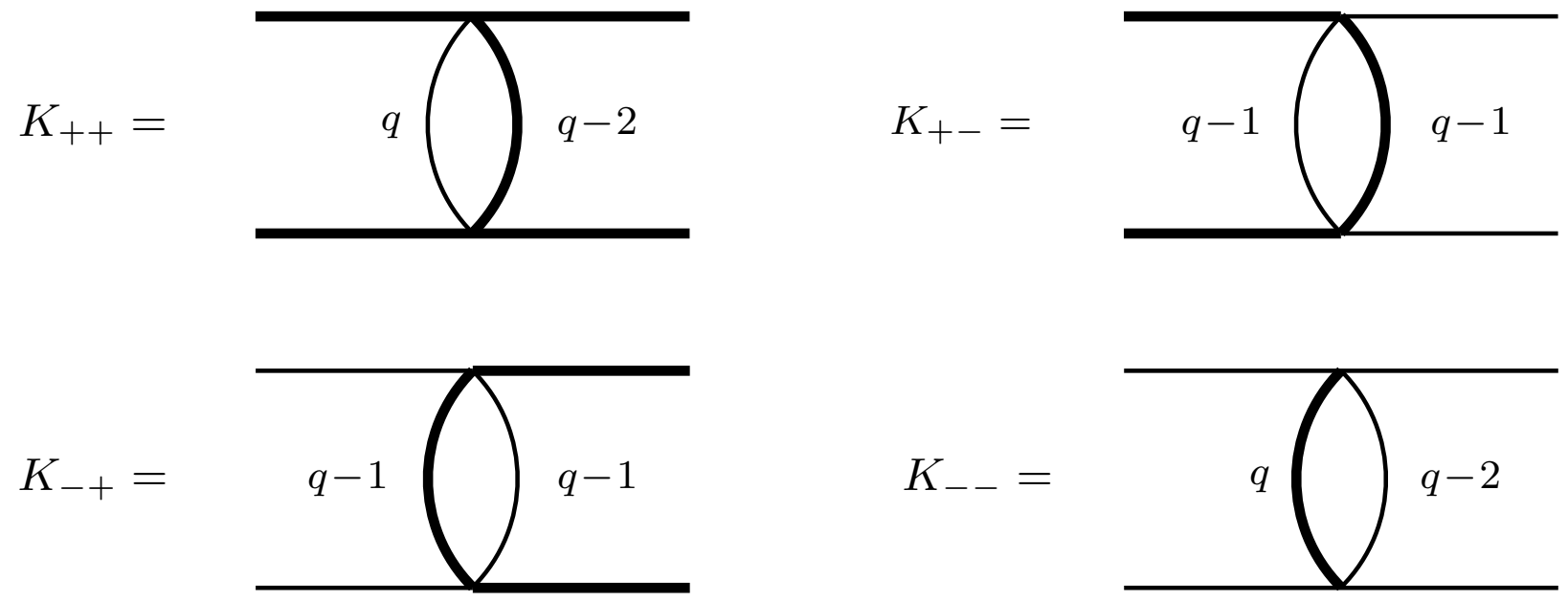
\begin{figure}[ht]
 \begin{center}
  \begin{tikzpicture}[scale=0.75]
  \node[] at (-4.5,1) {\small $K_{++}=$};
  \draw [line width=0.7mm] (0-1,2) arc (45:-45:1.41);
  \draw [line width=0.3mm] (0-1,2) arc (90+45:180+45:1.41);
  \draw [line width=0.7mm] (-2-1,0) -- (2-1,0);
  \draw [line width=0.7mm] (-2-1,2) -- (2-1,2);
  \node[] at (1.15-1,1) {\footnotesize $q\! -\! 2$};
  \node[] at (-0.75-1,1) {\footnotesize $q$};
  \node[] at (3.5,1) {\footnotesize $K_{+-}=$};
  \draw [line width=0.7mm] (6+1,2) arc (45:-45:1.41);
  \draw [line width=0.3mm] (6+1,2) arc (90+45:180+45:1.41);
  \draw [line width=0.7mm] (4+1,2) -- (6+1,2);
  \draw [line width=0.3mm] (6+1,2) -- (8+1,2);
  \draw [line width=0.7mm] (4+1,0) -- (6+1,0);
  \draw [line width=0.3mm] (6+1,0) -- (8+1,0);
  \node[] at (8-2.1,1) {\footnotesize$q\! -\!1$};
  \node[] at (8.2,1) {\footnotesize $q\! -\!1$};
  \end{tikzpicture}

\bigskip
\bigskip

  \begin{tikzpicture}[scale=0.75]
  \node[] at (-4.5,1) {\small $K_{-+}=$};
\draw [line width=0.3mm] (6+1,2) arc (45:-45:1.41);
\draw [line width=0.7mm] (6+1,2) arc (90+45:180+45:1.41);
\draw [line width=0.3mm] (4+1,0) -- (8+1,0);
\draw [line width=0.3mm] (4+1,2) -- (8+1,2);
\node[] at (7+1.1,1) {\footnotesize$q\! -\! 2$};
\node[] at (7-0.75,1) {\footnotesize $q$};
  \node[] at (3.65,1) {\small $K_{--}=$};
\draw [line width=0.3mm] (0-1,2) arc (45:-45:1.41);
\draw [line width=0.7mm] (0-1,2) arc (90+45:180+45:1.41);
\draw [line width=0.3mm] (-2-1,2) -- (0-1,2);
\draw [line width=0.7mm] (0-1,2) -- (2-1,2);
\draw [line width=0.3mm] (-2-1,0) -- (0-1,0);
\draw [line width=0.7mm] (0-1,0) -- (2-1,0);
\node[] at (1.15-1,1) {\footnotesize $q\! -\!1$};
\node[] at (-1.15-1,1) {\footnotesize $q\! -\!1$};
\end{tikzpicture}
\end{center}
\vspace{0.1cm}
\caption{\small Diagrammatic definition of the four types of kernel that gives the four-point correlation function. The thick and thin lines represent the dressed propagators $G_+$ and $G_-$, respectively. The rungs contain multiple dressed propagators with their multiplicity as indicated.}
\label{figure:kernels}
\end{figure}

In this section, we investigate the four-point function at order $1/N$ in the strong coupling $\beta^2 J \gg 1$ limit. The interesting four-point functions always come with pairs of indices since the random couplings $J_{i_1 \dots j_q}$ have zero mean. There are four kinds of such functions
\beal
 \label{eq:defn_4pt}
 \mathbf{F}_{ab}(\xx_1,\xx_2,\xx_3,\xx_4) &= \frac{1}{N^2} \sum_{i,j=1}^{N} \la \mathbb{T} \L \psi^i_a(\xx_1) \psi^i_a(\xx_2) \psi^j_b(\xx_3) \psi^j_b(\xx_4) \R \ra \\
  &= G_a(\xx_{12})G_b(\xx_{34}) +\frac{1}{N} \mathf_{ab}(\xx_1,\xx_2,\xx_3,\xx_4) + \mathcal{O}\l N^{-2} \r,
\eeal
where $\mathbb{T}$ represents the path ordering discussed in Sec. \ref{subsec:tcft}. The disconnected piece comes from a contraction with the dressed propagators. The $1/N$ piece $\mathf_{ab}$ contains contributions from the ladder diagrams built from the averaged two-point functions as shown in Fig. \ref{figure:kernels} and expressed symbolically as
\beq
 \mathf_{ab}(\xx_1,\xx_2,\xx_3,\xx_4) = \sum_{n=0}^{\infty} \mathf_{ab}^{(n)}(\xx_1,\xx_2,\xx_3,\xx_4),
\eeq
where $\mathf_{ab}^{(n)}$ denotes the contribution of $n$-rung ladder. The zero-rung ladder is just a product of the propagators,
\beq
 \mathf_{ab}^{(0)}(\xx_1,\xx_2,\xx_3,\xx_4) = \L -G_a(\xx_{13})G_a(\xx_{24}) + G_a(\xx_{14})G_a(\xx_{23})\R \delta_{ab}.
\eeq
This is the order-$1/N$ part of the four-point function of $N$ fermions whose correlators are all given by Wick contractions of the two-point functions $G_\pm$ of Eq. \eqref{eq:two-point_fn}, thus describing a theory of generalised free fields (GFFs) \cite{Gross_2017}. It is free in the sense that every correlator factorises into products of two-point functions, but the weights $\l\frac{\dpm}{2},\frac{\Delta_\mp}{2}\r$ of the fermions differ from those of a free fermion. The operators that appear in this four-point function, i.e. in the zero-rung ladder, are the bilinears of the GFF (Sec. \ref{subsec:operators}).

The standard technique to obtain the sum over ladders involves an integration kernel $K$, illustrated in Fig. \ref{figure:kernels}, that generates the $n+1$-rung ladder from the $n$-rung ladder,
\beq
\label{eq:ladder_kernel}
 \mathf_{ab}^{(n+1)}(\xx_1,\xx_2,\xx_3,\xx_4) = \sum_{c=\pm} \int d^2\xx \int d^2 \xx' K_{ac}(\xx_1,\xx_2,\xx,\xx')  \mathf_{cb}^{(n)}(\xx,\xx',\xx_3,\xx_4),
\eeq
where the integration kernel
\beal
 K_{ab}(\xx_1,\xx_2,\xx_3,\xx_4) &= J^2 (q-\delta_{ab})G_a(\xx_{13})G_a(\xx_{24})L_{ab}(\xx_{34}) \\
 \mathrm{with} \quad &L_{ab}(\xx) = \frac{\L G_+(\xx)G_-(\xx) \R^q}{G_a(\xx)G_b(\xx)}.
\eeal
 Symbolically we shall express the convolution in eq. \eqref{eq:ladder_kernel} together as $\mathf^{(n+1)}_{ab} = \sum_c K_{ac}*\mathf^{(n)}_{cb}$. Considering the convolution in Eq. \eqref{eq:ladder_kernel} as matrix multiplication, the sum over ladders can be obtained from a matrix inversion by summing over the geometric series,
\beq
 \label{eq:ladder_sum}
 \mathf_{ab} = \sum_{n=0}^{\infty} \sum_{c=\pm} \l K^n \r _{ac} * \mathf^{(0)}_{cb} = \sum_{c=\pm} \l \frac{1}{1-K} \r _{ac} * \mathf^{(0)}_{cb}.
\eeq
To perform the inversion one must show that $K$ is diagonalisable and that its eigenvalues are less than unity.

\subsection{Conformal symmetry and reflection sector}\label{subsec:conf_symmetry}

 The diagrammatics and integration kernels discussed so far are valid for any energy and coupling strength. We now focus on the IR limit where an analytic expression of the kernels can be obtained from the conformal ansatz of two-point function. In the conformal limit the four kernels can be reduced to a factorised form of chiral functions using the conformal two-point functions,
\beq
 K_{ab} = \frac{1}{\alpha_{ab}} K^{+}_{ab}(x_i^+) K^{-}_{ab}(x_i^-) \quad \mathrm{with} \quad \frac{1}{\alpha_{ab}} = (q-\delta_{ab}) \alpha_0^2.
\eeq
 The functional forms of these chiral factors are presented in appendix \ref{app:conf_kernel}. Notice that the kernels are independent of the SYK interaction strength $J$ since $\alpha_0$ has no explicit dependence on $J$. Our goal is to compute the four-point function using these chiral factors of the kernels in the conformal limit. We have to bring each $K_{ab}$ to a diagonal form in order to perform the sum in Eq. \eqref{eq:ladder_sum}. To achieve this, we must find a complete basis of eigenfunctions for the conformal kernel and express $\mathf_{ab}$ as a sum over these functions.
 
 The two-point functions expressed in Eq. \eqref{eq:two-point_fn} are invariant under conformal transformations. In two dimensions, the global conformal group acts as \iso up to a discrete quotient. We can choose the set of generators as
 \beq
  P_{\pm} = \partial_{\pm}, \quad D_\pm = - x^{\pm} \partial_{\pm} - \frac{\dpm}{2}, \quad \text{and} \quad K_\pm = (x^\pm)^2 \partial_{\pm} + x^\pm \dpm,
 \eeq
 where $\partial_{\pm}$ denotes $\partial/\partial x^\pm$. The generators annihilate the two-point functions and satisfy the commutation relations
 \beq
  [P_\pm,D_\pm] = -P_\pm , \quad [P_\pm,K_\pm] = -2 D_\pm \quad \text{and} \quad [D_\pm,K_\pm] = -K_\pm .
 \eeq
 The kernels are invariant operators under the global conformal group, which can be argued from the fact that the generators, acting on two spacetime points, commute with the kernels upto total derivative terms
 \beq
 \label{eq:kernel_commutes_conf}
  L^{(1+2)} * K_{ab}(\xx_1,\xx_2,\xx_3,\xx_4) = K_{ab}(\xx_1,\xx_2,\xx_3,\xx_4) * L^{(3+4)} ,
 \eeq
 where $L^{(i+j)} = L(\xx_i) + L(\xx_j)$ with $L$ representing the generators $P_\pm,D_\pm$ and $K_\pm$. Since $\mathf^{(0)}$ is also \iso invariant, the ladder contributions respect the conformal symmetry and can be expressed as simple powers times a function of the cross ratios,
 \beq
  \chpm = \frac{x^{\pm}_{12} x^{\pm}_{34}}{x^{\pm}_{13} x^{\pm}_{24}}.
 \eeq
 This allows us to consider the kernels as functions of cross ratios $K_{ab}(\chp,\chm;\tilde{\chi}^+,\tilde{\chi}^- )$. Since the kernels can be factorised into chiral parts $K^{\pm}_{ab}$, we can write
 \beq
  K_{ab}(\chp,\chm;\tilde{\chi}^+,\tilde{\chi}^- ) = K^{+}_{ab}(\chp,\tilde{\chi}^+) K^{-}_{ab}(\chm,\tilde{\chi}^-).
 \eeq
 The Casimir operator $C^{(1+2)}$ constructed from the generators acting on two spacetime points $\xx_1$ and $\xx_2$ is expressed as,
 \beal
 \label{eq:Casimir_1+2}
  &\quad C^{(1+2)} = C^{(1+2)}_+ + C^{(1+2)}_- , \quad \text{where} \\
  C^{(1+2)}_\pm = \l D^{(1)}_\pm + D^{(2)}_\pm \r ^2 &-\frac{1}{2} \l K^{(1)}_\pm + K^{(2)}_\pm \r \l P^{(1)}_\pm + P^{(2)}_\pm \r -\frac{1}{2} \l P^{(1)}_\pm + P^{(2)}_\pm \r \l K^{(1)}_\pm + K^{(2)}_\pm \r .
 \eeal
 Naturally, the Casimir is also separable into chiral parts. Each of the chiral differential operators $C^{(1+2)}_\pm $ admits a family of eigenfunctions expressed as simple powers times a function of $\chpm$. Unlike the 1D case, the spectrum of the Casimir in 2D is degenerate. We need the largest set of operators that commute with the kernels to determine a complete set of eigenfunctions representing the sum over ladders. In Sec. \ref{subsec:casimir_eigenfunctions}, we determine a complete set from the simultaneous eigenfunctions of the Casimir and the chiral reflection operator.

 In the conformal limit, the ladder contribution $\mathf^{(n)}_{ab}$ transforms as a four-point function of fields with conformal weights $(\dpl/2,\dmi/2)$ on $(x^a,x^{\bar{a}})$,
 \beq
  \label{eq:F_cross_ratio}
  \mathf_{ab}^{(n)}(\xx_1,\dots,\xx_4) = G_a(\xx_{12})G_b(\xx_{34}) \mathf_{ab}^{(n)}(\chp,\chm).
 \eeq
 The \iso invariance, the antisymmetry under $\xx_1 \leftrightarrow \xx_2$ and $\xx_3 \leftrightarrow \xx_4$, and the symmetry under $(\xx_1,\xx_2) \leftrightarrow (\xx_3,\xx_4)$ together allow us to set $x_1^\pm = 0$, $x_3^\pm = 1$, and $x_4^\pm \to \infty$, leaving $x_2^\pm = \chpm >0 $. Furthermore, the antisymmetry under $\xx_1 \leftrightarrow \xx_2$ translates to the symmetry of $\mathf_{ab}(\chp,\chm)$ under the simultaneous substitution $\chpm \to \frac{\chpm}{\chpm - 1}$,
  \beq
   \mathf_{ab}^{(n)}(\chp,\chm) =  \mathf_{ab}^{(n)} \l \frac{\chp}{\chp - 1},\frac{\chm}{\chm - 1} \r .
  \eeq
   This transformation maps the regions $0<\chpm<1$ and $1<\chpm<2$ to the regions $-\infty<\chpm<0$ and $2<\chpm<\infty$, respectively, on both lightcones at once. Its only fixed point with $\chpm>0$ is $\chp=\chm=2$; the lines $\chp=2$ and $\chm=2$ are mapped into themselves, but not pointwise. Since the two cross ratios are reflected together, the region $0<\chpm<2$ is not a fundamental domain: the regions where exactly one of $\chpm$ lies outside $(0,2)$ are mapped onto each other rather than onto $0<\chpm<2$. It is therefore convenient to decompose the ladder contributions at every order into
 \beq
 \label{eq:sectors}
  \mathf_{ab}=\mathf^{\ee}_{ab}+\mathf^{\oo}_{ab},\qquad
  \mathf^{\sigma}_{ab}(\chp,\chm)=\frac{1}{2}\L \mathf_{ab}(\chp,\chm) +\tau \mathf_{ab}\l\frac{\chp}{\chp-1},\chm\r\R ,
 \eeq
where $\tau = +1$ for $\sigma=\ee$ and $\tau = -1$ for $\sigma=\oo$. By the symmetry above, reflecting $\chm$ instead of $\chp$ gives the same functions. $\mathf^{\ee}_{ab}$ ($\mathf^{\oo}_{ab}$) is even (odd) under the reflection $R_{\pm}:\chpm\to\chpm/(\chpm-1)$ of each cross ratio separately,
\beq
\label{eq:reflection_criteria}
 R_{\pm} \mathf_{ab}^{\ee} = \mathf_{ab}^{\ee} \quad \text{and} \quad R_{\pm} \mathf_{ab}^{\oo} = -\mathf_{ab}^{\oo},
\eeq
and each of them is determined by its values in the region $0<\chpm\le2$. The reflection of a single cross ratio keeps the line $\chpm=2$ fixed pointwise, so that
 \beq
 \label{eq:boundary_condition_at2}
  \partial_{\chpm}\mathf^{\ee}_{ab}\big|_{\chpm=2}=0 \qquad \text{(Neumann)}, \qquad \mathf^{\oo}_{ab}\big|_{\chpm=2}=0 \qquad \text{(Dirichlet)} .
 \eeq
These properties translate to the eigenfunctions of the Casimir operator expressed as functions of the cross ratios. The kernels $K_{ab}$ commute with each reflection operator $R_\pm$ separately (Appendix \ref{app:kernel_cross_ratio}), so the two sectors evolve independently, and both are sourced by the zero-rung ladder. Furthermore, the integration kernels $K_{ab}$ become functions of fewer variables, leading to a simpler form of Eq. \eqref{eq:ladder_kernel} as
 \beq
 \label{eq:ladder_cross_ratio}
  \mathf^{(n+1),\sigma}_{ab}(\chp,\chm) = \int_0^2 d \chi'^+ \int_0^2 d \chi'^- \sum_{c=\pm} K^\sigma_{ac}(\chp,\chm;\chi'^+,\chi'^-)\, \mathf^{(n),\sigma}_{cb}(\chi'^+,\chi'^-),
 \eeq
 where the two-sector kernels as functions of cross ratios are presented in Appendix \ref{app:kernel_cross_ratio}.

\subsection{Eigenfunctions of the Casimir}\label{subsec:casimir_eigenfunctions}
 We proceed to determine the eigenfunctions of the kernels that satisfy the properties discussed in Sec. \ref{subsec:conf_symmetry}. From Eq. \eqref{eq:Casimir_1+2}, we can observe that the Casimir acting on two spacetime coordinates $C^{(1+2)}$ acts as
 \beq
  C^{(1+2)} \l \frac{1}{|x_{12}^+|^{\dpm} |x_{12}^-|^{\Delta_\mp}}\r \Psi(\chp,\chm) = \frac{1}{|x_{12}^+|^{\dpm} |x_{12}^-|^{\Delta_\mp}} \cc \; \Psi(\chp,\chm),
 \eeq
 on functions of cross ratios, where
 \beq
  \cc = \cc_+ + \cc_- \quad \text{with} \quad \cc_\pm = (\chpm)^2 (1-\chpm)\partial_{\chpm}^2 -(\chpm)^2 \partial_{\chpm}.
 \eeq
 The functions $\Psi(\chp,\chm)$ are simultaneous eigenfunctions of the chiral Casimir operators $\cc_\pm $,
 \beq
\label{eq:casimir_ode}
  \cc_+ \Psi = \hpl(\hpl -1) \Psi \quad \text{and} \quad \cc_- \Psi = \hmi(\hmi -1) \Psi.
 \eeq
 The differential equation with $\cc_+$ has two linearly independent solutions $F_{\hpl}(\chp)$ and $F_{1-\hpl}(\chp)$, where
 \beq
 \label{eq:first_ff}
  F_h(\chi) = \chi^h \ff(h,h,2h,\chi).
 \eeq
 Likewise, the equation for $\cc_-$ has linearly independent solutions $F_{\hmi}(\chm)$ and $F_{1-\hmi}(\chm)$. The general solution of $\Psi$ is a linear combination of $F_{\hpl}(\chp)F_{\hmi}(\chm)$ and three other terms with $\hpl$ replaced by $1-\hpl$, $\hmi$ replaced by $1-\hmi$, or both.

 \paragraph{Reflection sector.} As discussed in Sec. \ref{subsec:conf_symmetry}, the four-point function is invariant only under simultaneous reflection $\l \chp,\chm \r \to \l \frac{\chp}{\chp-1},\frac{\chm}{\chm-1} \r$. The chiral reflection operators $R_\pm$ commute with the Casimir $\cc_\pm$. The four-point function has even and odd sectors under each chiral reflections. The kernels commute with $R_{\pm}$ as well. The two sectors evolve independently under the kernels (Appendix \ref{app:kernel_cross_ratio}), and both are sourced by the zero-rung ladder. The eigenbasis of the kernels is therefore the simultaneous eigenbasis of $\cc_+,\cc_-$ and $R_\pm$,
 \beq
 \label{eq:eigenfunction_factorised}
  \Psi^{\sigma}_{(\hpl,\hmi)}(\chp,\chm) = \varphi_{\hpl}^\sigma (\chp) \varphi_{\hmi}^\sigma (\chm), \qquad \sigma =\ee,\oo ,
 \eeq
 where we have expressed the eigenfunctions in factorised forms.

 \paragraph{Boundary conditions.}
 The eigenfunctions should be normalisable with respect to the inner product defined as
 \beq
 \label{eq:def_inner_product}
  \la \Psi^{\sigma}, \Phi^{\sigma} \ra = \int_0^2 \frac{d\chp}{{\chp}^2} \int_0^2 \frac{d\chm}{{\chm}^2} {\Psi^{\sigma}}^* (\chp,\chm) \Phi^{\sigma}(\chp,\chm),
 \eeq
 whose integration region is restricted to $0 < \chpm \le 2$ by Eq. \eqref{eq:ladder_cross_ratio}, and the functions of different sectors are orthogonal to each other. This inner product makes the Casimir $\cc$ Hermitian up to boundary terms. Imposing that those boundary terms vanish therefore makes $\cc$ self-adjoint on the resulting domain, and its eigenfunctions form a complete set. Since $\cc_+$ and $\cc_-$ are independent, Hermiticity is imposed on each of them, which restricts the eigenvalues $\hpm(\hpm-1)$ to be real, and requires
 \beq
   0 = \la \Psi^{\sigma}, \cc_a \Phi^{\sigma} \ra - \la \cc_a \Psi^{\sigma}, \Phi^{\sigma} \ra = \int_0^2 \frac{d\chi^{\bar{a}}}{{\chi^{\bar{a}}}^2} \int_0^2 d\chi^a \partial_{\chi^a}\Big[ (1-\chi^a) \l {\Psi^{\sigma}}^* \partial_{\chi^a}\Phi^{\sigma} - \partial_{\chi^a}{\Psi^{\sigma}}^{*} \Phi^{\sigma} \r \Big] .
 \eeq
 The boundary terms arise at $\chpm=0,1^\mp \text{ and } 2 $. It follows from Eq. \eqref{eq:boundary_condition_at2} that the boundary terms vanish for both sectors at $\chpm=2$. Near $\chpm = 0$ the boundary terms vanish if the eigenfunctions decay at least as fast as $(\chp \chm)^{1/2}$. The singular contributions at $\chp = 1$ and at $\chm=1$ will cancel out provided that the logarithmic and constant terms match as we approach $\chpm = 1^-$ and $\chpm = 1^+$.

 \paragraph{The eigenfunctions.} We find the solutions for $\Psi^{\sigma}$ that satisfy these boundary conditions. It is convenient to express the solution in factorised form as Eq. \eqref{eq:eigenfunction_factorised}. For $\chi>1$ the even and odd solutions are
 \beq
  \tilF_h(\chi) = \ff \l \frac{h}{2},\frac{1-h}{2},\frac{1}{2},\frac{(2-\chi)^2}{{\chi}^2} \r , \quad \tilG_h(\chi) = \frac{2-\chi}{\chi} \; \ff \l \frac{1+h}{2},1-\frac{h}{2},\frac{3}{2},\frac{(2-\chi)^2}{{\chi}^2} \r,
 \eeq
 which satisfy the Neumann and Dirichlet boundary conditions at $\chi=2$, respectively. Matching the logarithmic and constant terms at $\chi=1$ to the solutions in Eq. \eqref{eq:first_ff} on $0<\chi<1$ gives the chiral factors
 \beq
 \label{eq:varphi_expr_even}
  \varphi_{h}^{\ee}(\chi) = \begin{cases}
      \frac{1}{\sqrt{\pi}} \; \Gamma \l \frac{h}{2} \r \Gamma \l \frac{1}{2} - \frac{h}{2} \r \tilF_h (\chi) \quad \text{for} \quad \chi >1 , \\
      A^\ee(h) F_h(\chi) + B^\ee(h) F_{1-h}(\chi) \quad \text{for} \quad \chi <1 ,
  \end{cases}
 \eeq
 \beq
 \label{eq:varphi_expr_odd}
  \varphi_{h}^{\oo}(\chi) = \begin{cases}
      \frac{2}{\sqrt{\pi}} \; \Gamma \l \frac{1+h}{2} \r \Gamma \l 1 - \frac{h}{2} \r \tilG_h (\chi) \quad \text{for} \quad \chi >1 , \\
      A^\oo(h) F_h(\chi) + B^\oo(h) F_{1-h}(\chi) \quad \text{for} \quad \chi <1 ,
  \end{cases}
 \eeq
 where the coefficients are
 \beq
  A^\ee(h) = \frac{1}{2} \tan (\pi h) \cot \l \frac{\pi h}{2} \r \frac{\Gamma(h)^2}{\Gamma(2h)} \quad \text{and} \quad B^\ee(h) = -\frac{1}{2} \tan (\pi h) \tan \l \frac{\pi h}{2} \r \frac{\Gamma(1-h)^2}{\Gamma(2-2h)},
 \eeq
 \beq
  A^\oo(h) = -\frac{1}{2} \tan (\pi h) \tan \l \frac{\pi h}{2} \r \frac{\Gamma(h)^2}{\Gamma(2h)} \quad \text{and} \quad B^\oo(h) = \frac{1}{2} \tan (\pi h) \cot \l \frac{\pi h}{2} \r \frac{\Gamma(1-h)^2}{\Gamma(2-2h)}.
 \eeq
  The detailed computation of $\Psi$ and the expressions of the coefficients are presented in Appendix \ref{app:eigenbasis_of_Casimir}. The even sector reproduces the eigenfunctions of 1D SYK model presented in reference \cite{Maldacena_2016}. The relation $A^\sigma(1-h) = B^\sigma(h)$ imply that in each sector,
 \beq
  \Psi^{\sigma}_{(\hpl,\hmi)} = \Psi_{(\hpl,1-\hmi)}^{\sigma} =\Psi_{(1-\hpl,\hmi)}^{\sigma} = \Psi_{(1-\hpl,1-\hmi)}^{\sigma}.
 \eeq

 \paragraph{The eigenbasis.} The vanishing of the boundary terms at $\chpm = 0$ requires $\varphi_{\hpm}^\sigma(\chpm)$ to decay at least as fast as ${\chpm}^{1/2}$. As in the 1D case there are two kinds of solutions. The continuum $h = \frac{1}{2} + is $ is marginally allowed in both sectors. The other kind of states are discrete states with real $h \geq 1$ where the ${\chi}^{1-h}$ term is absent. These are $B^\ee(h) = 0$, at $h = 2n$ in the even sector, and $B^\oo(h) = 0 $ at $h = 2n - 1$ in the odd sector. The complete basis of eigenfunctions therefore consists of two sectors, each with four kinds of states,
 \beal
 \label{eq:eigenbasis}
   \ee : (\hpl,\hmi) &\in \Big\{ \l \tfrac{1}{2}+is_+ ,\tfrac{1}{2}+is_- \r,\l \tfrac{1}{2}+is_+ ,2n_- \r,\l 2n_+ ,\tfrac{1}{2}+is_- \r, (2n_+,2n_-) \Big\}, \\
   \oo : (\hpl,\hmi) &\in \Big\{ \l \tfrac{1}{2}+is_+ ,\tfrac{1}{2}+is_- \r,\l \tfrac{1}{2}+is_+ ,2n_- - 1 \r,\l 2n_+ -1 ,\tfrac{1}{2}+is_- \r, (2n_+ -1,2n_- -1) \Big\},
 \eeal
 where $\mathf_{ab}^{\ee}$ and $\mathf_{ab}^{\oo}$ are expanded in the first and second set, respectively, with $n_\pm = 1,2,3,\dots $. These states are different from the states obtained in reference \cite{Murugan_2017}, where Euclidean coordinates were used with the symmetry group SL$(2,\mathbb{C})$. Finally, we note for later use that there is an integral representation of the chiral factors $\varphi_{h}(\chi)$ for the continuum states,
 \beal
  \label{eq:eigen_integral_rep}
  \varphi_h^{\ee}(\chi) &= \frac{1}{2} \int_{-\infty}^{\infty} dy \; \frac{ |\chi|^{h}}{|y|^{h}|\chi - y|^{h}|1-y|^{1-h}}, \\
  \varphi_h^{\oo}(\chi) &= \frac{1}{2} \sgn{\chi} \int_{-\infty}^{\infty} dy \; \frac{ \sgn{y} \sgn{\chi-y} \sgn{1-y} |\chi|^{h}}{|y|^{h}|\chi - y|^{h}|1-y|^{1-h}},
 \eeal
 which works for all $\chi>0$ but does not converge for the discrete states.

 \paragraph{Legendre forms.}
 It is convenient to use Legendre $P_{\nu}(y)$ and $Q_{\nu}(y)$ functions to represent the eigenfunctions in some scenarios discussed later in this paper. In terms of $y=2/\chi-1$ one has $\chi^2\partial_\chi=-2\partial_y$ and $(1-\chi)\partial_\chi=\frac12(1-y^2)\partial_y$, so that $\cc_\pm=-\partial_y(1-y^2)\partial_y$ and Eq. \eqref{eq:casimir_ode} becomes Legendre's differential equation of degree $h-1$,
 \beq
 \label{eq:legendre_de}
  (1-y^2)\,\varphi''-2y\,\varphi'+h(h-1)\,\varphi=0 .
 \eeq
 The reflection $\chi\to\chi/(\chi-1)$ is $y\to-y$, $d\chi/\chi^2=-dy/2$, and $\chi=0,1,2,\infty$ are $y=\infty,1,0,-1$. For $\chi>1$, i.e. $-1<y<1$, the solutions of Eq. \eqref{eq:legendre_de} that are regular at $y=1$ and at $y=-1$ are $P_{h-1}(y)=\ff\l h,1-h,1,\frac{\chi-1}{\chi}\r$ and $P_{h-1}(-y)=\ff\l h,1-h,1,\frac1\chi\r$ \cite{NIST:DLMF}. They are exchanged by $y\to-y$, so Eq. \eqref{eq:reflection_criteria} requires
 \beq
 \label{eq:eigen_legendre_ge1}
  \varphi^\sigma_h(\chi)=\frac{\pi}{\sin\pi h}\L P_{h-1} \l 1- \frac{2}{\chi} \r +\tau\,P_{h-1} \l \frac{2}{\chi} - 1 \r \R, \quad \chi>1, \quad \tau=+1 \, (-1) \; \text{for} \; \sigma=\ee \, (\oo) .
 \eeq
 where the prefactor makes the coefficient of $-\log(\chi-1)$ at $\chi\to1^+$ equal to one, which is expected from the $\chi>1$ expressions of $\varphi^\sigma_h(\chi)$ in Eqs. \eqref{eq:varphi_expr_even} and \eqref{eq:varphi_expr_odd}. For $0<\chi<1$, i.e. $y>1$, the solution that decays at $\chi=0$ is $Q_{h-1}(y)$. Matching the logarithmic and constant terms at $\chi=1$ gives
 \beal
 \label{eq:varphi_Legendre_form}
  \varphi^\ee_h(\chi) &= 2 \, Q_{h-1} \l \frac{2}{\chi} - 1 \r + \pi \tan \l \frac{\pi h}{2} \r P_{h-1}\l \frac{2}{\chi} - 1 \r , \\
  \varphi^\oo_h(\chi) &= 2 \, Q_{h-1} \l \frac{2}{\chi} - 1 \r - \pi \cot \l \frac{\pi h}{2} \r P_{h-1}\l \frac{2}{\chi} - 1 \r , \quad 0<\chi<1 .
 \eeal
 The Legendre function $Q_{h-1}(y)$ has a branch cut along $-1<y<1$. The expression of $\varphi^\sigma_h$ in Eq. \eqref{eq:varphi_Legendre_form} can be extended for all $\chi>0$ replacing $Q_{h-1}$ by $\text{Re}\,[Q_{h-1}(y)] = \text{Re}\,[Q_{h-1}(y \pm \ii 0)]$. For the discrete states in the eigenbasis, $\varphi^\sigma_h$ can be reduced to
 \beq
 \label{eq:varphi_discrete_states}
  \varphi^\sigma_h (\chi) = 2 \, \text{Re}\, \L Q_{h-1} \l \frac{2}{\chi}-1 \r \R, \quad h=2n \; \text{for} \; \sigma=\ee, \quad h=2n-1 \; \text{for} \; \sigma=\oo , \quad n \in \mathbb{N}.
 \eeq
 For the even sector, this matches exactly with the result stated in reference \cite{Maldacena_2016} for the 1D SYK model.

 \paragraph{Normalisation.}
 Normalisation of the eigenfunctions $\la \Psi^\sigma_{(\hpl,\hmi)}, \Psi^\sigma_{(\hpl',\hmi')} \ra$ is a product of the normalisation of its chiral factors,
 \beq
  \label{eq:defn_chiral_norm}
  \la \varphi^\sigma_{h}, \varphi^\sigma_{h'} \ra = \int_0^2 \frac{d\chi}{\chi^2} {\varphi^\sigma_h}^* \varphi^\sigma_{h'},
 \eeq
 so that Eq. \eqref{eq:def_inner_product} factorises into chiral inner products. The normalisation of the continuum states is fixed by its singular behaviour near $\chi = 0$, as in reference \cite{Maldacena_2016}. For the discrete states, the chiral factors $\varphi^\sigma_h$ are expressed in terms of Legendre function $Q$ (Eq. \eqref{eq:varphi_discrete_states}), and the inner product is obtained using standard integral formulas for $Q$. In both sectors, the normalisations of the chiral factors are
 \beq
  \label{eq:normalisation_continuum}
  \la \varphi^\sigma_{h}, \varphi^\sigma_{h'} \ra = \frac{\pi \tan \pi h}{4h -2} \, 2\pi \, \delta(s-s') , \qquad h=\frac{1}{2} + \ii s , \quad h' =\frac{1}{2} + \ii s' \quad \text{with} \; s,s'>0,
 \eeq
 \beq
  \label{eq:normalisation_discrete}
  \la \varphi^\sigma_{h}, \varphi^\sigma_{h'} \ra = \frac{\pi^2}{4h -2} \delta_{hh'} , \quad h,h' = 2 n \; (\sigma = \ee), \quad h,h' =2n - 1 \; (\sigma = \oo), \quad \text{with} \; n=1,2,3,\dots .
 \eeq

\subsection{Eigenvalues of the kernel}\label{subsec:eigenvalues}

 In the previous subsection we found that the eigenfunctions of the Casimir factorise into chiral functions. We now compute the eigenvalues of the kernels. In 1D SYK model, the eigenfunctions of the Casimir acting on two points have the form of a conformal three-point function of two fermions and an operator of dimension $h$, and since this eigenfunction is unique, it is also an eigenfunction of the kernel. In two dimensions the first statement still holds, but the second does not. For given conformal dimension $(\hpl,\hmi)$, there are two conformally covariant three-point functions,
 \beal
 \label{eq:3pt_eigenfunctions}
  f^{a,\ee}_{\hpl,\hmi}(\xx_1,\xx_2,\xx_0) &= \frac{\sgn{x_{12}^{a}}}{|x_{10}^a|^{h_a} |x_{20}^a|^{h_a} |x_{12}^a|^{\dpl - h_a}} \frac{1} {|x_{10}^{\bar{a}}|^{h_{\bar{a}}} |x_{20}^{\bar{a}}|^{h_{\bar{a}}} |x_{12}^{\bar{a}}|^{\dmi - h_{\bar{a}}}} , \\
  f^{a,\oo}_{\hpl,\hmi}(\xx_1,\xx_2,\xx_0) &= \frac{\sgn{x_{10}^{a}x_{20}^{a}}} {|x_{10}^a|^{h_a} |x_{20}^a|^{h_a} |x_{12}^a|^{\dpl - h_a}} \frac{\sgn{x_{10}^{\bar{a}} x_{20}^{\bar{a}} x_{12}^{\bar{a}}}}{|x_{10}^{\bar{a}}|^{h_{\bar{a}}} |x_{20}^{\bar{a}}|^{h_{\bar{a}}} |x_{12}^{\bar{a}}|^{\dmi - h_{\bar{a}}}},
 \eeal
 where $\bar{a}=-a$. Both of these three-point functions are antisymmetric under the exchange of points $\xx_1 \leftrightarrow \xx_2$ but they differ in their parity under the exchange of only one lightcone coordinate. $f^{a,\ee}$ is odd under $x_1^a \leftrightarrow x_2^a$ and even under $x_1^{\bar{a}} \leftrightarrow x_2^{\bar{a}}$, whereas $f^{a,\oo}$ is even under $x_1^a \leftrightarrow x_2^a$ and odd under $x_1^{\bar{a}} \leftrightarrow x_2^{\bar{a}}$. The chiral factors of the kernels in Eq. \eqref{eq:chiral_kernels} commute or anticommute with these exchanges, $\scP_\pm K^\pm_{ab}=(-1)^{1-\delta_{ab}}K^\pm_{ab}\scP_\pm$, as shown in Appendix \ref{app:eigenvalues}. Since the kernels commute with the conformal generators (Eq. \eqref{eq:kernel_commutes_conf}), $K_{ab}f^{b,\sigma}$ is a conformal three-point function with the weights of $f^{a}$, i.e. a combination of $f^{a,\ee}$ and $f^{a,\oo}$, and the parities allow only the one of the same sector. Hence each of the three-point functions is an eigenfunction of the kernels,
\beq
\label{eq:eigenvalue_3pt}
 \int d^2\xx_3\,d^2\xx_4\;K_{ab}(\xx_1,\xx_2,\xx_3,\xx_4)\,f^{b,\sigma}_{\hpl,\hmi}(\xx_3,\xx_4,\xx_0)=\mathk^\sigma_{ab}(\hpl,\hmi)\,f^{a,\sigma}_{\hpl,\hmi}(\xx_1,\xx_2,\xx_0).
\eeq
 In two dimensions, the operator that has to be added to the Casimir to make the spectrum nondegenerate is the exchange $\scP_\pm$ of one lightcone coordinate. In cross ratios this is the reflection of one cross ratio in Sec. \ref{subsec:casimir_eigenfunctions}, and we show below that the two three-point functions generate the two reflection sectors.

 The eigenvalues depend only on $\hpm$, since the \iso\ symmetry allows us to move $\xx_0$. In particular, we take $x_0^\pm\to\infty$, where $x_{10}$ and $x_{20}$ have the same sign, so that $\sgn{x_{10}x_{20}}\to1$ and $\sgn{x_{10}x_{20}x_{12}}\to\sgn{x_{12}}$. The three-point functions become
 \beal
 \label{eq:3pt_limit}
  f^{a,\ee}_{\hpl,\hmi}(\xx,\xx',\xx_0)&\;\longrightarrow\;\frac{1}{|x_0^+|^{2\hpl}|x_0^-|^{2\hmi}}\; \frac{\sgn{x^a-x'^a}}{|x^a-x'^a|^{\dpl-h_a}\,|x^{\bar a}-x'^{\bar a}|^{\dmi-h_{\bar a}}},\\
  f^{a,\oo}_{\hpl,\hmi}(\xx,\xx',\xx_0)&\;\longrightarrow\;\frac{1}{|x_0^+|^{2\hpl}|x_0^-|^{2\hmi}}\; \frac{\sgn{x^{\bar a}-x'^{\bar a}}}{|x^a-x'^a|^{\dpl-h_a}\,|x^{\bar a}-x'^{\bar a}|^{\dmi-h_{\bar a}}} .
 \eeal
 We insert these in Eq. \eqref{eq:eigenvalue_3pt} with the factorised kernel of Eq. (3.7), set $x_1^\pm=1$, $x_2^\pm=0$ and rename $x_3\to x$, $x_4\to x'$. The common factor $|x_0^+|^{-2\hpl}|x_0^-|^{-2\hmi}$ cancels, and the eigenvalues in the two sectors are
 \beal
  \label{eq:eigenvalue_int}
   \mathk^\ee_{ab}(\hpl,\hmi)&=\frac{1}{\alpha_{ab}}\int d^2\xx\,d^2\xx'\;K^+_{ab}(1,0,x^+,x'^+)\,K^-_{ab}(1,0,x^-,x'^-)\;\frac{\sgn{x^b-x'^b}}{|x^b-x'^b|^{\dpl-h_b}\,|x^{\bar b}-x'^{\bar b}|^{\dmi-h_{\bar b}}},\\
   \mathk^\oo_{ab}(\hpl,\hmi)&=\frac{1}{\alpha_{ab}}\int d^2\xx\,d^2\xx'\;K^+_{ab}(1,0,x^+,x'^+)\,K^-_{ab}(1,0,x^-,x'^-)\;\frac{\sgn{x^{\bar b}-x'^{\bar b}}}{|x^b-x'^b|^{\dpl-h_b}\,|x^{\bar b}-x'^{\bar b}|^{\dmi-h_{\bar b}}} ,
 \eeal
 where $\bar b=-b$. As discussed in Appendix \ref{app:conf_kernel}, on the lightcone coordinate $x^a$ the diagonal kernels $(a=b)$ carry $\sgn{x^a_{13}}\sgn{x^a_{24}}$ and $\dpl$, and on the lightcone coordinate $x^{\bar a}$ it carries $\dmi$ and no such signs. The off-diagonal kernels $(b={\bar a})$ carry in addition $\sgn{x_{34}}$ and the exponent $2-2\Delta$ on both lightcone coordinates, and these are compensated by the sign and the weight of $f^b$. In all cases the exponent of $|x-x'|$ is $2-\dpl-h_a$ on the lightcone $x^a$ and $2-\dmi-h_{\bar a}$ on the lightcone $x^{\bar a}$, and the factor $\sgn{x-x'}$ appears on the lightcone $x^a$ in the even sector and on the lightcone $x^{\bar a}$ in the odd sector. Therefore, only four chiral double integrals constitute the kernel eigenvalues,
 \beq
  \mathk^\sigma_{ab}(\hpl,\hmi) = \frac{1}{\alpha_{ab}}\,k^\sigma_{\dpl}(h_a)\,\tilde{k}^\sigma_{\dmi}(h_{\bar a}),
 \eeq
 where the chiral double integrals in the even sector are
 \beal
\label{eq:eigen_double_int_even}
 k^\ee_\Delta(h)&=\int_{-\infty}^{\infty}dx\int_{-\infty}^{\infty}dx'\;\frac{\sgn{1-x}\sgn{-x'}\sgn{x-x'}}{|1-x|^{\Delta}\,|x'|^{\Delta}\,|x-x'|^{2-\Delta-h}},\\
 \tilde{k}^\ee_\Delta(h)&=\int_{-\infty}^{\infty}dx\int_{-\infty}^{\infty}dx'\;\frac{1}{|1-x|^{\Delta}\,|x'|^{\Delta}\,|x-x'|^{2-\Delta-h}},
\eeal
 and in the odd sector they are
\beal
\label{eq:eigen_double_int_odd}
 k^\oo_\Delta(h)&=\int_{-\infty}^{\infty}dx\int_{-\infty}^{\infty}dx'\;\frac{\sgn{1-x}\sgn{-x'}}{|1-x|^{\Delta}\,|x'|^{\Delta}\,|x-x'|^{2-\Delta-h}},\\
 \tilde{k}^\oo_\Delta(h)&=\int_{-\infty}^{\infty}dx\int_{-\infty}^{\infty}dx'\;\frac{\sgn{x-x'}}{|1-x|^{\Delta}\,|x'|^{\Delta}\,|x-x'|^{2-\Delta-h}} .
\eeal
 The integrals on the lightcone $x^a$ are $k^\sigma_{\dpl}(h_a)$ and those on the lightcone $x^{\bar a}$ are $\tilde{k}^\sigma_{\dmi}(h_{\bar a})$. We evaluate them using the Fourier transform identities
\beq
\label{eq:Fourier_identity1}
 \frac{\sgn{x}}{|x|^r} = \int_{-\infty}^{\infty} \frac{dk}{2\pi}\,e^{-\ii kx}\,c(r)\,|k|^{r-1} \sgn{k}, \quad c(r) = 2^{1-r}\, \ii \sqrt{\pi}\, \frac{\Gamma \l 1-\frac{r}{2} \r }{\Gamma \l \frac{1}{2}+\frac{r}{2} \r}=2\ii\,\Gamma(1-r)\cos\frac{\pi r}{2} ,
\eeq
\beq
\label{eq:Fourier_identity2}
 \frac{1}{|x|^r} = \int_{-\infty}^{\infty} \frac{dk}{2\pi}\,e^{-\ii kx}\,d(r)\,|k|^{r-1}, \quad d(r) = 2^{1-r} \sqrt{\pi}\, \frac{\Gamma \l \frac{1}{2}-\frac{r}{2} \r }{\Gamma \l \frac{r}{2} \r}=2\,\Gamma(1-r)\sin\frac{\pi r}{2} .
\eeq
 The integrals converge for $1-\Delta<\text{Re}(h)<\Delta$. For $\Delta < \tfrac{1}{2}$ they are defined by analytic continuation from $\Delta > \tfrac{1}{2}$, as in the 1D SYK model. We obtain the factors $k$ and $\tilde{k}$ in the two sectors as
\beal
\label{eq:factors_eigen}
 k^\ee_\Delta(h)&=-\frac{c(2-\Delta-h)}{c(\Delta-h)}\L c(\Delta)\R^2, &\qquad \tilde{k}^\ee_\Delta(h)&=\frac{d(2-\Delta-h)}{d(\Delta-h)}\L d(\Delta)\R^2,\\
 k^\oo_\Delta(h)&=-\frac{d(2-\Delta-h)}{d(\Delta-h)}\L c(\Delta)\R^2, &\qquad \tilde{k}^\oo_\Delta(h)&=\frac{c(2-\Delta-h)}{c(\Delta-h)}\L d(\Delta)\R^2 .
\eeal
 The prefactors $-[c(\Delta)]^2$ and $[d(\Delta)]^2$ come from the kernel and are the same in both sectors. Moving $\sgn{x-x'}$ from one lightcone to the other exchanges the ratio of $c$'s for the ratio of $d$'s. Therefore, the eigenvalues in the two sectors are
\beal
\label{eq:eigenvalues}
 \mathk^\sigma_{ab}(\hpl,\hmi) &= \frac{1}{\alpha_{ab}}\,k^\sigma_{\dpl}(h_a)\,\tilde{k}^\sigma_{\dmi}(h_{\bar a})\\
 &=\alpha_0^2 \begin{pmatrix}
    (q-1)\,k^\sigma_{\dpl}(\hpl)\,\tilde{k}^\sigma_{\dmi}(\hmi) & q \, k^\sigma_{\dpl}(\hpl)\, \tilde{k}^\sigma_{\dmi}(\hmi) \\
    q \, k^\sigma_{\dpl}(\hmi)\,\tilde{k}^\sigma_{\dmi}(\hpl) & (q-1)\, k^\sigma_{\dpl}(\hmi)\, \tilde{k}^\sigma_{\dmi}(\hpl)
  \end{pmatrix} .
\eeal
 A detailed computation of these results is presented in Appendix \ref{app:eigenvalues}. Since $c(\Delta)^2/d(\Delta)^2=-\cot^2\frac{\pi\Delta}{2}$, the functions of the two sectors are related by
\beq
\label{eq:odd_even_relation}
 k^\oo_\Delta(h)=\cot^2\frac{\pi\Delta}{2}\;\tilde{k}^\ee_\Delta(h),\qquad \tilde{k}^\oo_\Delta(h)=\tan^2\frac{\pi\Delta}{2}\;k^\ee_\Delta(h).
\eeq
 For $j=0$ this gives $\mathk^\oo_{ab}=\mathk^\ee_{\bar a\bar b}$, so that for non-chiral fermions the two sectors have the same eigenvalues and the same determinant. At integer $h$ one also has $k^\oo_\Delta(h)=-k^\ee_\Delta(h)$ and $\tilde{k}^\oo_\Delta(h)=-\tilde{k}^\ee_\Delta(h)$, so that $\mathk^\oo_{ab}=\mathk^\ee_{ab}$ whenever $\hpl$ and $\hmi$ are both integers.

 The three-point functions in Eq. \eqref{eq:3pt_eigenfunctions} are related to the eigenfunctions of the Casimir, in both sectors, by
\beq
\label{eq:rel_eigenfunc}
 G_a(\xx_{12})\,G_b(\xx_{34})\,\Psi^\sigma_{(\hpl,\hmi)}(\chp,\chm)=\frac{\mathtt{b}^2}{4}\int d^2\xx_0\;f^{a,\sigma}_{\hpl,\hmi}(\xx_1,\xx_2,\xx_0)\,f^{b,\sigma}_{1-\hpl,1-\hmi}(\xx_3,\xx_4,\xx_0) .
\eeq
 The right-hand side transforms under \iso\ in the same way as the four-point function of $\psi_a\psi_a\psi_b\psi_b$, so it can be evaluated in the frame $x^\pm_1=0$, $x^\pm_2=\chpm$, $x^\pm_3=1$, $x^\pm_4\to\infty$, $x^\pm_0=y^\pm$. In this frame the $x_0$ integral on each lightcone is the integral representation of $\varphi^\sigma_{\hpm}(\chpm)$ in Eq. \eqref{eq:eigen_integral_rep}. The two three-point functions therefore generate the two reflection sectors. Applying the kernel to the legs $(\xx_1,\xx_2)$ of Eq. \eqref{eq:rel_eigenfunc} and using Eq. \eqref{eq:eigenvalue_3pt},
 \beq
 \label{eq:kernel_on_Psi}
  \sum_{a}K_{ca}*\Big[G_a(\xx_{12})\,G_b(\xx_{34})\,\Psi^\sigma_{(\hpl,\hmi)}\Big]=\sum_a \mathk^\sigma_{ca}(\hpl,\hmi)\;G_c(\xx_{12})\,G_b(\xx_{34})\,\Psi^\sigma_{(\hpl,\hmi)} .
 \eeq
 Thus $\Psi^\sigma_{(\hpl,\hmi)}$ are eigenfunctions of the kernels, which act on the chirality indices through the matrix $\mathk_{ab}^\sigma(\hpl,\hmi)$, and different $(\hpl,\hmi)$ and different sectors are not mixed. The integral representation converges only for $0<\text{Re}\,\hpm<1$. For the discrete states, $\hpm=2n_\pm$ in the even sector and $\hpm=2n_\pm-1$ in the odd sector, the eigenvalues are the analytic continuation of Eq. \eqref{eq:factors_eigen}. Applying the kernel instead to the legs $(\xx_3,\xx_4)$ gives the eigenvalue at $(1-\hpl,1-\hmi)$, and consistency requires
 \beq
 \label{eq:eigen_symmetry}
  \mathk^\sigma_{ab}(1-\hpl,1-\hmi)=\mathk^\sigma_{ab}(\hpl,\hmi).
 \eeq
 This holds since each of the functions in Eq. \eqref{eq:factors_eigen} is invariant under $h\to1-h$ (Appendix \ref{app:eigenvalues}).

 For $(\hpl,\hmi)=(1,1)$, and equally at $(0,0)$, $(1,0)$ and $(0,1)$, we find that in both sectors
 \beq
 \label{eq:eigen_00}
  \left. \mathk^\sigma_{ab} \right|_{(\hpl,\hmi)=(0,0)}=-\begin{pmatrix}q-1&q\\ q&q-1\end{pmatrix},
 \eeq
 with the eigenvalues $1$ and $1-2q$. In the odd sector, the state corresponding to $(\hpl,\hmi) = (1,1)$ is a basis state, $\Psi^\oo_{(1,1)} = \log |1-\chp| \log |1-\chm|$. For $(\hpl,\hmi)=(2,0)$ and $(0,2)$, we find in both sectors
\beq
\label{eq:eigen_20}
 \left. \mathk^\sigma_{ab} \right|_{(\hpl,\hmi)=(2,0)} = \begin{pmatrix} \frac{(1-\Delta)(\Delta + j)}{\Delta(2-\Delta-j)} & \frac{\Delta +j}{\Delta(2-\Delta-j)} \\ \frac{\Delta - j}{\Delta(2-\Delta+j)}  & \frac{(1-\Delta)(\Delta - j)}{\Delta(2-\Delta+j)}  \end{pmatrix},
\eeq
\beq
\label{eq:eigen_02}
 \left. \mathk^\sigma_{ab} \right|_{(\hpl,\hmi)=(0,2)} = \begin{pmatrix} \frac{(1-\Delta)(\Delta - j)}{\Delta(2-\Delta+j)} & \frac{\Delta -j}{\Delta(2-\Delta+j)} \\ \frac{\Delta + j}{\Delta(2-\Delta -j)}  & \frac{(1-\Delta)(\Delta + j)}{\Delta(2-\Delta -j)}  \end{pmatrix}.
\eeq
 These three states with $(\hpl,\hmi) = (1,1),(2,0),(0,2)$ and their images under single or simultaneous substitution $\hpm \to 1- \hpm$ satisfy $\det(\mathbb{I}-\mathk^\sigma)=0$ identically. The modes $(2,0)$ and $(0,2)$ correspond to chiral reparametrisations, and $(1,1)$ corresponds to local Lorentz transformation (Sec. \ref{subsec:lorentz_mode}), which is treated as a gauge symmetry in reference \cite{Turiaci_2017}. Of these states only $(\hpl,\hmi)=(1,1)$ in the odd sector belongs to the basis of Eq. \eqref{eq:eigenbasis}. In Sec. \ref{subsec:operators} we show that there are no other discrete states at which $\det(\mathbb{I}-\mathk^\sigma)$ vanishes.

 \paragraph{Closed forms and special values.} Inserting the first forms of $c$ and $d$, we obtain the closed-form expressions,
 \beal
 \label{eq:chiral_k_tildek}
  k^\ee_\Delta(h)&=\frac{\pi\,\Gamma\l1-\frac\Delta2\r^2}{\Gamma\l\frac{1+\Delta}2\r^2}\,
 \frac{\Gamma\l\frac{h+\Delta}2\r\Gamma\l\frac{1-h+\Delta}2\r}{\Gamma\l\frac{3-h-\Delta}2\r\Gamma\l\frac{2+h-\Delta}2\r},&\quad
 \tilde{k}^\ee_\Delta(h)&=\frac{\pi\,\Gamma\l\frac{1-\Delta}2\r^2}{\Gamma\l\frac\Delta2\r^2}\,
 \frac{\Gamma\l\frac{h+\Delta-1}2\r\Gamma\l\frac{\Delta-h}2\r}{\Gamma\l\frac{2-h-\Delta}2\r\Gamma\l\frac{1+h-\Delta}2\r},\\
 k^\oo_\Delta(h)&=\frac{\pi\,\Gamma\l1-\frac\Delta2\r^2}{\Gamma\l\frac{1+\Delta}2\r^2}\,
 \frac{\Gamma\l\frac{h+\Delta-1}2\r\Gamma\l\frac{\Delta-h}2\r}{\Gamma\l\frac{2-h-\Delta}2\r\Gamma\l\frac{1+h-\Delta}2\r},&\quad
 \tilde{k}^\oo_\Delta(h)&=\frac{\pi\,\Gamma\l\frac{1-\Delta}2\r^2}{\Gamma\l\frac\Delta2\r^2}\,
 \frac{\Gamma\l\frac{h+\Delta}2\r\Gamma\l\frac{1-h+\Delta}2\r}{\Gamma\l\frac{3-h-\Delta}2\r\Gamma\l\frac{2+h-\Delta}2\r}.
 \eeal
 Under $h\to1-h$ the two Gamma functions in the numerator of each $h$-dependent factor are exchanged, and so are the two in the denominator, which proves Eq. \eqref{eq:eigen_symmetry}. The $h$-dependent factor of $k^\ee_\Delta$ and $\tilde{k}^\oo_\Delta$ has poles at $h=\Delta+2n+1$, $h=-\Delta-2n$ and zeros at $h=-\Delta+2m+3$, $h=\Delta-2m-2$, and that of $\tilde{k}^\ee_\Delta$ and $k^\oo_\Delta$ has poles at $h=\Delta+2n$, $h=-\Delta-2n+1$ and zeros at $h=-\Delta+2m+2$, $h=\Delta-2m-1$, with integers $n,m \geq 0$. 
 
 At $h=1$, and equally at $h=0$, the chiral functions of both sectors satisfy
 \beq
  \label{eq:alpha_0_as_k}
  \alpha_0^2\,k^\sigma_{\dpl}(1)\,\tilde{k}^\sigma_{\dmi}(1) = -1 .
 \eeq
 This is the normalisation $\alpha_0^2$ in Eq. \eqref{eq:alpha0} written in terms of the chiral functions. It allows us to eliminate $\alpha_0^2$ from the eigenvalue matrix in Eq. \eqref{eq:eigenvalues}. In what follows, $\mathk^\sigma_{ab}$ depends on $k^\sigma_\Delta(h)$ and $\tilde{k}^\sigma_\Delta(h)$ only through the ratios $k^\sigma_\Delta(h)/k^\sigma_\Delta(1)$ and $\tilde{k}^\sigma_\Delta(h)/\tilde{k}^\sigma_\Delta(1)$.

\subsection{Sum of all ladders}\label{subsec:summ_ladders}

 We can write the sum over all ladders in the basis of eigenfunctions of the Casimir determined in Sec. \ref{subsec:casimir_eigenfunctions}. The ladder contributions split into the parts $\mathf^{\ee}_{ab}$ and $\mathf^{\oo}_{ab}$ of Eq. \eqref{eq:sectors}, and the kernel does not mix them. Each part is expanded in the eigenbasis of its own sector, and from Eq. \eqref{eq:ladder_sum} the order $1/N$ contribution to the four-point function can be expressed as
 \beq
 \label{eq:ladder_sum0}
  \tilde{\mathf}_{ab} = \sum_{\sigma=\ee,\oo}\ \sum_{\hpl,\hmi}\Psi^\sigma_{(\hpl,\hmi)}\l\frac{1}{\mathbb{I}-\mathk^\sigma}\r_{ab}
  \frac{\la \mathf^{(0),\sigma}_{b},\Psi^\sigma \ra}{\la \Psi^\sigma,\Psi^\sigma \ra},
 \eeq
 where we have used $\mathf^{(0)}_{ab} = \delta_{ab}G_b G_b \mathf^{(0)}_{b}$ as the zero-rung ladder. The $h_\pm$ sums run over the complete basis (Eq. \eqref{eq:eigenbasis}) except the state $(\hpl,\hmi) = (1,1)$ in the odd sector which has a vanishing $\det(\mathbb{I}-\mathk^\oo)$. We introduced $\tilde{\mathf}$ to distinguish the sum in Eq. \eqref{eq:ladder_sum0} from the full $1/N$ contribution $\mathf$. We shall treat the state $(\hpl,\hmi) = (1,1)$ separately in Sec. \ref{subsec:lorentz_mode}. $\mathk^\sigma_{ab}$ are the eigenvalue matrices (Eq. \eqref{eq:eigenvalues}). The normalisation of $\Psi^\sigma_{(\hpl,\hmi)}$ is fixed by the normalisations of its chiral components $\varphi^\sigma_{\hpm}$ (Eqs. \eqref{eq:normalisation_continuum} and \eqref{eq:normalisation_discrete}).
 
 To determine the inner product $\la \mathf^{(0),\sigma}_{b},\Psi^\sigma\ra$, we write the zero-rung ladder as a function of the cross ratios,
 \beq
  \mathf^{(0)}_{bb}(\xx_1,\dots,\xx_4) = \mathtt{b}^2 \frac{\sgn{x_{12}^b}\sgn{x_{34}^b}}{|x_{12}^b|^{\dpl} |x_{34}^b|^{\dpl} |x_{12}^{\bar{b}}|^{\dmi} |x_{34}^{\bar{b}}|^{\dmi} } \mathf^{(0)}_{b}(\chp,\chm) \quad \text{with} \quad \bar{b} = -b,
 \eeq
 where
 \beq
 \label{eq:zero_rung_ladder}
  \mathf^{(0)}_{b}(\chp,\chm) = -\sgn{\chi^b} |\chi^b|^{\dpl} |\chi^{\bar{b}}|^{\dmi} - \sgn{\frac{\chi^b}{\chi^b -1}} \left| \frac{\chi^b}{\chi^b -1} \right|^{\dpl} \left| \frac{\chi^{\bar{b}}}{\chi^{\bar{b}}-1} \right|^{\dmi}.
 \eeq
 The two terms are exchanged by the simultaneous reflection $\chpm\to\chpm/(\chpm-1)$, but a single reflection maps each of them to a different function. So $\mathf^{(0)}_b$ has both an even and an odd part,
 \beal
 \label{eq:zero_rung_split}
  \mathf^{(0),\ee}_b(\chp,\chm)&=\frac{1}{2} \L \mathf^{(0)}_b(\chp,\chm)+\mathf^{(0)}_b\l\frac{\chp}{\chp-1},\chm\r \R,\\
  \mathf^{(0),\oo}_b(\chp,\chm)&=\frac{1}{2} \L \mathf^{(0)}_b(\chp,\chm)-\mathf^{(0)}_b\l\frac{\chp}{\chp-1},\chm\r \R,
 \eeal
 and $\la \mathf^{(0),\sigma}_b,\Psi^\sigma\ra$ in \eqref{eq:ladder_sum0} is the inner product of $\Psi^\sigma$ with the part of its own sector,
 \beq
 \label{eq:inner_product_sector}
  \la \mathf^{(0),\sigma}_b,\Psi^\sigma\ra=\int_0^2\frac{d\chp}{(\chp)^2}\int_0^2\frac{d\chm}{(\chm)^2}\,\mathf^{(0),\sigma}_b(\chp,\chm)\, \Psi^\sigma_{(\hpl,\hmi)}(\chp,\chm).
 \eeq
 The integrals in Eq. \eqref{eq:inner_product_sector} are evaluated in Appendix \ref{app:zero-rung_ladder}. In both sectors the result is given by the chiral eigenvalue functions,
 \beq
 \label{eq:inner_product_result}
  \la \mathf^{(0),\sigma}_{\pm}, \Psi^\sigma \ra = - \frac{1}{8} \; k^{\sigma}_{\dpl}(\hpm)\,\tilde{k}^{\sigma}_{\dmi}(h_{\mp})
 \eeq
 Thus, the inner product reduces to the kernel eigenvalue, similar to the 1D SYK model. This holds for the projection onto the eigenfunctions of the same sector, which is what enters in Eq. \eqref{eq:ladder_sum0}. Substituting \eqref{eq:inner_product_result} into \eqref{eq:ladder_sum0}, the order $1/N$ contribution to the four-point function becomes
 \beq
  \tilde{\mathf}_{ab} = -\frac{1}{8} \sum_{\sigma=\ee,\oo}\ \sum_{\hpl,\hmi}\Psi^\sigma_{(\hpl,\hmi)}\l\frac{1}{\mathbb{I} - \mathk^\sigma}\r_{ab} \frac{k^\sigma_{\dpl}(h_b)\,\tilde{k}^\sigma_{\dmi}(h_{\bar b})}{\la\Psi^\sigma,\Psi^\sigma\ra}.
 \eeq
 With the normalisations in Eqs. \eqref{eq:normalisation_continuum} and \eqref{eq:normalisation_discrete}, the sum over $(\hpl,\hmi)$ consists of four kinds of terms in each sector,
\beal
 \label{eq:ladder_sum_explicit}
 \tilde{\mathf}_{ab}=-\frac{1}{8} \sum_{\sigma} &\L \l \int_0^\infty \frac{ds_+}{2\pi}\frac{4\hpl-2}{\pi\tan\pi\hpl} +\sum_{\hpl}\frac{4\hpl-2}{\pi^2} \r  \l \int_0^\infty \frac{ds_-}{2\pi}\frac{4\hmi-2}{\pi\tan\pi\hmi}
 + \sum_{\hmi} \frac{4\hmi-2}{\pi^2} \r \R \\
 &\qquad \qquad \qquad \times\l\frac{1}{\mathbb{I}-\mathk^\sigma}\r_{ab}k^\sigma_{\dpl}(h_b)\tilde{k}^\sigma_{\dmi}(h_{\bar b})\,\Psi^\sigma_{(\hpl,\hmi)}(\chp,\chm),
\eeal
 where the sums run over the discrete states, $\hpm = 2n_\pm$ for $\sigma = \ee$ and $\hpm = 2n_\pm -1$ for $\sigma = \oo$ with $n_\pm = 1,2,3,\dots$. Since the chiral functions in $\mathk^\sigma_{ab}$ depend only on the index $a$, the matrix in Eq. \eqref{eq:ladder_sum_explicit} takes a simple form,
\beq
 \label{eq:resolvent_column}
 \l \frac{1}{\mathbb{I}-\mathk^\sigma}\r_{ab}k^\sigma_{\dpl}(h_b)\tilde{k}^\sigma_{\dmi}(h_{\bar b})=
 \frac{\delta_{ab}\,k^\sigma_{\dpl}(h_a)\tilde{k}^\sigma_{\dmi}(h_{\bar a})+\L q-(2q-1)\delta_{ab}\R \alpha_0^2 \prod_{c=\pm}k^\sigma_{\dpl}(h_c)\tilde{k}^\sigma_{\dmi}(h_c)}{\det\l\mathbb{I}-\mathk^\sigma\r},
\eeq
with
\beq
 \label{eq:det_ladder}
 \det\l\mathbb{I}-\mathk^\sigma\r=1-(q-1)\alpha_0^2\sum_{c=\pm}k^\sigma_{\dpl}(h_c)\tilde{k}^\sigma_{\dmi}(h_{\bar c})-(2q-1)\alpha_0^4\prod_{c=\pm}k^\sigma_{\dpl}(h_c)\tilde{k}^\sigma_{\dmi}(h_c).
\eeq
The matrix \eqref{eq:resolvent_column} is symmetric, so that $\mathf_{+-}=\mathf_{-+}$, as required by the symmetry under $(\xx_1,\xx_2)\leftrightarrow(\xx_3,\xx_4)$. The determinant is invariant under the single or simultaneous transformations $\hpm \to 1-\hpm$ and under the exchange $\hpl \leftrightarrow \hmi$. Furthermore, $j \to -j$ moves the determinant to the other sector (Appendix \ref{app:eigenvalues}),
 \beq
  \left. \det\l\mathbb{I}-\mathk^\ee \r \right|_j = \left. \det\l\mathbb{I}-\mathk^\oo \r \right|_{-j}.
 \eeq

 The integrals over continuum states and the sums over discrete states can be combined into contour integrals, separately, for $\hpl$ and for $\hmi$. The integrand in Eq. \eqref{eq:ladder_sum_explicit} is invariant under $\hpl \to 1 - \hpl$, or $\hmi \to 1-\hmi$, or both; so the integrals over $s_\pm$ can be extended to the whole real line. We split the factor $\frac{2h-1}{\pi \tan \pi h}$ into two parts,
 \beq
  \frac{2h-1}{\pi \tan \pi h} = \frac{1}{2} \L \mu^\ee (h) + \mu^\oo (h) \R, \qquad \mu^\ee (h) = \frac{2h-1}{\pi \tan \frac{\pi h}{2}} = \mu^\oo (1-h).
 \eeq
 The two terms are exchanged by $s\to -s$, so either one may be kept in the integrals over the whole real $s_\pm$ line. In each sector we keep its own measure, $\mu^{\ee}$ for $\sigma=\ee$ and $\mu^{\oo}$ for $\sigma=\oo$. The reason is the location of their poles. $\mu^{\ee}(h)$ has simple poles at $h=2n$ and $\mu^{\oo}(h)$ at $h=2n+1$, $n\in\mathbb{Z}$, both with residue $(4h-2)/\pi^2$. Therefore, the poles of $\mu^\sigma$ with $\text{Re}\,h>\frac12$ are situated where the discrete states of the corresponding sector lie, and the residues are their inverse norms (Eq. \eqref{eq:normalisation_discrete}). The sums over the discrete states can then be written as sums over the residues of $\mu^\sigma$, provided the rest of the integrand is holomorphic at these points.

 This is the case, as we now show factor by factor. By Eqs. \eqref{eq:varphi_expr_even} and \eqref{eq:varphi_expr_odd}, the only poles of $\varphi^\sigma_h$ with $\text{Re}\,h>0$ are simple poles that come from $\tan\frac{\pi h}{2}$ in the even sector and from $\cot\frac{\pi h}{2}$ in the odd sector. In the sum over discrete states, these poles lie at the integers of the other sector, where $\mu^\sigma$ has a simple zero, so $\mu^\sigma(h)\varphi^\sigma_h$ stays finite. At the discrete states of the sector $\sigma$, $\varphi^\sigma_h$ is holomorphic. Since $0<\dpm<1$, the poles of $k^\sigma_{\dpl}(h)$ and $\tilde{k}^\sigma_{\dmi}(h)$ are never at integer values of $h$, and they lie on $\text{Re}\,h=\frac{1}{2}$ only if $\dpm=\frac{1}{2}$. The determinant $\det(\mathbb{I}-\mathk^\sigma)$ does not vanish when both weights are discrete states of the sector, except at $(\hpl,\hmi)=(1,1)$ in the odd sector. When one weight is discrete and the other lies on the continuum, the determinant has no zeros in the even sector. In the odd sector it vanishes at $(1,\frac{1}{2} \pm\ii s_*)$ for certain $s_*$, when the spin $j$ exceeds a threshold $j_c(q)>0$ (Sec. \ref{subsec:operators}). In addition, $\det(\mathbb{I}-\mathk^\sigma)$ must not vanish when both weights lie on the continuum. This fails when the largest eigenvalue of $\mathk^\sigma$ on the continuum, $(2q-1)\alpha_0^2k^\sigma_{\dpl}(\frac12)\tilde{k}^\sigma_{\dmi}(\frac{1}{2})$, reaches one, which happens for $q\le 32$ in a range of $j$ that includes $j=0$ (Sec. \ref{subsec:operators} and Fig. \ref{fig:validity_region}). Restricting ourselves to the region where neither kind of zero occurs, the four kinds of terms in $\tilde{\mathf}_{ab}$ can be combined into
 \beq
  \label{eq:contour_form}
  \tilde{\mathf}_{ab}(\chp,\chm) = -\frac{1}{8} \sum_{\sigma} \int_{\mathscr{C}_\sigma} \frac{dh_+}{2\pi \ii} \int_{\mathscr{C}_\sigma} \frac{dh_-}{2\pi \ii} \mu^\sigma (\hpl) \mu^\sigma (\hmi) \l\frac{1}{\mathbb{I}-\mathk^\sigma}\r_{ab}k^\sigma_{\dpl}(h_b)\tilde{k}^\sigma_{\dmi}(h_{\bar b})\,\Psi^\sigma_{(\hpl,\hmi)}(\chp,\chm) ,
 \eeq
 where the contour $\mathscr{C}_\sigma$ runs along $\text{Re}\,(\hpm) = \frac{1}{2}$, encircles counter-clockwise the poles of $\mu^\sigma (\hpm)$ with $\text{Re}\,(\hpm) > \frac{1}{2}$ except the point $(\hpl,\hmi)=(1,1)$ of the odd sector, and is deformed so that the poles of $\tilde{k}^\ee_{\Delta}(h)$ at $h=\dmi$ and of $k^\oo_{\Delta}(h)$ at $h=\dpl$ lie to its right and their images $h=1-\dmi$ and $h=1-\dpl$ to its left.

 The deformation is needed because the poles at $h=\dpm$ lie to the left of the $\text{Re}\,(h)=\frac{1}{2}$ line for $\dpm < \frac{1}{2}$, and the zero-rung ladder, which behaves as $|\chi|^{\Delta_\pm}$ at small $\chi$, is not square integrable with the measure $d\chi/\chi^2$. The projections in Eq. \eqref{eq:inner_product_result} converge on the line only for $\dpm > \frac{1}{2}$ and are defined by analytic continuation from there. As $\dpm$ decreases through $\frac{1}{2}$, the poles at $h=\dpm$ cross the line to the left and their images cross it to the right. Eq. \eqref{eq:contour_form} with the deformed contour is the continuation from $\dpm > \frac{1}{2}$. At small $\chi$, when the deformed contour is closed to the right, it picks up the pole at $h=\dpm$ and reproduces the $|\chi|^{\Delta_\pm}$ behaviour of the zero-rung ladder. In Sec. \ref{subsec:operators} we show that these poles belong to light bilinears of the GFF and that the deformed contour reproduces the zero-rung ladder.

\subsection{Operators of the model}\label{subsec:operators}

 The operators that appear in the product of two fermions can be read off from the expansion of the four-point function in the small $\chi$ limit as in reference \cite{Maldacena_2016}. An analogous expansion for $\tilde{\mathf}$ can be obtained from Eq. \eqref{eq:contour_form} by pushing either $\hpl$ or $\hmi$ contour from the $s$ axis rightward to infinity. In this process, the residues of the discrete states cancel, leaving the residues originating from the rest of the integrand.

 The function $F_{1-h}(\chi)=\chi^{1-h}\ff(1-h,1-h,2-2h,\chi)$ grows as $\chi^{1-\text{Re}\,h}$ for small $\chi$, so the contour cannot be moved to large $\text{Re}\,h$ in this form. For $0<\chp<1$ we write $\varphi^\sigma_{\hpl}=A^\sigma(\hpl)F_{\hpl}+A^\sigma(1-\hpl)F_{1-\hpl}$. On the line $\text{Re}\,\hpl=\frac12$, the rest of the integrand with the symmetric measure $\frac12\L\mu^{\ee}(\hpl)+\mu^{\oo}(\hpl)\R$ is invariant under $\hpl\to1-\hpl$, and this substitution maps the second term onto the first. Since $\l\mu^{\ee}(h)+\mu^{\oo}(h)\r A^\sigma(h)=\mu^\sigma(h)\,\Gamma(h)^2/\Gamma(2h)$, and $\varphi^\sigma_h=\frac{\Gamma(h)^2}{\Gamma(2h)}F_h$ at the discrete states, Eq. \eqref{eq:contour_form} becomes
 \beal
  \label{eq:contour_form_chi<1}
  \tilde{\mathf}_{ab}(\chp,\chm)=-\frac{1}{8}\sum_{\sigma}&\int_{\mathscr{C}_\sigma}\frac{d\hmi}{2\pi\ii}\,\mu^\sigma(\hmi)\,\varphi^\sigma_{\hmi}(\chm)\\
  &\times\int_{\mathscr{C}_\sigma}\frac{d\hpl}{2\pi\ii}\,\mu^\sigma(\hpl)\,\frac{\Gamma(\hpl)^2}{\Gamma(2\hpl)}\,F_{\hpl}(\chp)\l\frac{1}{\mathbb{I}-\mathk^\sigma}\r_{ab}k^\sigma_{\dpl}(h_b)\tilde{k}^\sigma_{\dmi}(h_{\bar b}) ,
 \eeal
 for $0<\chp<1$.

 \paragraph{Zero-rung ladder and GFF bilinears.}
 
 Eq. \eqref{eq:contour_form_chi<1} must reproduce the zero-rung ladder when the kernels are switched off, $\mathk^\sigma \to 0$. In this limit the determinant has no zero at the odd sector state $(1,1)$, which can be included in the contour. The integrand factorises into the two lightcones, and for $0 < \chpm <1$ both contours can be pushed to the right. The residues of $\mu^\sigma$ at the discrete states cancel. Only the poles of the chiral eigenvalue functions $k^\sigma_{\Delta}$ and $\tilde{k}^\sigma_{\Delta}$ to the right of $\mathscr{C}_\sigma$ contribute. By Eq. \eqref{eq:chiral_k_tildek}, the poles lie at $h=\Delta +n$, with odd integer $n \geq 1$ for $k^\ee_\Delta$ and $\tilde{k}^\oo_\Delta$, and even integer $n \geq 0$ for $k^\oo_\Delta$ and $\tilde{k}^\ee_\Delta$. Their residues are
 \beq
 \label{eq:chiral_residues}
  \mathop{\text{Res}}_{h=\Delta+n}\L\mu^\sigma(h)\frac{\Gamma(h)^2}{\Gamma(2h)}k^\sigma_\Delta(h)\R= 4 a_n(\Delta) = -\mathop{\text{Res}}_{h=\Delta+n}\L\mu^\sigma(h)\frac{\Gamma(h)^2}{\Gamma(2h)}\tilde{k}^\sigma_\Delta(h)\R, \quad a_n(\Delta) = \frac{(\Delta)_n^2}{n!\,(2\Delta+n-1)_n },
 \eeq
 where $(\Delta)_n$ represents the Pochhammer symbol, and the zero-rung ladder can be expressed as,
 \beq
  \label{eq:zero_rung_gff}
  \mathf^{(0),\sigma}_b(\chp,\chm)=2\sum_{n_b,n_{\bar b}}a_{n_b}(\dpl)\,a_{n_{\bar b}}(\dmi)\,F_{\dpl+n_b}(\chi^b)\,F_{\dmi+n_{\bar b}}(\chi^{\bar b}),
 \eeq
 where $n_b$ is odd and $n_{\bar b}$ even in the even sector, and $n_b$ even and $n_{\bar b}$ odd in the odd sector. This is the 2D analogue of the representation of the zero-rung ladder by the poles of $k_c(h)$ in the 1D model (Appendix C of reference \cite{Maldacena_2016}). Since the zero-rung ladder is the four-point function of the GFF at order $1/N$, Eq. \eqref{eq:zero_rung_gff} is its expansion in conformal blocks. The coefficients are positive, and $(\dpl+n_b,\dmi+n_{\bar b})$ are the weights of the bilinears of the GFF. Bilinears with $n_b=0$ or $n_{\bar b}=0$ have the weight $\dpl$ or $\dmi$, smaller than $\frac{1}{2}$, on one lightcone. We call them light GFF bilinears. They come from the poles of $\tilde k^{\ee}_{\dmi}(h)$ at $h=\dmi$ and of $k^{\oo}_{\dpl}(h)$ at $h=\dpl$, which lie to the left of the line $\text{Re}\,h=\frac{1}{2}$ and are kept to the right of $\mathscr{C}_\sigma$ by the contour deformation defined in Sec. \ref{subsec:summ_ladders}. We have checked Eq. \eqref{eq:zero_rung_gff}, and the contour form in Eq. \eqref{eq:contour_form} with $\mathk^\sigma \to 0$, numerically against Eq. \eqref{eq:zero_rung_split}.

 \paragraph{Sum of ladders.} We now push the $\hpl$ contour in Eq. \eqref{eq:contour_form_chi<1}. The residues of $\mu^\sigma(\hpl)$ at the discrete states cancel as before. Unlike in the zero-rung ladder, the poles of the chiral eigenvalue functions do not contribute. Where one of the products $k^\sigma_{\dpl}(h_a)\tilde{k}^\sigma_{\dmi}(h_{\bar a})$ diverges and the other stays finite, numerator and denominator of Eq. \eqref{eq:resolvent_column} grow linearly but their ratio stays finite for generic $\hmi$, and the same holds when both diverge. This is analogous to the fact that $k_c/(1-k_c)\to-1$ at the poles of $k_c$ in the 1D model. What remains are the zeros $\hpl^*(\hmi)$ of $\det(\mathbb{I}-\mathk^\sigma)$ to the right of the contour,
 \beal
  \label{eq:ope_one_lightcone}
  \tilde{\mathf}_{ab}(\chp,\chm)=\frac{1}{8}\sum_{\sigma}&\int_{\mathscr{C}_\sigma}\frac{d\hmi}{2\pi\ii}\,\mu^\sigma(\hmi)\,\varphi^\sigma_{\hmi}(\chm)\\
  &\times\sum_{\hpl^*}\mathop{\text{Res}}_{\hpl=\hpl^*}\L\mu^\sigma(\hpl)\frac{\Gamma(\hpl)^2}{\Gamma(2\hpl)}\l\frac{1}{\mathbb{I}-\mathk^\sigma}\r_{ab}k^\sigma_{\dpl}(h_b)\tilde{k}^\sigma_{\dmi}(h_{\bar b})\R F_{\hpl^*}(\chp) .
 \eeal
 The zeros come in pairs $\hpl^*$, $1-\hpl^*$. Of each pair the sum contains the one with $\text{Re}\,\hpl^* > \frac{1}{2}$. The exception is the pair that continues a light GFF bilinear, i.e. the pair that tends to $\hpl=\Delta_\mp$ ($\sigma=\ee,\oo$) as $\mathk^\sigma \to 0$. Of that pair the sum contains the member that tends to $\dmi$ ($\sigma=\ee$) or $\dpl$ ($\sigma=\oo$) as $\text{Im}\,\hmi\to\infty$. Eq. \eqref{eq:ope_one_lightcone} expands the four-point function in conformal blocks of one lightcone, with weights $\hpl^*$ that depend on the weight $\hmi$ of the other lightcone, and the residue plays the role of the squared OPE coefficient.

\begin{figure}[ht]
 \centering
 \includegraphics[width=0.85\textwidth]{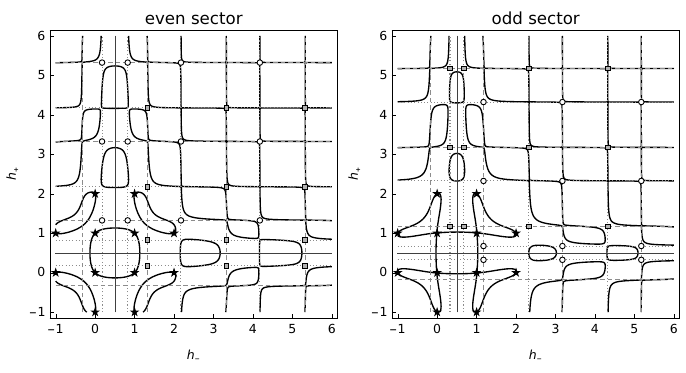}
 \caption{ Real section of the spectral curve \eqref{eq:spectral_curve} for $q=4$, $j=0.3\Delta$ ($\dpl=0.325$, $\dmi=0.175$), even (left) and odd (right) sector. These parameters lie outside the region of Fig. \ref{fig:validity_region}. Dashed and dotted lines: poles of $\hat{k}$ and $\hat{\tilde k}$ in $\lambda^\sigma_\pm$. Open circles and grey squares: points where both factors of $\lambda^\sigma_+$ or of $\lambda^\sigma_-$ have a pole, at the weights of the GFF bilinears. Stars: the zero modes $(1,1)$, $(2,0)$, $(0,2)$ and their images. Thin lines: $\text{Re}\,\hpm=\tfrac{1}{2}$.}
 \label{fig:spectral_curve}
\end{figure}

 \paragraph{The spectral curve.}
 In the 1D SYK model, the poles of the integrand are isolated points, $k_c(h_m)=1$. Here they lie on the curve $h^*_+(\hmi)$ that satisfies
 \beq
  \label{eq:spectral_curve}
  \det (\mathbb{I}-\mathk^\sigma) = 0,
 \eeq
 The eigenvalue matrix in Eq. \eqref{eq:eigenvalues} can also be expressed as
 \beq
  \label{eq:eigenvalues_diag}
  \mathk^\sigma_{ab}(\hpl,\hmi) = \begin{pmatrix}
      \lambda^\sigma_+ & 0 \\ 0 & \lambda^\sigma_-
  \end{pmatrix} \begin{pmatrix}
      q-1 & q \\ q & q-1 
  \end{pmatrix} ,
 \eeq
 where $\lambda^\sigma_\pm(\hpl,\hmi) = \alpha_0^2 \, k^\sigma_{\dpl}(\hpm) \tilde{k}^\sigma_{\dmi}(h_\mp) $, and the determinant of $\mathbb{I}-\mathk^\sigma$ has the factorised form
 \beq
  \label{eq:determinant_factorised}
  \det (\mathbb{I}-\mathk^\sigma) = q (1+\lambda^\sigma_+) (1+\lambda^\sigma_-) \L \Delta - \frac{\lambda^\sigma_+}{1+\lambda^\sigma_+} - \frac{\lambda^\sigma_-}{1+\lambda^\sigma_-} \R.
 \eeq
 The zeros of $\det(\mathbb{I}-\mathk^\sigma)$ are the points where
 \beq
 \label{eq:determinant_zero_satisfy}
  \lambda^\sigma_+ = \lambda^\sigma_- = -1   \qquad \text{or} \qquad \frac{\lambda^\sigma_+}{1+\lambda^\sigma_+} + \frac{\lambda^\sigma_-}{1+\lambda^\sigma_-} = \Delta .
 \eeq
 Eliminating $\alpha_0^2$ using Eq. \eqref{eq:alpha_0_as_k}, the functions $\lambda^\sigma_\pm$ become
 \beq
  \label{eq:lambda_pm}
  \lambda^\ee_\pm(\hpl,\hmi) = - \hat{k}_{\dpl}(\hpm) \hat{\tilde{k}}_{\dmi}(h_\mp) , \qquad  \lambda^\oo_\pm(\hpl,\hmi) = - \hat{\tilde{k}}_{\dpl}(\hpm) \hat{k}_{\dmi}(h_\mp),
 \eeq
 where by Eq. \eqref{eq:chiral_k_tildek},
 \beal
 \label{eq:khat}
  \hat{k}_{\Delta}(h) &= \frac{k^\ee_\Delta(h)}{k^\ee_\Delta(1)} = \frac{\tilde{k}^\oo_\Delta(h)}{\tilde{k}^\oo_\Delta(1)} = - \frac{\Gamma(2-\Delta)\Gamma(h+\Delta-1)}{\Gamma(\Delta)\Gamma(h-\Delta+1)} \frac{\cos \frac{\pi(h+\Delta)}{2}}{\cos \frac{\pi(h-\Delta)}{2}}, \\
  \hat{\tilde k}_{\Delta}(h) &= \frac{k^\oo_\Delta(h)}{k^\oo_\Delta(1)} = \frac{\tilde{k}^\ee_\Delta(h)}{\tilde{k}^\ee_\Delta(1)} = \frac{\Gamma(2-\Delta)\Gamma(h+\Delta-1)}{\Gamma(\Delta)\Gamma(h-\Delta+1)} \frac{\sin \frac{\pi(h+\Delta)}{2}}{\sin \frac{\pi(h-\Delta)}{2}}.
 \eeal
 It is difficult to obtain a closed-form analytic solution of the spectral curve $\hpl^*(\hmi)$ for generic values of $\hpm$. Here, we analyse the spectral curve analytically for some special values of $\hpm$ and use numerical method for generic $\hpm$. Fig. \ref{fig:spectral_curve} shows the real section of the spectral curve in both sectors. It passes through the zero modes $(2,0)$, $(0,2)$ and $(1,1)$ and their images under $\hpm \to 1-\hpm$. Away from them it consists of branches that run along the lines where $\lambda^\sigma_+$ or $\lambda^\sigma_-$ has a pole, i.e.\ along the weights of the GFF bilinears.

\begin{figure}[ht]
 \centering
 \includegraphics[width=0.95\textwidth]{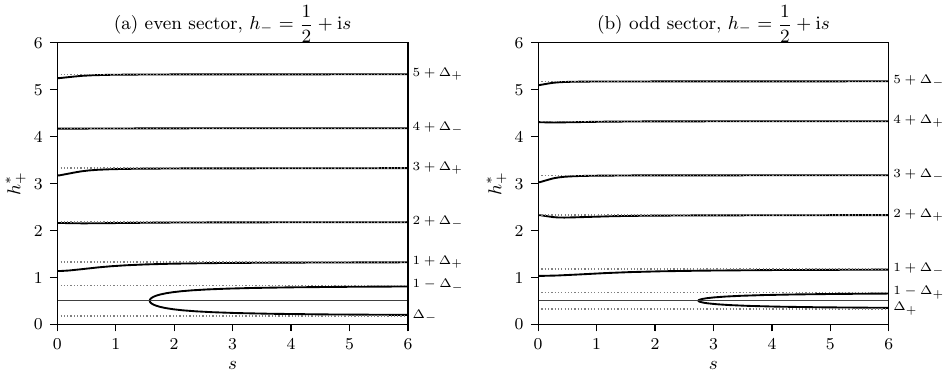}
 \caption{\small Real zeros $\hpl^*$ of $\det(\mathbb{I}-\mathk^\sigma)$ for $\hmi=\frac12+\ii s$, with $q=4$ and $j=0.3\Delta$ ($\dpl=0.325$, $\dmi=0.175$), in the even (a) and odd (b) sector. Dotted lines: the limits $s\to\infty$, which are the weights of the GFF bilinears, and the image $1-\Delta_\mp$ of the light one. Thin line: $\text{Re}\,\hpl=\frac12$. These parameters lie outside the region of Fig. \ref{fig:validity_region}.}
 \label{fig:spectral_bands}
\end{figure}

 \paragraph{Bands.} Fig. \ref{fig:spectral_bands} shows the real zeros $\hpl^*$ for $\hmi=\frac12+\ii s$. As $s\to\infty$, $\hat{k}_\Delta(\hmi)$ and $\hat{\tilde k}_\Delta(\hmi)$ vanish, and the zeros approach the poles of $\lambda^\sigma_\pm$ in $\hpl$, i.e. the weights of the GFF bilinears, $1+\dpl+2m$ and $2+\dmi+2m$ in the even sector and $1+\dmi+2m$ and $2+\dpl+2m$ in the odd sector, $m=0,1,2,\dots$. In the integral over $s$ in Eq. \eqref{eq:ope_one_lightcone}, each weight $\hpl^*$ therefore sweeps a band that ends at the weight of a GFF bilinear, and the spin $\hpl-\hmi$ of these contributions is not quantised. The light GFF bilinear gives one more pair of zeros, which tends to $\Delta_\mp$ and $1-\Delta_\mp$ for $\sigma=\ee$ ($\oo$).

 \paragraph{Range of validity of the contour form.}
 Eq. \eqref{eq:ope_one_lightcone} requires that no zero $h^*_+(\hmi)$ reaches the $\hpl$ contour while $\hmi$ runs over $\mathscr{C}_\sigma$. At integer weights $n \geq 1$, the trigonometric ratios in Eq. \eqref{eq:khat} are $\pm 1$, hence the factors in $\lambda_\pm$ reduce to
 \beq
  \hat{k}_{\Delta}(n) = \hat{\tilde k}_{\Delta}(n) = (-1)^{n+1} \frac{(\Delta)_{n-1}}{(2-\Delta)_{n-1}},
 \eeq
 whose modulus decreases strictly from one as we increase $n$. At $(\hpl,\hmi) = (1,1)$ and its images under $\hpm \to 1-\hpm$, $\lambda_\pm = -1$ and the determinant is zero. Since $\hat{k}_{\Delta}(0) = \hat{\tilde k}_{\Delta}(0) = 1$ and $\hat{k}_{\Delta}(2) = \hat{\tilde k}_{\Delta}(2) = -\Delta/(2-\Delta)$, at $(\hpl,\hmi) = (0,2),(2,0)$, the square bracket in Eq. \eqref{eq:determinant_factorised} vanishes. For the state $(2,2)$ and $(n_+,n_-)$, where at least one of $n_\pm \geq 3$, none of the factors in Eq. \eqref{eq:determinant_factorised} vanish for $\Delta \le 1/2$. Hence the only zeros at integer weights are $(1,1),(2,0),(0,2)$ and their images under single or simultaneous substitution $\hpm \to 1-\hpm$.

 \begin{figure}[ht]
 \centering
 \includegraphics[width=0.8\textwidth]{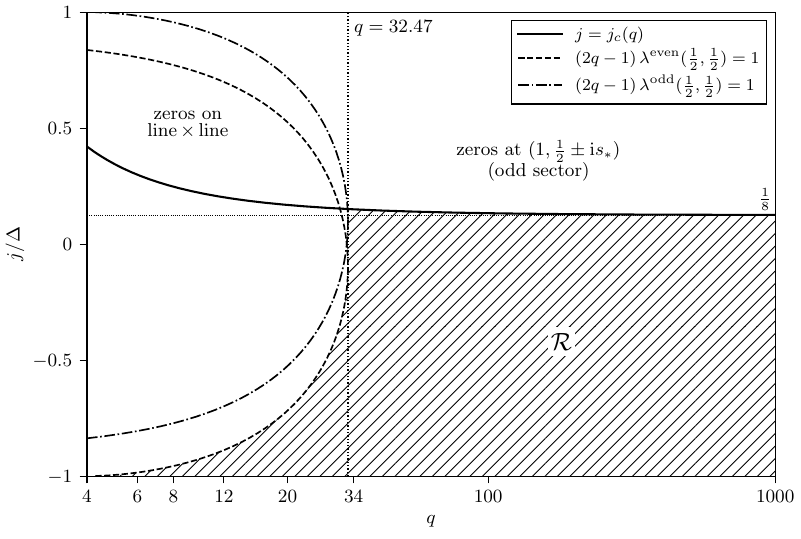}
 \caption{ The region of the $(q,j)$ plane in which $\det(\mathbb{I}-\mathk^\sigma)$ has no zero on the basis states, apart from the Lorentz mode; here $\Delta=1/q$ and $q$ is treated as continuous. Solid line: $j=j_c(q)$; above it $\det(\mathbb{I}-\mathk^\oo)$ vanishes at $(\hpl,\hmi)=(1,\frac12\pm\ii s_*)$. Dashed (dash-dotted) line: $(2q-1)\lambda^\sigma(\frac12,\frac12)=1$ for $\sigma=\ee$ ($\sigma=\oo$), where $\lambda^\sigma(\frac12,\frac12)=\alpha_0^2\,k^\sigma_{\dpl}(\frac12)\,\tilde{k}^\sigma_{\dmi}(\frac12)$ and $(2q-1)\lambda^\sigma(\frac12,\frac12)$ is the largest eigenvalue of $\mathk^\sigma$ on the continuum; between the two curves of one sector, $\det(\mathbb{I}-\mathk^\sigma)$ vanishes with both weights on the continuum. Dotted lines: $q=32.47$, the largest $q$ with such zeros, and $j/\Delta=\frac{1}{8}$, the large-$q$ limit of $j_c/\Delta$.}
 \label{fig:validity_region}
\end{figure}

 For the states on the line $ h = \frac{1}{2}+ \ii s $, Eqs. \eqref{eq:chiral_k_tildek} and \eqref{eq:khat} lead to
 \beal
  \hat{k}_{\Delta} \l \tfrac{1}{2}+\ii s \r &= \frac{\Gamma\l 1-\frac{\Delta}{2} \r \Gamma\l \frac{3-\Delta}{2} \r}{\Gamma\l \frac{1+\Delta}{2} \r \Gamma\l \frac{\Delta}{2} \r} \frac{\left| \Gamma \l \frac{1}{4} + \frac{\Delta}{2} + \frac{\ii s}{2} \r \right|^2}{\left| \Gamma \l \frac{5}{4} - \frac{\Delta}{2} + \frac{\ii s}{2} \r \right|^2}, \\
  \hat{\tilde k}_{\Delta} \l \tfrac{1}{2}+ \ii s \r &= \frac{\Gamma\l \frac{1-\Delta}{2} \r \Gamma\l 1- \frac{\Delta}{2} \r}{\Gamma\l \frac{\Delta}{2} \r \Gamma\l \frac{\Delta-1}{2} \r} \frac{\left| \Gamma \l \frac{\Delta}{2} - \frac{1}{4} + \frac{\ii s}{2} \r \right|^2}{\left| \Gamma \l \frac{3}{4} - \frac{\Delta}{2} + \frac{\ii s}{2} \r \right|^2}.
 \eeal
 $\hat{k}_{\Delta} \l \tfrac{1}{2}+\ii s \r$ is positive, whereas $\hat{\tilde k}_{\Delta} \l \tfrac{1}{2}+ \ii s \r$ is negative since $\Gamma \l \tfrac{\Delta-1}{2} \r < 0 $ for $0 < \Delta < 1$. Both functions decay in modulus as $|s|^{-2(1-\Delta)}$, and for real $s$ they satisfy the bound
 \beq
  0 < \hat{k}_{\Delta} \l \tfrac{1}{2}+\ii s \r \le \hat{k}_{\Delta} \l \tfrac{1}{2} \r, \qquad \hat{\tilde k}_{\Delta} \l \tfrac{1}{2} \r \le \hat{\tilde k}_{\Delta} \l \tfrac{1}{2}+ \ii s \r < 0 \, ,
 \eeq
 which follow from the product formula \cite{NIST:DLMF}
 \beq
  \left| \frac{\Gamma(x)}{\Gamma(x+\ii y)} \right|^2 = \prod_{k \geq 0} \l 1 + \frac{y^2}{(x+k)^2} \r , \quad x \neq 0,-1,-2,\dots .
 \eeq
 Consider the states in the odd sector with $(\hpl,\hmi) = (1,\tfrac{1}{2}+\ii s) $. From Eq. \eqref{eq:lambda_pm}, $\lambda^\oo_+ = - \hat{k}_{\dmi}(\tfrac{1}{2}+\ii s)$ and $\lambda^\oo_- = - \hat{\tilde k}_{\dpl}(\tfrac{1}{2}+\ii s)$, and the zeros of the determinant are at the points where
 \beq
  L \l 1,\tfrac{1}{2} + \ii s \r := \frac{\hat{k}_{\dmi}(\tfrac{1}{2}+\ii s)}{1-\hat{k}_{\dmi}(\tfrac{1}{2}+\ii s)} + \frac{\hat{\tilde k}_{\dpl}(\tfrac{1}{2}+\ii s)}{1-\hat{\tilde k}_{\dpl}(\tfrac{1}{2}+\ii s)} = -\Delta.
 \eeq
 For $j>j_c(q)$, defined by $L(1,\tfrac{1}{2}) = -\Delta$, the function $L(1,\tfrac{1}{2} + \ii s)$ is minimal at $s=0$ and increases monotonically with $|s|$. For $j<j_c(q)$ we find numerically $L(1,\tfrac{1}{2} + \ii s)>-\Delta$ for all $s$. Since $L(1,\tfrac{1}{2})$ decreases strictly in $j$, a pair of zeros $\pm s_*$ exists if and only if $j>j_c(q)$, and it is then unique. Numerical analysis suggests that the ratio $j_c/\Delta$ decreases strictly with $q$, and $j_c/\Delta \to \tfrac{1}{8}$ for large $q$ ($\Delta \to 0$). The analysis for the states $(\hpl,\hmi) = (\tfrac{1}{2}+\ii s,1)$ is exactly same since $L\l 1,\tfrac{1}{2} + \ii s \r = L\l \tfrac{1}{2} + \ii s,1 \r$.

 For the states $(\hpl,\hmi) = (n,\tfrac{1}{2}+\ii s),(\tfrac{1}{2}+\ii s,n)$ with integer $n \geq 2 $, the moduli of $\hat{k}_{\Delta}(n)$ and $\hat{\tilde k}_{\Delta}(n)$ are smaller than one, hence the condition for zeros is stronger than $j>j_c(q)$. With both weights on the continuum, $(\hpl,\hmi) = (\tfrac{1}{2}+\ii s_+,\tfrac{1}{2}+\ii s_-)$, the determinant vanishes only when $(2q-1)\lambda^\sigma(\tfrac{1}{2},\tfrac{1}{2}) \ge 1$. This happens only for $q \le 32 $, and at $q=4$ for every $j$ in at least one of the two sectors. Fig. \ref{fig:validity_region} shows the region in which neither kind of zero occurs. Although the spectral curve is well defined beyond this region, Eq. \eqref{eq:ope_one_lightcone} does not hold as written. At a zero of the determinant on the principal series the kernel has eigenvalue one on a normalisable state, so the sum of ladders is not defined there. Further study is needed to understand whether these poles of $(\mathbb{I}-\mathk^\sigma)^{-1}$ on the principal series signal an instability of the 2D model in the conformal limit.

\subsection{The Lorentz mode}\label{subsec:lorentz_mode}

 The sum over ladders in Eq. \eqref{eq:ladder_sum0} leaves out the state $(\hpl,\hmi)=(1,1)$ of the odd sector, where $\det(\mathbb{I}-\mathk^\oo)$ vanishes. In this subsection, we show that this state is the odd part of the local Lorentz transformations of the conformal solution, and we compute its contribution in the conformal limit, following the treatment of the $h=2$ states of the 1D model in reference \cite{Maldacena_2016}.

 The kernels act on the chirality index through the matrix $\mathk^\sigma_{ab}$ (Eq. \eqref{eq:kernel_on_Psi}). A fluctuation $\delta G_a(\xx_1,\xx_2)/G_a(\xx_{12})=w_a\,\Phi(\xx_1,\xx_2)$, with the same $\Phi$ for both chiralities, is characterised by its chirality vector $w=(w_+,w_-)$. At $(\hpl,\hmi)=(1,1)$, Eqs. \eqref{eq:eigen_00} and \eqref{eq:alpha_0_as_k} give
 \beq
 \label{eq:chirality_matrix}
  \mathk^\sigma_{ab}(1,1)=\delta_{ab}-q\,u_au_b ,\qquad u=(1,1),
 \eeq
 with the eigenvalue $1-2q$ on $u$ and the eigenvalue one on $v=(1,-1)$. Writing Eq. \eqref{eq:ds1} as $\Sigma_a=-J^2(G_+G_-)^q/G_a$, a fluctuation with chirality vector $w$ changes the self-energies by $\delta\Sigma_a/\Sigma_a=q(w_++w_-)\Phi-w_a\Phi$. The first term is the change of the rung $(G_+G_-)^q$ and gives $-q\,u_au_b$. It vanishes along $v$, since $u\cdot v=0$, and only $\delta_{ab}$ acts there. The components $v_a=\pm 1$ are the Lorentz charges of $\psi_\pm$ in Sec. \ref{subsec:tcft}, and $u\cdot v=0$ states that the interaction, with $q$ fermions of each chirality, is Lorentz invariant. We show below that every Lorentz mode has chirality vector $v$.

 From Eq. \eqref{eq:varphi_discrete_states} we have $\varphi^\oo_1(\chi)=-\log|1-\chi|$. Using Eqs. \eqref{eq:inner_product_result} and \eqref{eq:alpha_0_as_k}, the projection of the zero-rung ladder on this state is $\la\mathf^{(0),\oo}_\pm,\Psi^\oo_{(1,1)}\ra=1/(8\alpha_0^2)$. Moreover, Eq. \eqref{eq:normalisation_discrete} gives the normalisation of the state as $\la\Psi^\oo_{(1,1)},\Psi^\oo_{(1,1)}\ra=\pi^4/4$. We split the source $\delta_{cb}$ into the eigenvectors, $\delta_{cb}=\frac12u_cu_b+\frac12v_cv_b$, on which $(\mathbb{I}-\mathk^\oo)^{-1}$ acts as $1/(2q)$ and $1/(1-k_L)$, where $k_L$ is the eigenvalue on $v$. The $(1,1)$ term of Eq.~\eqref{eq:ladder_sum0} is then
 \beq
 \label{eq:lorentz_11_term}
  \mathf_{(1,1),ab}(\chp,\chm)=\frac{1}{2\pi^4\alpha_0^2}\l\frac{1}{4q}+\frac{v_av_b}{2(1-k_L)}\r\log|1-\chp|\log|1-\chm| .
 \eeq
 The first term comes from $u$. It is finite and belongs to the four-point function in the conformal limit. The second term, the Lorentz channel, comes from $v$ and diverges at $k_L=1$, the value in Eq. \eqref{eq:chirality_matrix}. It is proportional to $v_av_b$ and drops out of $\sum_b\mathf_{(1,1),ab}$. In Eq. \eqref{eq:contour_form} the point $(1,1)$ of the odd sector is excluded: the pole of $\mu^\oo(\hmi)$ there coincides with the zero of $\det(\mathbb{I}-\mathk^\oo)$, so in the residue at $\hpl=1$ the $\hmi$ contour must not encircle $\hmi=1$, and vice versa.

 \paragraph{Local Lorentz transformations.} Under the local Lorentz transformation $\psi_\pm \to\Lambda^{\pm 1} \psi_\pm$ (Sec. \ref{subsec:tcft}), the bilinears transform as
 \beq
 \label{eq:lorentz_bilinears}
  G_\pm(\xx_1,\xx_2)\to\Lambda^{\pm1}(\xx_1)\Lambda^{\pm1}(\xx_2)\,G_\pm(\xx_1,\xx_2),\qquad
  \Sigma_\pm(\xx_1,\xx_2)\to\Lambda^{\mp1}(\xx_1)\Lambda^{\mp1}(\xx_2)\,\Sigma_\pm(\xx_1,\xx_2).
 \eeq
 Eq. \eqref{eq:ds1} is consistent with this, since $G_\pm^{q-1}G_\mp^q$ acquires $\Lambda^{\pm(q-1)}\Lambda^{\mp q}=\Lambda^{\mp1}$ at each point. In $G_\pm*\Sigma_\pm$ the factors at the integration point cancel, and $\Lambda^{\pm1}(\xx_1)\Lambda^{\mp1}(\xx_2)=1$ on the support of $\delta^2(\xx_{12})$. Thus, the IR relations \eqref{eq:ds1} and \eqref{eq:ds2_IR} remain invariant for arbitrary $\Lambda(\xx)$. By contrast, the conformal solution \eqref{eq:two-point_fn} generally does not, aside from the case of a constant $\Lambda=e^{j\eta}$ accompanied by the boost $x^\pm\to e^{\pm\eta}x^\pm$. This is precisely how the solution carries the spin $j$. With $\Lambda=e^\lambda$, the linearised transformation
 \beq
 \label{eq:lorentz_linearised}
  \delta_\lambda G_a(\xx_1,\xx_2)=v_a\L\lambda(\xx_1)+\lambda(\xx_2)\R G_a(\xx_{12})
 \eeq
 has chirality vector $v$ and $\Phi=\lambda(\xx_1)+\lambda(\xx_2)$. Since linearised transformations should also solve the conformal DS equations, varying Eq. \eqref{eq:ds2_IR} gives $\delta G_a=G_a*\delta\Sigma_a*G_a$, varying Eq. \eqref{eq:ds1} gives $\delta\Sigma_a=-J^2\sum_b(q-\delta_{ab})L_{ab}\,\delta G_b$, and together $\delta G_a=\sum_bK_{ab}*\delta G_b$ with the kernel of Eq. \eqref{eq:ladder_kernel}. Therefore, every linearised Lorentz transformation is an eigenfunction of the kernels with eigenvalue one and chirality vector $v$, which agrees with the eigenvalue one of Eq. \eqref{eq:eigen_00}.

The generators of \iso in Sec. \ref{subsec:conf_symmetry} annihilate $G_a(\xx_{12})$. The Casimir operator in Eq. \eqref{eq:Casimir_1+2} acts on $G_a(\xx_{12})\,\Phi$ as
 \beq
 \label{eq:casimir_on_ratio}
  C^{(1+2)}_\pm\L G_a(\xx_{12})\,\Phi\R=-G_a(\xx_{12})\l x^\pm_{12}\r^2\partial_{x^\pm_1}\partial_{x^\pm_2}\Phi .
 \eeq
 For $\Phi=\lambda(\xx_1)+\lambda(\xx_2)$ this vanishes, so the Lorentz modes have $\hpm(\hpm-1)=0$, i.e. the weights $(0,0)$, equivalent to $(1,1)$ by Eq. \eqref{eq:eigen_symmetry}. Conversely, a $\Phi$ annihilated by both operators is a sum of four functions, each of one $x^+$ and one $x^-$. Since $\delta G_a$ is antisymmetric under $\xx_1\leftrightarrow\xx_2$, $\Phi$ is symmetric, so $\Phi=\lambda(\xx_1)+\lambda(\xx_2)+\tilde\lambda(x^+_1,x^-_2)+\tilde\lambda(x^+_2,x^-_1)$. The exchange operator $\scP_+:x^+_1\leftrightarrow x^+_2$ exchanges $\lambda$ and $\tilde\lambda$. It commutes with the kernels, and acts as the reflection operator $R_+$ of Eq. \eqref{eq:reflection_criteria} on functions of the cross ratios. The odd sector has $\tilde\lambda=-\lambda$, so the odd $(1,1)$ eigenfunctions are
 \beq
 \label{eq:lorentz_odd_mode}
  (1-\scP_+)\L\lambda(\xx_1)+\lambda(\xx_2)\R=\lambda(x^+_1,x^-_1)+\lambda(x^+_2,x^-_2)-\lambda(x^+_2,x^-_1)-\lambda(x^+_1,x^-_2) .
 \eeq
 Eq. \eqref{eq:lorentz_odd_mode} is the alternating sum of $\lambda$ over the corners of the lightcone rectangle with opposite corners $\xx_1$ and $\xx_2$, so
 \beq
 \label{eq:odd_mode_rectangle}
  (1-\scP_+)\L\lambda(\xx_1)+\lambda(\xx_2)\R=\int_{x^+_2}^{x^+_1}dy^+\int_{x^-_2}^{x^-_1}dy^-\;\partial_+\partial_-\lambda(y^+,y^-) .
 \eeq
 The odd mode depends on $\lambda$ only through $\partial_+\partial_-\lambda$. A chiral sum $\lambda=\lambda_+(x^+)+\lambda_-(x^-)$ therefore has no odd part. Conversely, if Eq. \eqref{eq:lorentz_odd_mode} vanishes for all $\xx_1,\xx_2$, fixing $\xx_2$ shows that $\lambda$ is a chiral sum. For a chiral sum, $\Lambda$ is a product of chiral factors, so the transformation \eqref{eq:lorentz_bilinears} keeps the factorised form \eqref{eq:two-point_fn}, and its $\Phi$ is invariant under $\scP_+$ and lies in the even sector. The odd $(1,1)$ modes are thus labelled by $\lambda$ modulo chiral sums,
 \beq
 \label{eq:lorentz_modulo}
  \lambda\sim\lambda+\lambda_+(x^+)+\lambda_-(x^-),
 \eeq
 or equivalently by $\partial_+\partial_-\lambda$. For a chiral product $\lambda=f(x^+)g(x^-)$ the odd mode factorises,
 \beq
 \label{eq:chiral_product}
  (1-\scP_+)\L\lambda(\xx_1)+\lambda(\xx_2)\R=\L f(x^+_1)-f(x^+_2)\R\L g(x^-_1)-g(x^-_2)\R ,
 \eeq
 into chiral factors odd under $\scP_+$ and $\scP_-$ respectively, as in Eq. \eqref{eq:eigenfunction_factorised}. A chiral sum is the case where $f$ or $g$ is constant, and its factor vanishes. Any $\lambda$ is a superposition of chiral products, for example by Fourier transform. As $\xx_2\to\xx_1$, Eq. \eqref{eq:odd_mode_rectangle} gives $x^+_{12}x^-_{12}\,\partial_+\partial_-\lambda(\xx_1)$. This corresponds to the decay $\chp\chm$ in the frame where $x^\pm_1=0$, $x^\pm_2=\chpm$, $x^\pm_3=1$, and $x^\pm_4\to\infty$, and in this setup the odd mode is normalisable. By contrast, the even component approaches $4\lambda(\xx_1)$ and is therefore not normalisable, consistent with the fact that Eq. \eqref{eq:eigenbasis} contains no even $(1,1)$ state. In the same frame the chiral product $\lambda(\xx)=\log|1-x^+|\log|1-x^-|$ vanishes at $\xx_1$, and Eq. \eqref{eq:chiral_product} gives
 \beq
 \label{eq:lorentz_Psi11}
  (1-\scP_+)\L\lambda(\xx_1)+\lambda(\xx_2)\R=\log|1-\chp|\log|1-\chm|=\Psi^\oo_{(1,1)}(\chp,\chm) .
 \eeq

 \paragraph{Three- and four-point functions.} The three-point function of Eq. \eqref{eq:3pt_eigenfunctions} at $(1,1)$ is a Lorentz mode of the same kind. Dividing by $G_a(\xx_{12})$ the signs combine into $x^\pm_{12}/(x^\pm_{10}x^\pm_{20})$, and by Eq. \eqref{eq:chiral_product}
 \beq
 \label{eq:lorentz_3pt}
  f^{a,\oo}_{1,1}(\xx_1,\xx_2,\xx_0)=\frac{G_a(\xx_{12})}{\mathtt{b}}\,\frac{x^+_{12}x^-_{12}}{x^+_{10}x^+_{20}x^-_{10}x^-_{20}}=\frac{G_a(\xx_{12})}{\mathtt{b}}\,(1-\scP_+)\L\lambda_0(\xx_1)+\lambda_0(\xx_2)\R ,
 \eeq
 with the chiral product $\lambda_0(\xx)=1/[(x^+-x^+_0)(x^--x^-_0)]$. For the four-point function, $1-\chpm=x^\pm_{14}x^\pm_{23}/(x^\pm_{13}x^\pm_{24})$ gives
 \beq
 \label{eq:lorentz_4pt}
  \Psi^\oo_{(1,1)}=\l1-\scP^{(12)}_+\r\l1-\scP^{(34)}_+\r\sum_{i=1,2}\sum_{j=3,4}\log|x^+_{ij}|\log|x^-_{ij}| ,
 \eeq
 where $\scP^{(12)}_+=\scP_+$ and $\scP^{(34)}_+$ exchanges $x^+_3\leftrightarrow x^+_4$. This is the odd part of the exchange of a field $\lambda$, with two-point function $\la\lambda(\xx)\lambda(\xx')\ra$ equal to $\log|x^+-x'^+|\,\log|x^--x'^-|$, between two transformations \eqref{eq:lorentz_linearised}, since $\la\delta_\lambda G_a(\xx_1,\xx_2)\,\delta_\lambda G_b(\xx_3,\xx_4)\ra=v_av_b\,G_a(\xx_{12})G_b(\xx_{34})\sum_{i,j}\la\lambda(\xx_i)\lambda(\xx_j)\ra$. This two-point function is a chiral product defined only modulo chiral sums, since rescaling $x^\pm$ adds terms of one lightcone coordinate, and by Eq. \eqref{eq:lorentz_modulo} these drop out. For plane waves $\lambda=e^{\ii(k_+x^++k_-x^-)}$, Eq. \eqref{eq:chiral_product} gives $\prod_{a=\pm}(e^{\ii k_a x^a_1}-e^{\ii k_a x^a_2})$, which vanishes for the chiral sums $k_+=0$ or $k_-=0$. With $\int_{-\infty}^\infty\frac{dk}{2|k|}\sum_ic_ie^{\ii ka_i}=-\sum_ic_i\log|a_i|$ for $\sum_ic_i=0$, where the condition again removes the constants, Eq. \eqref{eq:lorentz_4pt} becomes
 \beq
 \label{eq:lorentz_plane_wave}
  \Psi^\oo_{(1,1)}=\int_{-\infty}^{\infty}\frac{dk_+}{2|k_+|}\int_{-\infty}^{\infty}\frac{dk_-}{2|k_-|}\prod_{a=\pm}\l e^{\ii k_a x^a_1}-e^{\ii k_a x^a_2}\r\l e^{-\ii k_a x^a_3}-e^{-\ii k_a x^a_4}\r .
 \eeq
 This is the analogue of the expansion of the $h=2$ contribution in reparametrisation modes in reference \cite{Maldacena_2016}. The factor $1/|k_+k_-|$ is proportional to the inverse norm of the modes, since $\int dy\,|1-e^{\ii ky}|^2/y^2=2\pi|k|$ on each lightcone with the measure of Eq. \eqref{eq:def_inner_product}.

 \paragraph{Away from the conformal limit.} In the 1D model the reparametrisations are broken explicitly by the UV term, which lifts the $h=2$ eigenvalues to $1-O(|n|/\beta J)$ and gives the enhanced contribution of order $\beta J$ \cite{Maldacena_2016}. The local Lorentz transformations are instead a symmetry of the full model. Absorbing the UV term of Eq. \eqref{eq:ds2} into $\hat\Sigma_a=\Sigma_a+2\ii\,\epsilon^{\mu\nu}e^a_\mu\partial_\nu$, the effective action splits into a part that depends only on $G_a$ and $\hat\Sigma_a$, which is invariant under Eq. \eqref{eq:lorentz_bilinears} with $\hat\Sigma$ in place of $\Sigma$, and a UV part that is a local function of $e^\pm_\mu$ and of the coincident derivatives of $G_a$, whose stationarity condition is Eq. \eqref{eq:ds3}. The coincident derivatives of $G_a$ transform with $\Lambda^{2v_a}(\xx)$, which compensates $e^a_\mu\to\Lambda^{-2v_a}e^a_\mu$ in the UV term. The full quadratic action therefore annihilates $\delta_\lambda G$ together with $\delta_\lambda e^\pm_\mu$ at any coupling. After the fluctuations of $e^\pm_\mu$ are eliminated, this is $(\mathbb{I}-K-K_{\rm UV})\,\delta_\lambda G=0$, where $K_{\rm UV}$ is a local rung that acts only through the first coincident derivatives. Nothing here corresponds to the explicit breaking that lifts $h=2$ in 1D, consistent with the local Lorentz symmetry being a gauge symmetry \cite{Turiaci_2017}.
 The odd $(1,1)$ state is nevertheless not a gauge mode. If Eq. \eqref{eq:lorentz_odd_mode} were of the form $\lambda'(\xx_1)+\lambda'(\xx_2)$, setting $\xx_1=\xx_2$ would give $\lambda'=0$, so fixing the gauge does not remove it. The eigenvalue one of the kernel on the state $(1,1)$ rests on the fact that $\delta_\lambda G$ is a zero mode, and so is $P_+\delta_\lambda G$, because $P_+$ commutes with the conformal kernels (Eq. \eqref{eq:P_commutes_K}). By Eq. \eqref{eq:odd_mode_rectangle}, the odd mode vanishes together with its first derivatives at coincident points, so $K_{\rm UV}$ annihilates it.
 
 The Lorentz channel can be written in a form that does not depend on what lifts the eigenvalue. We assume that the lifting is translation invariant and it shifts only the eigenvalue of each mode, leaving the eigenfunctions and the source in Eq. \eqref{eq:lorentz_11_term} unchanged. At fixed $(k_+,k_-)$ the odd mode with eigenvalue one is unique, so the corrected kernel acts on it by a number $k_L(k_+,k_-)$. Inserting Eq. \eqref{eq:lorentz_plane_wave} into the second term of Eq. \eqref{eq:lorentz_11_term}, dividing each mode by its own $1-k_L$, and reversing the sign algebra of Eq. \eqref{eq:lorentz_4pt}, the Lorentz channel becomes
 \beal
 \label{eq:lorentz_channel_exchange}
  \mathf^{\rm L}_{ab}(\xx_1,\dots,\xx_4)&=\frac{v_av_b}{4\pi^4\alpha_0^2}\l1-\scP^{(12)}_+\r\l1-\scP^{(34)}_+\r\sum_{i=1,2}\sum_{j=3,4}\scD_\lambda(\xx_i-\xx_j),\\
  \scD_\lambda(\xx)&=\int_{-\infty}^{\infty}dk_+\int_{-\infty}^{\infty}dk_-\;\frac{e^{\ii(k_+x^++k_-x^-)}}{4|k_+k_-|\l1-k_L(k_+,k_-)\r} .
 \eeal
 Like $\mathf_{ab}(\chp,\chm)$ in Eq. \eqref{eq:F_cross_ratio}, $\mathf^{\rm L}_{ab}$ is divided by $G_a(\xx_{12})G_b(\xx_{34})$, but it need not be a function of the cross ratios only. Eq. \eqref{eq:lorentz_channel_exchange} is Eq. \eqref{eq:lorentz_4pt} with $\log|x^+|\log|x^-|/(1-k_L)$ replaced by $\scD_\lambda$: the exchange of $\lambda$ with $\la\lambda(\xx)\lambda(0)\ra=\scD_\lambda(\xx)$, so that $4|k_+k_-|(1-k_L)$ is, up to normalisation, the quadratic action of the odd Lorentz modes. By Eq. \eqref{eq:lorentz_modulo} this action depends on $\lambda$ only through $\partial_+\partial_-\lambda$, so the terms in $\scD_\lambda$ that are constant or depend on only one lightcone coordinate drop out, and boost invariance makes $k_L$ a function of $k_+k_-$. The conformal limit $k_L=1$ is the divergence of Eq. \eqref{eq:lorentz_11_term}. Including the contributions of the state $(1,1)$ in the ladder sum, the order $1/N$ four-point function is expressed as
 \beq
 \label{eq:four_point_lorentz}
  \mathf_{ab}=\tilde{\mathf}_{ab}+\frac{1}{8q\pi^4\alpha_0^2}\log|1-\chp|\log|1-\chm|+\mathf^{\rm L}_{ab},
 \eeq
 with $\tilde{\mathf}_{ab}$ of Eq. \eqref{eq:contour_form} and the Lorentz channel $\mathf^{\rm L}_{ab}$ of Eq. \eqref{eq:lorentz_channel_exchange}. Its only remaining free ingredient is the propagator $\scD_\lambda$ for the odd Lorentz modes away from the conformal limit. Working out this propagator beyond the conformal regime is left for future work.

\section{Reparametrisation modes and towards a bulk dual}\label{sec:repara_modes}

 In the 1D SYK model, the reparametrisation modes connect the spectrum of the bilinears with the dual gravitational description. They are eigenfunctions of the conformal kernel with eigenvalue one. The explicit breaking of reparametrisation symmetry lifts their eigenvalue. Their effective action, the Schwarzian, is also the boundary action of nearly AdS$_2$ dilaton gravity \cite{kitaevtalks,Maldacena_2016,Almheiri:2014cka,Jensen:2016pah,maldacena2016conf,Engels_y_2016,Kitaev_2017,Mertens_2022}. In this section we discuss the corresponding modes of the 2D model in the light of the spectrum obtained in Sec. \ref{sec:four_point_fn}, and what they imply for a 3D bulk dual. The effective action of these modes and its relation to AdS$_3$ gravity were first discussed in reference \cite{Turiaci_2017}. We do not derive new effective action here. We explain how the results of Sec. \ref{sec:four_point_fn} would fit into that discussion, and we list what would be needed to establish a duality.

 \subsection{Reparametrisation modes}\label{subsec:reparametrisation_modes}

 Under an infinitesimal chiral reparametrisation, $x^+ \to x^+ + \vare(x^+)$, the two-point functions transform according to Eq. \eqref{eq:bilinear_transformations}. The variation in $G_a$ with respect to $\vare$ is
 \beq
 \label{eq:var_G}
  \delta_{\vare} G_a(\xx_1,\xx_2) = \L \frac{\Delta_a}{2} (\vare'(x_1^+) + \vare'(x_2^+)) + \vare(x_1^+)\partial_{x_1^+} + \vare(x_2^+)\partial_{x_2^+} \R G_a(\xx_1,\xx_2),
 \eeq
 where $\vare'(x^+_i) = \left. \partial_{x^+} \vare \right|_{x^+ = x^+_i}$. The linearised variations of the IR DS equations in Eqs. \eqref{eq:ds1} and \eqref{eq:ds2_IR} lead to
 \beq
  \delta_\vare G_a = \sum_b K_{ab} * \delta_\vare G_b ,
 \eeq
 so $\delta_\vare G$ is an eigenfunction of the kernel with eigenvalue one. By Eq. \eqref{eq:casimir_on_ratio}, the ratio $\delta_\vare G_a/G_a$ has the weights $(\hpl,\hmi) = (2,0)$: $C_-^{(1+2)}$ annihilates it because it has no dependence on $x^-$, and it has eigenvalue 2 under $C_+^{(1+2)}$. Its chirality vector $(\dpl,\dmi)$ is the eigenvector of the matrix in Eq. \eqref{eq:eigen_20} with eigenvalue one. Likewise, an infinitesimal chiral reparametrisation $x^- \to x^- + \tilde{\vare}(x^-)$ leads to the reparametrisation mode $(\hpl,\hmi) = (0,2)$ with eigenvalue one of the matrix in Eq. \eqref{eq:eigen_02} for the chirality vector $(\dmi,\dpl)$.

 The DS equations in the IR limit are invariant under a larger group. A reparametrisation of the form $x^+ \to \phi_+(x^+,x^-)$ with $\partial_{x^+} \phi_+ > 0$ preserves the IR DS equations provided $x^-$ coordinate remains unchanged. This is the diffeomorphism invariance of $S_{\text{IR}}$ with the fermions transforming as densities of weight $1/2q$ \cite{Turiaci_2017}, combined with the local Lorentz transformation of Sec. \ref{subsec:lorentz_mode}. Therefore, Eq. \eqref{eq:var_G} with $\vare(x^+,x^-)$ depending on both lightcone coordinates is also an eigenfunction of the conformal kernel with eigenvalue one. The ratio $\delta_\vare G_a/G_a$ is annihilated by $C_-^{(1+2)}$, and $C_+^{(1+2)}$ obtains the eigenvalue 2. So, the conformal weights of such reparametrisations are still $(2,0)$.

 Unlike the 1D SYK model, where the $h=2$ reparametrisation mode is part of the complete set of eigenfunctions, the reparametrisation modes $(2,0)$ and $(0,2)$ do not belong to the eigenbasis in Eq. \eqref{eq:eigenbasis} in the 2D model. In terms of cross ratios, the $(2,0)$ and $(0,2)$ modes do not decay as $\chm \to 0$ and $\chp \to 0$, respectively. Thus, the reparametrisation modes in the 2D model are not normalisable. The only normalisable zero mode is the odd Lorentz mode of Sec. \ref{subsec:lorentz_mode}, and its eigenvalue is protected because the local Lorentz symmetry is not broken explicitly.
 
 \subsection{Effective action of the reparametrisations}\label{subsec:eff_action}

 The Schwarzian action of the 1D model is the action of the configurations obtained by reparametrising the conformal solution. Locality and the SL$(2,\mathbb{R})$ invariance of the conformal solution fix this action to be the Schwarzian derivative, and the kernel enters only through its coefficient \cite{Maldacena_2016,Kitaev_2017}. The reparametrisation mode is a coordinate on the coset Diff$(S^1)$/SL$(2,\mathbb{R})$, a coadjoint orbit of the Virasoro group, and the Schwarzian is the natural action on this orbit \cite{Alekseev_1988,Stanford_2017,Mandal_2017}.

 The same route was followed for the 2D model in reference \cite{Turiaci_2017}. Shifting $\Sigma_a$ by the UV term of Eq. \eqref{eq:ds2} moves all the dependence on the frame fields into a local term that contains the coincident derivatives of $G_a$ appearing in Eq. \eqref{eq:ds3}. This term is evaluated on the chiral reparametrisations $\l x^+(u),x^-(v)\r$ of the conformal solution, with the singular part of the coincident limit subtracted. In the chiral limit and near $q=2$ it becomes \cite{Turiaci_2017}
 \beq
 \label{eq:rb_tv_action}
  \frac{S}{N}=\frac{\mathtt{b}}{12}\int du\,dv\L e^+_v\{x^+,u\}+e^-_u\{x^-,v\}\R-\int d^2x\,\epsilon^{\mu\nu}e^+_\mu e^-_\nu ,
 \eeq
 where $\{x,u\}=x'''/x'-\frac32\l x''/x'\r^2$ is the Schwarzian derivative. The off-diagonal frame components $e^+_v$ and $e^-_u$ are sources for the Schwarzian derivatives. Eliminating the frame fields sets $e^-_u$ and $e^+_v$ proportional to $\{x^+,u\}$ and $\{x^-,v\}$, and gives the double Schwarzian action,
 \beq
 \label{eq:rb_double_schwarzian}
  S=\frac{N\alpha_S}{J}\int du\,dv\,\{x^+,u\}\{x^-,v\} .
 \eeq
 The coefficient $\alpha_S$ was obtained in reference \cite{Turiaci_2017} in the chiral limit $j \to \Delta$, to second order in $\epsilon_q = 1- \frac{2}{q}$ around $q=2$, as
 \beq
 \label{eq:alpha_S_q=2}
  \alpha_S = \frac{\alpha_0}{144}(1-2\epsilon_q^2) + \mathcal{O}(\epsilon_q^4).
 \eeq
 Although $\alpha_0$ diverges in the chiral limit, $\alpha_0/J$ and the coefficient $N\alpha_S/J$ of the action in Eq. \eqref{eq:rb_double_schwarzian} are finite.

 Since $x^+$ depends only on $u$ and $x^-$ only on $v$, the action \eqref{eq:rb_double_schwarzian} is the product of two one-dimensional Schwarzian actions,
 \beq
 \label{eq:rb_factorised}
  S=\frac{N\alpha_S}{J}\L\int du\,\{x^+,u\}\R\L\int dv\,\{x^-,v\}\R .
 \eeq
 For $x^+=u+\varepsilon_+(u)$ and $x^-=v+\varepsilon_-(v)$, integration by parts gives $\int du\,\{x^+,u\}=-\frac12\int du\,(\varepsilon_+'')^2+O(\varepsilon_+^3)$, and likewise for $x^-$. Hence
 \beq
 \label{eq:rb_quartic}
  S=\frac{N\alpha_S}{4J}\int du\,(\vare_+'')^2\int dv\,(\vare_-'')^2+\dots ,
 \eeq
 where the dots represent higher order terms. Each factor of Eq. \eqref{eq:rb_quartic} starts at second order in $\vare_\pm$, so around the zero temperature vacuum on two-dimensional Minkowski space, $\vare_\pm = 0$, the expansion of the action starts at fourth order. There is no term quadratic in $\vare_\pm$ that could act as a kinetic term, and therefore there is no propagator for the chiral reparametrisations and no enhanced contribution at order $1/N$ in this state. The same holds in a 2D CFT: a holomorphic reparametrisation is a symmetry, and the quadratic action of the reparametrisation mode comes entirely from its dependence on the other lightcone coordinate \cite{Haehl_2018,Cotler_2018}. A quadratic action for $\vare_+$ appears when the $x^-$ sector carries energy, $ \int dv \left\lbrace x^-, v \right\rbrace \neq 0$. In a thermal state, when the $x^-$ sector is at inverse temperature $\beta_-$,
 \beq
  x^- = e^{2\pi v/ \beta_-} \qquad \text{and} \qquad \left\lbrace x^- , v \right\rbrace = - \frac{2\pi^2}{\beta_-^2}.
 \eeq
 We take the $v$ direction to have a finite length $L_-$, which regulates the infinite lightcone, so that $\int dv \lbrace x^-, v \rbrace = -2\pi^2 L_-/ \beta_-^2$. Then Eq. \eqref{eq:rb_factorised} becomes one-dimensional Schwarzian action for $x^+$,
 \beq
 \label{eq:rb_chiral_schwarzian}
  S = -\frac{2\pi^2 N \alpha_S L_-}{J \beta_-^2} \int du \lbrace x^+, u \rbrace.
 \eeq
 This is the reduction to a single Schwarzian action noted in reference \cite{Turiaci_2017}. A universal Schwarzian sector of this kind also appears in 2D CFTs with a large central charge, in near-extremal states where one chirality is at low temperature \cite{Ghosh_2019}. The coupling in Eq. \eqref{eq:rb_chiral_schwarzian} is proportional to $L_-$, so the fluctuations of $\varepsilon_+$ are suppressed by $1/L_-$.

 Away from the chiral limit and from $q=2$, the coefficient $\alpha_S$ is not known. Non-conformal corrections to the kernels shift their eigenvalues on the states $(2,0)$ and $(0,2)$. The shift has to be evaluated for chiral modes on a circle or in a thermal state, or for the modes with $\vare(x^+,x^-)$ of Sec. \ref{subsec:reparametrisation_modes}, whose action would be the two-dimensional generalisation of the Schwarzian. This requires the solution beyond the conformal limit, as for the Lorentz mode of Sec. \ref{subsec:lorentz_mode}, and we leave it for future work.

 \subsection{Relation to the stress tensor}\label{subsec:stress_tensor}

 The weights $(2,0)$ and $(0,2)$ are those of the chiral components $T(x^+)$ and $\bar{T}(x^-)$ of a stress tensor, and reference \cite{Turiaci_2017} identified the $(2,0)$ mode with the stress tensor. The relation can be made precise through the conformal Ward identity. In a 2D CFT the stress tensor generates the reparametrisations $x^+\to x^++\vare(x^+)$. Its OPE with a field of weight $\Delta_a/2$ on the $x^+$ lightcone is
 \beq
 \label{eq:rb_T_ope}
  T(x^+_0)\,\psi_a(\xx_1)\sim\frac{\Delta_a/2}{(x^+_{01})^2}\,\psi_a(\xx_1)+\frac{1}{x^+_{01}}\,\partial_{x^+_1}\psi_a(\xx_1) ,
 \eeq
 so that each field changes by $\l\frac{\Delta_a}{2}\vare'+\vare \,\partial_+\r\psi_a$, and the two-point function changes by the first line of Eq. \eqref{eq:var_G}. The same Ward identity fixes the three-point function
 \beq
 \label{eq:rb_T_3pt}
  \la T(x^+_0)\,\psi_a(\xx_1)\,\psi_a(\xx_2)\ra=\frac{\Delta_a}{2}\,\frac{(x^+_{12})^2}{(x^+_{01})^2(x^+_{02})^2}\,G_a(\xx_{12}) ,
 \eeq
 whose coefficient is proportional to the weight $\Delta_a$. The eigenvector $(\dpl,\dmi)$ of the matrix in Eq. \eqref{eq:eigen_20} with eigenvalue one has the same property: the $(2,0)$ mode couples to $\psi_+$ and $\psi_-$ in the ratio of their weights on the $x^+$ lightcone, as the stress tensor does. The reparametrisation $\vare(x^+)$ itself has the weights $(-1,0)$, same as the weights of the shadow of $T$ \cite{Haehl_2018,Haehl_2019}.

 $T$ and $\bar{T}$ are not operators of the UV theory, which is a topological CFT with gauged Virasoro symmetry (Sec. \ref{subsec:tcft}). In the IR they are composite fields of the reparametrisation modes. The Schwarzian derivatives $\{x^+,u\}$ and $\{x^-,v\}$ are the stress tensors induced by the conformal map $(u,v)\to(x^+,x^-)$, and in Eq. \eqref{eq:rb_tv_action} the frame components $e^+_v$ and $e^-_u$ couple to them as metric perturbations couple to $T$ and $\bar{T}$. Eliminating the frame fields produces the product $T\bar{T}$. This is the origin of the interpretation of the leading deviation from conformal symmetry as a $T\bar{T}$ deformation \cite{Turiaci_2017,Zamolodchikov_2004,Smirnov_2016,Cavaglia_2016}, in the same way as the $T\bar{T}$ deformation of a 2D QFT is obtained by coupling it to a dynamical frame with an action quadratic in the frame fields \cite{McGough_2018,Cardy_2018,Dubovsky_2017,Dubovsky_2018}.

 For $j<j_0(q)$, where $j_0$ is the root of $\partial_{\hpl}\det(\mathbb{I}-\mathk^{\ee})=0$ at $(0,2)$ ($j_0 \simeq -\Delta^2/2$ at large $q$), the conformal four-point function contains no block with the weights of $T$ or $\bar{T}$. The points $(2,0)$ and $(0,2)$ lie on the spectral curve in Eq. \eqref{eq:spectral_curve}, but they are not basis states. Their reflections $(2,1)$ and $(1,2)$ are zeros of $\det(\mathbb{I}-\mathk^\sigma)$ in both sectors. In Eq. \eqref{eq:ope_one_lightcone} the $\hmi$ contour keeps the discrete states of its sector, $\hmi=1$ in the odd sector and $\hmi=2$ in the even sector, so closing the $\hpl$ contour reaches $(2,1)$ in the odd sector and $(1,2)$ in the even sector. There the measure has a simple zero, $\mu^{\oo}(2)=\mu^{\ee}(1)=0$, which cancels the pole. For $j>j_0(q)$, the zero that continues the light GFF bilinear at the even discrete state $\hmi=2$ ends at $(0,2)$. There it meets the double pole of $\mu^{\ee}(\hpl)\Gamma(\hpl)^2/\Gamma(2\hpl)$ at $\hpl=0$, which lies on the other side of the contour. The contour is pinched, and the residue diverges as the kernel approaches its conformal form. By the symmetry $\hpl\leftrightarrow\hmi$ the same happens at $(2,0)$. This is similar to the 1D model, where the energy enters the four-point function through the enhanced contribution of the $h=2$ modes and not as an operator of the conformal OPE \cite{Maldacena_2016}. The normalisation of $T$, i.e. the central charge, is therefore fixed by the explicit breaking. Reference \cite{Turiaci_2017} compared the free energy of the action \eqref{eq:rb_double_schwarzian} on a cylinder of circumference $L$ with that of a CFT with a $T\bar{T}$ deformation of coupling $\mu$, and found
 \beq
 \label{eq:rb_c_mu}
  \frac{c}{24\pi}=\frac{N\alpha_S}{4J},\qquad \mu=\frac{24\pi}{c},
 \eeq
 in units where $L=2\pi$. Since $J$ has mass dimension two, restoring the units gives $c\propto N\alpha_S/(JL^2)$. The central charge defined in this way depends on the scale, like the coefficient $N\alpha_S/(\beta J)$ of the Schwarzian action that determines the specific heat of the 1D model \cite{Maldacena_2016}. It measures the explicit breaking rather than a property of the IR fixed point.

 \subsection{Towards a bulk dual}\label{subsec:bulk_dual}

 The relation of the 1D model to nearly AdS$_2$ gravity relies on two facts. The soft sector, governed by the Schwarzian action, is the boundary dynamics of Jackiw-Teitelboim gravity \cite{Maldacena_2016,Jensen:2016pah,maldacena2016conf,Engels_y_2016}, and the bilinears form a tower of bulk fields whose couplings follow from the three-point functions \cite{Maldacena_2016,Jevicki_2016,Polchinski_2016,Gross_2017}. A duality for the 2D model can be approached along the same two lines, supplemented by dynamical tests.

 For the soft sector, reference \cite{Turiaci_2017} showed that the double Schwarzian action \eqref{eq:rb_double_schwarzian} equals the on-shell Einstein action of AdS$_3$ with a finite radial cutoff if $1/16\pi G=N\alpha_S/4J$ in units of the AdS radius. With Eq. \eqref{eq:rb_c_mu} this gives the Brown-Henneaux central charge $c=3/2G$  \cite{Brown_1986}, and $\mu=24\pi/c$ is the relation $\mu=16\pi G/r_c^2$ of reference \cite{McGough_2018} with the cutoff at the AdS radius, $r_c=1$. This identification concerns pure gravity. With bulk matter, a sharp cutoff requires further double-trace deformations \cite{Kraus_2018}, and the $T\bar T$ deformation alone corresponds to mixed boundary conditions at the asymptotic boundary \cite{Guica_2021}. The double Schwarzian also leaves out the reparametrisations $x^+\to\phi_+(x^+,x^-)$ of Sec.~\ref{subsec:reparametrisation_modes}. In AdS$_3$ gravity the boundary gravitons are reparametrisations of this kind, governed by two copies of the Alekseev-Shatashvili action with the Brown-Henneaux central charge as coupling \cite{Alekseev_1988,Cotler_2018}. At quadratic order in $\vare=\phi_+-x^+$ this action is proportional to $c\int d^2x\,\partial_-\vare\,\partial_+^3\vare$, which vanishes for a purely chiral $\vare(x^+)$ as Eq.~\eqref{eq:rb_quartic} does, and its exchange gives the stress-tensor contribution at order $1/c$ \cite{Haehl_2018,Cotler_2018}. Reference \cite{Altland_2025} derived this action from a chain of SYK grains coupled by a chiral kinetic term, and matched its coupling to that of the BTZ black hole. A dual with three-dimensional gravity therefore requires that the explicit breaking gives $\phi_+(x^+,x^-)$ an action of this form. Its coefficient would also fix the contribution at the weights $(2,0)$ and $(0,2)$ for $j>j_0(q)$ (Sec. \ref{subsec:stress_tensor}), and together with $\alpha_S$ at general $(q,j)$ it is the main missing input.

 Finally, there are dynamical tests. Einstein gravity gives the maximal Lyapunov exponent $2\pi/\beta$ \cite{Maldacena:2015waa} and, in 2D, a butterfly velocity equals to the speed of light \cite{Roberts2014}. A maximal growth for the 2D model has to be checked with the retarded kernel at finite temperature. By scale invariance, the conformal saddle point contributes to $\log Z$ a term proportional to $NL/\beta$, which defines a Cardy central charge of order $N$. In a holographic CFT it coincides with the normalisation of $T$, whereas $c$ in Eq. \eqref{eq:rb_c_mu} is set by the explicit breaking. A dual of the conformal IR is best defined in the shaded region of Fig. \ref{fig:validity_region}. Outside it the kernel has eigenvalue one on normalisable states of the principal series, which in 1D SYK-like models signals operators of complex dimension and an instability \cite{Murugan_2017,Giombi_2017,Kim_2019,Benedetti_2021}. The large-$q$ limit, where the 1D model reduces to Liouville quantum mechanics \cite{Maldacena_2016,Bagrets_2016} and where reference \cite{Turiaci_2017} suggested that the UV term could be treated exactly, is the natural place to compute $\alpha_S$ at general $j$ and the action of $\phi_+(x^+,x^-)$.

\section{Conclusion}\label{sec:conclusion}

 In this work we have studied the four-point function of the 2D generalisation of the SYK model proposed by Turiaci and Verlinde \cite{Turiaci_2017} at order $1/N$ in the conformal limit, following the approach of Maldacena and Stanford for the 1D model \cite{Maldacena_2016}. The UV theory is a tensor product of topological Ising CFTs, and its Minkowski correlation functions follow from the braiding of chiral vertex operators. In the IR, the real-time sign of the self-energy is required for a conformal solution with a real propagator and a positive variance of the couplings, which restricts the spin to $|j|<\Delta$.

The conformal kernels commute with the exchange of a single lightcone coordinate. As a consequence, the four-point function splits into an even and an odd sector under the reflection of one cross ratio. In each sector we found a complete basis of eigenfunctions of the Casimir and the eigenvalues of the kernels. The eigenvalues form $2 \times 2$ matrices in the chirality of the fermions, whose entries are products of chiral functions of the two conformal weights. The projections of the zero-rung ladder are given by the same functions, and the sum over ladders takes a closed form, which we have expressed as a double contour integral. We checked numerically that this representation reproduces the zero-rung ladder in both sectors, provided the contour is deformed to keep the poles $h=\dpm$ to its right and their images $h=1-\dpm$ to its left.

The main difference from the 1D model is that the poles of the ladder sum do not lie at isolated points but on a spectral curve in the plane of the two conformal weights. In the expansion in conformal blocks of one lightcone, the weights on that lightcone depend on the weight of the other. Our numerical study shows that, for weights of the other lightcone on the principal series, they sweep continuous bands that end at the weights of the generalised free field. The spin of the operators is not quantised, in line with the spin $j$ of the fermions. The contour representation, and hence this operator content, holds only in a region of the $(q,j)$ plane. For $q\ge34$ it requires $j<j_c(q)$. For $q\le32$ the kernel has eigenvalue one on states of the principal series unless $j$ is close to $-\Delta$, and at $q=4$ this happens for every $j$.

The kernel has eigenvalue one on the chiral reparametrisations with weights $(2,0)$ and $(0,2)$, and on the state $(1,1)$. Unlike in the 1D model, the reparametrisation modes are not normalisable, and the conformal four-point function contains no block with the weights of the stress tensor for $j<j_0(q)$. The odd $(1,1)$ state is normalisable and is the odd part of the local Lorentz transformations of the conformal solution. The eigenvalue one of this state is protected by the factorisation of the conformal propagators rather than by a symmetry that is broken explicitly, and its contribution is not regulated in the conformal limit. The double Schwarzian action of reference \cite{Turiaci_2017} starts at quartic order around the vacuum on the 2D Minkowski plane, so the chiral reparametrisations give no enhanced contribution there. A quadratic action for one lightcone coordinate appears when the other lightcone is at finite temperature.

Several questions still remain open. The most immediate is the solution beyond the conformal limit. It determines whether the Lorentz channel and the reparametrisation modes acquire enhanced contributions, the coefficient of the effective action at general $(q,j)$, and the scale-dependent central charge. Outside the region where the contour representation of the four-point function holds, the kernel has eigenvalue one on the principal series. In SYK-like models, the ladder kernel having eigenvalue one on the principal series corresponds to an operator with complex dimension \cite{Murugan_2017,Giombi_2017}, the analogue of a field below the Breitenlohner–Freedman bound, and signals an instability of the conformal solution \cite{Kim_2019,Benedetti_2021}. In the 2D model studied here, these zeros lie on bands rather than at the weights of operators, and whether they signal an instability remains an open question. A numerical solution of the Dyson-Schwinger equations could show what happens here. The retarded kernel at finite temperature would determine the Lyapunov exponent and the butterfly velocity, which a holographic dual with Einstein gravity requires to be maximal. Finally, it remains to be seen whether the model has a bulk dual and how the spectrum obtained here would be encoded in it. We leave these questions for future work.

\section*{Acknowledgements}

 A.C. and P.K.D. acknowledge financial support from the DST-INSPIRE fellowship (IF200258, IF200253), Department of Science and Technology, Government of India.

\appendix

\section{Conformal kernel of the ladder diagrams}\label{app:conf_kernel}
 This appendix contains the expressions of the chiral factors constituting the integration kernels of Eq. \eqref{eq:ladder_kernel}. In the conformal limit, the kernels are expressed in terms of the two-point functions as
 \beq
  K_{ab}(\xx_1,\xx_2,\xx_3,\xx_4) = J^2 (q-\delta_{ab})G_a(\xx_{13})G_a(\xx_{24})L_{ab}(\xx_{34}),
 \eeq
 where the rung factor involves
 \beq
  L_{ab}(\xx) = \frac{\L G_+(\xx)G_-(\xx) \R^q}{G_a(\xx)G_b(\xx)}.
 \eeq
 Since the two-point functions are factorised into chiral lightcone pieces, the kernels can also be brought to a factorised form
 \beq
  K_{ab} = \frac{1}{\alpha_{ab}} K^{+}_{ab}(x_i^+) K^{-}_{ab}(x_i^-) \quad \mathrm{with} \quad \frac{1}{\alpha_{ab}} = (q-\delta_{ab}) \alpha_0^2.
 \eeq
 The chiral factors follow directly from the two point function and takes the compact form
 \beal
 \label{eq:chiral_kernels}
  K_{aa}^{a}(x_i^a) &= \frac{\sgn{x_{13}^a}\sgn{x_{24}^a}}{|x_{13}^a|^{\dpl}|x_{24}^a|^{\dpl}|x_{34}^a|^{2-2\dpl}},
 &\quad K_{aa}^{\bar a}(x_i^{\bar a}) &= \frac{1}{|x_{13}^{\bar a}|^{\dmi}|x_{24}^{\bar a}|^{\dmi}|x_{34}^{\bar a}|^{2-2\dmi}}, \\
 K_{a\bar a}^{a}(x_i^a) &= \frac{\sgn{x_{13}^a}\sgn{x_{24}^a}\sgn{x_{34}^a}}{|x_{13}^a|^{\dpl}|x_{24}^a|^{\dpl}|x_{34}^a|^{2-2\Delta}},
 &\quad K_{a\bar a}^{\bar a}(x_i^{\bar a}) &= \frac{\sgn{x_{34}^{\bar a}}}{|x_{13}^{\bar a}|^{\dmi}|x_{24}^{\bar a}|^{\dmi}|x_{34}^{\bar a}|^{2-2\Delta}}.
 \eeal
 where $\bar{a}=-a$. In $K^c_{ab}$ the factor $\sgn{x_{13}} \sgn{x_{24}}$ and the exponent $\dpl$ appear exactly when the lightcone $c$ equals the first chirality index $a$. The factor $\sgn{x_{34}}$ appears on both lightcones in the off-diagonal kernels only. This is the same orientation structure that appears in the second three-point structure discussed in Sec. \ref{subsec:eigenvalues} and Appendix \ref{app:eigenvalues}.

\section{The kernels as functions of cross ratios}\label{app:kernel_cross_ratio}
 In this appendix we express the integration kernels of Eq. \eqref{eq:ladder_kernel} as a function of cross ratios, and fold it onto the region $0<\chpm<2$ used in Eq. \eqref{eq:ladder_cross_ratio}. The $(n+1)$-rung ladder is obtained by acting the kernels on the $n$-rung ladder as
 \beq
 \label{eq:C1}
  G_a(\xx_{12}) \mathf_{ab}^{(n+1)}(\chp,\chm) = \int d^2 \xx_5 d^2 \xx_6 \sum_{c=\pm} K_{ac}(\xx_1,\xx_2,\xx_5,\xx_6) G_c(\xx_{56}) \mathf_{cb}^{(n)} (\chi'^+,\chi'^-).
 \eeq
  The kernel of Eq. \eqref{eq:ladder_kernel} is accompanied by the factor $G_c(x_{56})$ coming from the conformal form \eqref{eq:F_cross_ratio} of the $n$-rung ladder. The product $K_{ac}\,G_c(x_{56})$ is independent of $c$. The additional $\sgn{x^a_{56}}$ and the exponent $2-2\Delta$ carried by $K_{a \bar{a}}$ relative to $K_{aa}$ in Eq. \eqref{eq:C1} are exactly compensated by the corresponding factors of $G_{\bar a}(x_{56})$ relative to $G_a(x_{56})$. Therefore, only the weight $1/\alpha_{ac}=(q-\delta_{ac})\alpha_0^2$ depends on the index $c$.
 
  The equation can be reduced to a two-dimensional integral equation using conformal symmetry. We can set $x_1^\pm = 0$, $x_3^\pm = 1$ and $x_4^\pm \to \infty$, so that $x_2^\pm = \chpm $. Trading the integration variables $x_6^\pm$ for $\chi'^\pm = \frac{x_{56}^\pm}{x_5^\pm -1}$ with the integration measure $dx_5^\pm dx_6^\pm = (1-x_5^\pm) dx_5^\pm d\chi' $, we find 
\beq
 \label{eq:int_p_pm}
 \mathf^{(n+1)}_{ab}(\chp,\chm) = \int_{-\infty}^{\infty} \frac{d\chi'^a}{|\chi'^a|^2} \int_{-\infty}^{\infty} \frac{d\chi'^{\bar{a}}}{|\chi'^{\bar{a}}|^2} \; p_{+}(\chi^a,\chi'^a) \; p_{-}(\chi^{\bar{a}},\chi'^{\bar{a}}) \sum_{c=\pm} \frac{1}{\alpha_{ac}}\mathf^{(n)}_{cb}(\chi'^+,\chi'^-) ,
\eeq
 with $\bar{a} = -a$, and the chiral kernels are
 \beq
  p_+(\chi,\chi') = \sgn{\chi \chi'} \l \frac{|\chi||\chi'|}{|\chi - \chi'|} \r^{\dpl} m_+(\chi,\chi'), \quad
  p_-(\chi,\chi') = \l \frac{|\chi||\chi'|}{|\chi - \chi'|} \r^{\dmi} m_-(\chi,\chi'),
 \eeq
 where the functions $m_\pm$ are given by the one-dimensional integrals
 \beal
  m_+(\chi^a,\chi'^a) &= \sgn{\chi^a -\chi'^a} \int_{-\infty}^{\infty} dx_5^+ \frac{\sgn{x_5^+} \sgn{1- x_5^+} \sgn{1- \frac{1-\chi'^a}{\chi^a - \chi'^a}x_5^+}}{|x_5^+|^{\dpl}|1-x_5^+|^{1-\dpl} \left| 1- \frac{1-\chi'^a}{\chi^a - \chi'^a}x_5^+ \right|^{\dpl}} , \\
  m_-(\chi^{\bar{a}},\chi'^{\bar{a}}) &= \int_{-\infty}^{\infty} dx_5^- \frac{1}{|x_5^-|^{\dmi}|1-x_5^-|^{1-\dmi} \left| 1- \frac{1-\chi'^{\bar{a}}}{\chi^{\bar{a}} - \chi'^{\bar{a}}}x_5^- \right|^{\dmi}}.
 \eeal
 The exponents in this expression follow the index $a$, so that $\dpl$ always accompanies $m_+$. The integrals over $x_5^\pm$ can be determined by dividing up the region of integration and using the Euler identity,
 \beq
  \label{eq:Euler_integral}
  \int_0^1 \frac{dx}{(1-\eta x)^a x^b (1-x)^c} = \frac{\Gamma(1-b)\Gamma(1-c)}{\Gamma(2-b-c)} \ff(a,1-b,2-b-c,\eta).
 \eeq
 The functions $m_{\pm}$ are obtained as
 \beal
m_\pm(\chi^a,\chi'^a)=
\begin{cases}
 \frac{2\pi}{\sin(\pi \dpm)} \ff(1-\dpm,\dpm,1,z_a) -B_{\dpm}\l\frac{1}{1-z_a}\r -B_{1-\dpm}\l\frac{1}{1-z_a}\r, & z_a \le 0, \\
\mp \frac{2\pi z_a^{-\dpm}}{\sin(\pi \dpm)} \ff \l \dpm,\dpm,1,\frac{z_a-1}{z_a} \r  +\frac{2\pi}{\sin(\pi \dpm)} \ff(\dpm,1-\dpm,1,z_a), & 0\le z_a \le 1, \\
\mp \frac{2\pi}{\sin(\pi \dpm)} \ff(1-\dpm,\dpm,1,1-z_a) +B_{\dpm}\l\frac{1}{z_a}\r +B_{1-\dpm}\l\frac{1}{z_a}\r, & z_a \ge 1,
\end{cases}
 \eeal
where
\beq
 z_{a} = \frac{1-\text{min}(\chi^a , \chi'^a)}{|\chi^a - \chi'^a|} \quad \text{and} \quad B_{\Delta}(z) = \frac{\Gamma(\Delta)^2}{\Gamma(2\Delta)} \; \ff (\Delta,\Delta,2\Delta,z).
\eeq

 We are interested in applying the kernel in the region $0<\chpm \le 2$ that supports the normalisation of the eigenfunctions and the inner products to compute the sum over ladder diagrams. The transformation $\chi \to \frac{\chi}{\chi-1}$ maps the interval $(-\infty,0)$ to $(0,1)$ and $\left[2,\infty \right)$ to $\left(1,2 \right]$. As noted in Sec. \ref{subsec:conf_symmetry}, the simultaneous reflection $(\chp,\chm) \to \left( \frac{\chp}{\chp-1},\frac{\chm}{\chm-1} \right)$ is a symmetry of the ladder contributions at every order,
 \beq
  \mathf_{ab}^{(n)} (\chp,\chm) = \mathf_{ab}^{(n)} \left( \frac{\chp}{\chp-1},\frac{\chm}{\chm-1} \right).
 \eeq
 A reflection of a single cross ratio is not a symmetry of $\mathf^{(n)}_{ab}$. Therefore, folding involves the reflected function besides $\mathf_{ab}^{(n)}$,
 \beq
  \mathf_{ab}^{(n),\text{ref}} (\chp,\chm) := \mathf_{ab}^{(n)} \left( \frac{\chp}{\chp-1},\chm \right) = \mathf_{ab}^{(n)} \left( \chp,\frac{\chm}{\chm-1} \right),
 \eeq
 which gives $\mathf_{ab}^{(n)}$ on the regions where exactly one of $\chpm$ lies outside $(0,2)$. We split each $\chpm$ integral in Eq. \eqref{eq:int_p_pm} at the fixed points and substitute $\chi \to \frac{\chi}{\chi-1}$ for the regions outside $(0,2)$. The integration measure is unchanged and each outer region gives the reflected kernel
 \beal
  \tilde{p}_+(\chi,\chi') &:= p_+ \l \chi, \frac{\chi'}{\chi'-1} \r = \sgn{\chi' -1} \l \frac{|\chi| |\chi'|}{|\chi +\chi' -\chi \chi'|} \r^{\dpl} m_+ \l \chi,\frac{\chi'}{\chi'-1} \r , \\
  \tilde{p}_-(\chi,\chi') &:= p_- \l \chi, \frac{\chi'}{\chi'-1} \r = \l \frac{|\chi| |\chi'|}{|\chi +\chi' -\chi \chi'|} \r^{\dmi} m_- \l \chi,\frac{\chi'}{\chi'-1} \r.
 \eeal
 The chiral kernels are invariant when the reflection acts on both of their arguments, $p_\pm \l \frac{\chi}{\chi-1}, \frac{\chi'}{\chi'-1} \r = p_\pm (\chi,\chi') $ and $\tilde{p}_\pm \l \frac{\chi}{\chi-1}, \frac{\chi'}{\chi'-1} \r = \tilde{p}_\pm (\chi,\chi') $, which implies $p_\pm \l \frac{\chi}{\chi-1}, \chi' \r = \tilde{p}_\pm (\chi,\chi') $ and $\tilde{p}_\pm \l \frac{\chi}{\chi-1}, \chi' \r = p_\pm (\chi,\chi') $. So, the $n$-rung ladder and its reflected function can be folded to the region $0 < \chpm \le 2$ as
 \beq
 \label{eq:matrix_valued_kernel}
  \begin{pmatrix}\mathf^{(n+1)}_{ab}\\ \mathf_{ab}^{(n+1),\text{ref}}\end{pmatrix}\!(\chp,\chm)
=\int_0^2\!\!\int_0^2 \frac{d\chi'^+ d\chi'^-}{(\chi'^+ \chi'^-)^2}\sum_{c=\pm}\frac{1}{\alpha_{ac}}
\begin{pmatrix} \mathcal{A}_a & \mathcal{B}_a\\ \mathcal{B}_a & \mathcal{A}_a\end{pmatrix}
\begin{pmatrix}\mathf^{(n)}_{cb} \\ \mathf_{cb}^{(n),\text{ref}}\end{pmatrix}\!(\chi'^+ ,\chi'^-),
 \eeq
 where, the folded kernels are
 \beal
  \mathcal{A}_a (\chi^a,\chi^{\bar{a}},\chi'^a,\chi'^{\bar{a}}) &= p_+(\chi^a,\chi'^a) p_-(\chi^{\bar{a}},\chi'^{\bar{a}}) + \tilde{p}_+(\chi^a,\chi'^a) \tilde{p}_-(\chi^{\bar{a}},\chi'^{\bar{a}}), \\
  \mathcal{B}_a (\chi^a,\chi^{\bar{a}},\chi'^a,\chi'^{\bar{a}}) &= p_+(\chi^a,\chi'^a) \tilde{p}_-(\chi^{\bar{a}},\chi'^{\bar{a}}) + \tilde{p}_+(\chi^a,\chi'^a) p_-(\chi^{\bar{a}},\chi'^{\bar{a}}).
 \eeal
 It is easy to see that the kernel remains invariant when acted with the chiral reflection operators $R_\pm$ defined in Sec. \ref{subsec:conf_symmetry}. Since $\mathcal{A}_a \pm \mathcal{B}_a = (p_+ \pm \tilde{p}_+)(p_- \pm \tilde{p}_-) $, the matrix valued equation \eqref{eq:matrix_valued_kernel} can be decoupled by taking the combinations
 \beq
  \mathf^{(n)}_{ab}=\mathf^{(n),\ee}_{ab}+\mathf^{(n),\oo}_{ab},\qquad
  \mathf^{(n),\sigma}_{ab} = \frac{1}{2}\l \mathf^{(n)}_{ab} +\tau \mathf_{ab}^{(n),\text{ref}} \r,
 \eeq
 where $\tau = +1$ for $\sigma=$ even and $\tau = -1$ for $\sigma=$ odd. $\mathf^{\ee}$ and $\mathf^{\oo}$ are even and odd, respectively, on the whole $\chpm$ plane under single chiral reflections. The folded and decoupled ladder recursion relations become
 \beq
  \mathf^{(n+1),\sigma}_{ab}(\chp,\chm) = \int_0^2\!\!\int_0^2 \frac{d\chi'^+ d\chi'^-}{(\chi'^+ \chi'^-)^2} (p_+ +\tau \tilde{p}_+)(p_- +\tau \tilde{p}_-) \sum_{c=\pm}\frac{1}{\alpha_{ac}} \mathf^{(n),\sigma}_{cb}(\chi'^+,\chi'^-).
 \eeq
 Both sectors are sourced by the zero-rung ladder in Eq. \eqref{eq:zero_rung_ladder}, which is decomposed into the two sectors and can be expressed for $0 < \chpm \le 2$ as
 \beq
  \mathf^{(0),\sigma}_{b}(\chp,\chm) = -\frac{1}{2} \L \l \chi^b \r^{\dpl} + \tau \; \sgn{\chi^b -1} \l \frac{\chi^b}{|\chi^b -1|} \r^{\dpl}  \R \L \l \chi^{\bar{b}} \r^{\dmi} + \tau \l \frac{\chi^{\bar{b}}}{|\chi^{\bar{b}} -1|} \r^{\dmi}  \R,
 \eeq
 with $\mathf^{(0),\sigma}_{ab} = \delta_{ab}\mathf^{(0),\sigma}_{b}$. Since the folded kernels satisfy $(p_\pm + \tau \tilde{p}_\pm) \l \frac{\chi}{\chi-1} ,\chi' \r = \tau (p_\pm + \tau \tilde{p}_\pm) \l \chi,\chi' \r$, each sector is preserved at every order.

\section{Complete basis of eigenfunctions of the Casimir}\label{app:eigenbasis_of_Casimir}

 In this appendix, we identify a complete basis of eigenfunctions of the Casimir imposing proper boundary conditions. As a differential operator of the cross ratios, the Casimir operator splits into two chiral parts,
 \beq
  \cc = \cc_+ + \cc_- , \quad \text{where} \quad \cc_\pm = (\chpm)^2 (1-\chpm)\partial_{\chpm}^2 -(\chpm)^2 \partial_{\chpm}.
 \eeq
 The eigenfunctions $\Psi(\chp,\chm)$ of the Casimir operator $\cc$ are simultaneous eigenfunctions of $\cc_+$ and $\cc_-$,
 \beq
  \cc_+ \Psi = \hpl(\hpl -1) \Psi \quad \text{and} \quad \cc_- \Psi = \hmi(\hmi -1) \Psi ,
 \eeq
 where the eigenvalue of $\cc$ can be identified as $\hpl(\hpl -1) + \hmi(\hmi -1) $. The eigenvalue equation for $\cc_\pm$ has two linearly independent solutions $F_{\hpm}(\chpm)$ and $F_{1-\hpm}(\chpm)$, where
 \beq
  F_{h}(\chi) = \l \chi \r ^{h} \ff(h,h,2h,\chi).
 \eeq
 The general solution for $\Psi$ can be written as a linear combination of $F_{\hpl}(\chp)F_{\hmi}(\chm)$ and three other terms with $\hpl$ replaced by $1-\hpl$, $\hmi$ replaced by $1-\hmi$, or both. We shall impose appropriate boundary conditions on $\Psi$ to determine a complete set of eigenfunctions of $\cc$.

 \paragraph{The eigenfunctions.} As noted in Sec. \ref{subsec:conf_symmetry}, an appropriate solution of $\Psi$ in the region $0 < \chpm <2$ determines the eigenfunction completely for all values of $\chpm$ except the singular points $\chpm = 0,1$. The hypergeometric functions in $F_{h}(\chi)$ and $F_{1-h}(\chi)$ are well suited to the regime $\chi < 1$. For $\chi >1$, it is convenient to consider the two linearly independent solutions
 \beq
 \label{eq:soln_for_chi>1}
  \tilF_h(\chi) = \ff \l \frac{h}{2},\frac{1-h}{2},\frac{1}{2},\frac{(2-\chi)^2}{{\chi}^2} \r , \quad \tilG_h(\chi) = \frac{2-\chi}{\chi} \ff \l \frac{1+h}{2},1-\frac{h}{2},\frac{3}{2},\frac{(2-\chi)^2}{{\chi}^2} \r ,
 \eeq
 which are even and odd, respectively, under $\chi \to \frac{\chi}{\chi-1}$. At $\chi=2$, $\tilF_h$ has a vanishing derivative satisfying Neumann boundary condition and $\tilG_h$ vanishes satisfying Dirichlet boundary condition. By Eq. \eqref{eq:boundary_condition_at2}, the eigenfunctions in the regime $\chi>1$ are
 \beq
 \label{eq:gen_soln_22}
  \Psi^{\ee}(\chp,\chm) = c^\ee \; \tilF_{\hpl}(\chp) \tilF_{\hmi}(\chm), \quad \Psi^{\oo}(\chp,\chm) = c^\oo \; \tilG_{\hpl}(\chp) \tilG_{\hmi}(\chm),
 \eeq
 where the normalisation constants are conveniently chosen as
 \beal
  c^\ee &= \frac{1}{\pi} \; \Gamma \l \frac{\hpl}{2} \r \Gamma \l \frac{1}{2}-\frac{\hpl}{2} \r \Gamma\l\frac{\hmi}{2}\r\Gamma\l \frac{1}{2}-\frac{\hmi}{2} \r , \\
  c^\oo &= \frac{4}{\pi} \; \Gamma \l \frac{1+\hpl}{2} \r \Gamma \l 1 - \frac{\hpl}{2} \r \Gamma\l\frac{1+\hmi}{2}\r\Gamma\l 1 - \frac{\hmi}{2} \r .
 \eeal
  With this choice the coefficient of $ \log |1-\chp| \log |1-\chm|$ is one in both sectors as $\chpm \to 1$.

  The coefficients in the other regions follow from comparing the logarithmic and constant terms at $\chpm = 1$. In the region where $\chp > 1 $ and $\chm < 1$, the general solution of the eigenfunction has the form
 \beal
  \label{eq:gen_soln_21}
  \Psi^\ee &= c_1^\ee \; \tilde{F}_{\hpl}(\chp) F_{\hmi}(\chm) + c_2^\ee \; \tilde{F}_{\hpl}(\chp) F_{1-\hmi}(\chm), \\
  \Psi^\oo &= c_1^\oo \; \tilde{G}_{\hpl}(\chp) F_{\hmi}(\chm) + c_2^\oo \; \tilde{G}_{\hpl}(\chp) F_{1-\hmi}(\chm).
 \eeal
 We shall use the expansion about $z=1$ \cite{NIST:DLMF},
  \beq
   \ff (a,b,a+b,z) = \frac{\Gamma (a+b)}{\Gamma(a)\Gamma(b)} \Big[ - \log |1-z| + 2\psi(1) - \psi(a) - \psi(b) \Big] + \mathcal{O}\Big[ (1-z)\log |1-z| \Big],
  \eeq
  together with the duplication and reflection formula for the digamma function $\psi$, to find the coefficients by comparing the logarithmic and constant terms as we approach $\chm \to 1^-$ in Eq. \eqref{eq:gen_soln_21} and $\chm \to 1^+$ in Eq. \eqref{eq:gen_soln_22},
 \beal
  \label{eq:coefficients_21}
  c_1^\ee &= \frac{1}{2} \tan \pi \hmi \cot \frac{\pi \hmi}{2} \frac{\Gamma(\hmi)^2}{\Gamma(2\hmi)} \frac{\Gamma \l \frac{\hpl}{2} \r \Gamma \l \frac{1}{2} - \frac{\hpl}{2} \r}{\sqrt{\pi}}, \\
  c_2^\ee &= -\frac{1}{2} \tan \pi \hmi \tan \frac{\pi \hmi}{2} \frac{\Gamma(1-\hmi)^2}{\Gamma(2-2\hmi)} \frac{\Gamma \l \frac{\hpl}{2} \r \Gamma \l \frac{1}{2} - \frac{\hpl}{2} \r}{\sqrt{\pi}}, \\
  c_1^\oo &= - \tan \pi \hmi \tan \frac{\pi \hmi}{2} \frac{\Gamma(\hmi)^2}{\Gamma(2\hmi)} \frac{\Gamma \l \frac{1}{2} + \frac{\hpl}{2} \r \Gamma \l 1- \frac{\hpl}{2} \r }{\sqrt{\pi}}, \\
  c_2^\oo &=  \tan \pi \hmi \cot \frac{\pi \hmi}{2} \frac{\Gamma(1-\hmi)^2}{\Gamma(2-2\hmi)} \frac{\Gamma \l \frac{1}{2} + \frac{\hpl}{2} \r \Gamma \l 1-\frac{\hpl}{2} \r }{\sqrt{\pi}}.
 \eeal
 Similarly, in the region with $\chp < 1$ and $\chm > 1 $ the eigenfunctions are
 \beal
  \label{eq:gen_soln_12}
  \Psi^\ee &= \tilde{c}_1^\ee \; \tilde{F}_{\hmi}(\chm) F_{\hpl}(\chp) + \tilde{c}_2^\ee \; \tilde{F}_{\hmi}(\chm) F_{1-\hpl}(\chp), \\
  \Psi^\oo &= \tilde{c}_1^\oo \; \tilde{G}_{\hmi}(\chm) F_{\hpl}(\chp) + \tilde{c}_2^\oo \; \tilde{G}_{\hmi}(\chm) F_{1-\hpl}(\chp),
 \eeal
 where $\tilde{c}_{1,2}^\ee$ and $\tilde{c}_{1,2}^\oo$ are obtained from Eq. \eqref{eq:coefficients_21} by exchanging $\hpl \leftrightarrow \hmi$. Finally, the eigenfunction for $ \chpm < 1 $ can be obtained from either Eq. \eqref{eq:gen_soln_21} or Eq. \eqref{eq:gen_soln_12} using the same procedure,
 \beq
  \Psi^\sigma = a_{11}^\sigma \; F_{\hpl} F_{\hmi} + a_{12}^\sigma \; F_{\hpl} F_{1-\hmi} + a_{21}^\sigma \; F_{1-\hpl}F_{\hmi} + a_{22}^\sigma \; F_{1-\hpl}F_{1-\hmi},
 \eeq
 with arguments $ F_{\hpl} \equiv F_{\hpl}(\chp)$ and $F_{\hmi} \equiv F_{\hmi}(\chm)$. The coefficients are obtained as
 \beal
  a_{11}^\ee &= \frac{1}{4} \tan \pi \hpl \tan \pi \hmi \cot \frac{\pi \hpl}{2} \cot \frac{\pi \hmi}{2} \frac{\Gamma(\hpl)^2 \Gamma(\hmi)^2}{\Gamma(2\hpl)\Gamma(2\hmi)}, \\
  a_{12}^\ee &= -\frac{1}{4} \tan \pi \hpl \tan \pi \hmi \cot \frac{\pi \hpl}{2} \tan \frac{\pi \hmi}{2} \frac{\Gamma(\hpl)^2 \Gamma(1-\hmi)^2}{\Gamma(2\hpl)\Gamma(2-2\hmi)}, \\
  a_{21}^\ee &= -\frac{1}{4} \tan \pi \hpl \tan \pi \hmi \tan \frac{\pi \hpl}{2} \cot \frac{\pi \hmi}{2} \frac{\Gamma(1-\hpl)^2 \Gamma(\hmi)^2}{\Gamma(2-2\hpl)\Gamma(2\hmi)}, \\
  a_{22}^\ee &= \frac{1}{4} \tan \pi \hpl \tan \pi \hmi \tan \frac{\pi \hpl}{2} \tan \frac{\pi \hmi}{2} \frac{\Gamma(1-\hpl)^2 \Gamma(1-\hmi)^2}{\Gamma(2-2\hpl)\Gamma(2-2\hmi)}, \\
  a_{11}^\oo &= \frac{1}{4} \tan \pi \hpl \tan \pi \hmi \tan \frac{\pi \hpl}{2} \tan \frac{\pi \hmi}{2} \frac{\Gamma(\hpl)^2 \Gamma(\hmi)^2}{\Gamma(2\hpl)\Gamma(2\hmi)}, \\
  a_{12}^\oo &= -\frac{1}{4} \tan \pi \hpl \tan \pi \hmi \tan \frac{\pi \hpl}{2} \cot \frac{\pi \hmi}{2} \frac{\Gamma(\hpl)^2 \Gamma(1-\hmi)^2}{\Gamma(2\hpl)\Gamma(2-2\hmi)}, \\
  a_{21}^\oo &= -\frac{1}{4} \tan \pi \hpl \tan \pi \hmi \cot \frac{\pi \hpl}{2} \tan \frac{\pi \hmi}{2} \frac{\Gamma(1-\hpl)^2 \Gamma(\hmi)^2}{\Gamma(2-2\hpl)\Gamma(2\hmi)}, \\
  a_{22}^\oo &= \frac{1}{4} \tan \pi \hpl \tan \pi \hmi \cot \frac{\pi \hpl}{2} \cot \frac{\pi \hmi}{2} \frac{\Gamma(1-\hpl)^2 \Gamma(1-\hmi)^2}{\Gamma(2-2\hpl)\Gamma(2-2\hmi)}.
 \eeal
 The even sector coefficients are those of the product of two 1D SYK eigenfunctions \cite{Maldacena_2016}. In the odd sector the factors of $\cot \frac{\pi h}{2}$ and $\tan \frac{\pi h}{2}$ are interchanged. The coefficients are factorised in every region, so that in the region $0<\chpm \le 2$ the eigenfunctions can be expressed in the factorised form,
 \beq
  \Psi_{(\hpl,\hmi)}^\sigma (\chp,\chm) = \varphi_{\hpl}^\sigma (\chp)\varphi_{\hmi}^\sigma (\chm) ,
 \eeq
 where
 \beq
  \varphi_{h}^{\ee}(\chi) = \begin{cases}
      \frac{1}{\sqrt{\pi}} \; \Gamma \l \frac{h}{2} \r \Gamma \l \frac{1}{2} - \frac{h}{2} \r \tilF_h (\chi) \quad \text{for} \quad \chi >1 , \\
      A^\ee(h) F_h(\chi) + B^\ee(h) F_{1-h}(\chi) \quad \text{for} \quad \chi <1 ,
  \end{cases}
 \eeq
 \beq
  \varphi_{h}^{\oo}(\chi) = \begin{cases}
      \frac{2}{\sqrt{\pi}} \; \Gamma \l \frac{1+h}{2} \r \Gamma \l 1 - \frac{h}{2} \r \tilG_h (\chi) \quad \text{for} \quad \chi >1 , \\
      A^\oo(h) F_h(\chi) + B^\oo(h) F_{1-h}(\chi) \quad \text{for} \quad \chi <1 ,
  \end{cases}
 \eeq
 with
 \beq
  A^\ee(h) = \frac{1}{2} \tan (\pi h) \cot \l \frac{\pi h}{2} \r \frac{\Gamma(h)^2}{\Gamma(2h)} \quad \text{and} \quad B^\ee(h) = -\frac{1}{2} \tan (\pi h) \tan \l \frac{\pi h}{2} \r \frac{\Gamma(1-h)^2}{\Gamma(2-2h)},
 \eeq
 \beq
  A^\oo(h) = -\frac{1}{2} \tan (\pi h) \tan \l \frac{\pi h}{2} \r \frac{\Gamma(h)^2}{\Gamma(2h)} \quad \text{and} \quad B^\oo(h) = \frac{1}{2} \tan (\pi h) \cot \l \frac{\pi h}{2} \r \frac{\Gamma(1-h)^2}{\Gamma(2-2h)}.
 \eeq
 The relations $A^\sigma(1-h) = B^\sigma(h)$, together with the manifest symmetry of the solutions in Eq. \eqref{eq:soln_for_chi>1} under $h \leftrightarrow 1-h$, imply that the eigenfunctions are symmetric under $\hpl \leftrightarrow 1-\hpl$ and (or) $\hmi \leftrightarrow 1-\hmi$ in each sector,
 \beq
  \Psi^{\sigma}_{(\hpl,\hmi)} = \Psi^{\sigma}_{(\hpl,1-\hmi)} =\Psi^{\sigma}_{(1-\hpl,\hmi)} = \Psi^{\sigma}_{(1-\hpl,1-\hmi)}.
 \eeq

 \paragraph{The complete basis.} The vanishing of boundary terms at $\chpm = 0$ requires the chiral factors to decay at least as fast as ${\chpm}^{1/2}$. The continuum $\hpm = \frac{1}{2} + \ii s_\pm $ is marginally allowed in both sectors. For real $\hpm>\frac12$ the term $F_{1-\hpm}$ must be absent, which imposes $B^\sigma(\hpm)=0$. Since $\Gamma(1-h)^2/\Gamma(2-2h)$ is finite and nonzero away from the integers, and has a simple pole at each integer, while $\tan(\pi h)$ has a simple zero there, $B^\ee(h)$ ($B^\oo(h)$) vanishes at even (odd) $h$ and has poles at odd (even) $h$. The discrete states are therefore $h=2n$ in the even sector and $h=2n-1$ in the odd sector with $n=1,2,3,\dots$. The complete basis in each sector is
 \beal
  \label{eq:basis}
  \ee:&\quad \hpm\in\big\{\tfrac12+\ii s_\pm,\ s_\pm>0\big\}\cup\{2,4,6,\dots\},\\
  \oo:&\quad \hpm\in\big\{\tfrac12+\ii s_\pm,\ s_\pm>0\big\}\cup\{1,3,5,\dots\},
 \eeal
 so that each sector contains the four kinds of states: (continuum, continuum), (continuum, discrete), (discrete, continuum) and (discrete, discrete).

\section{Eigenvalues of the kernel}\label{app:eigenvalues}

 In this appendix we present the detailed computation of the results stated in Sec. \ref{subsec:eigenvalues}. The appendix elaborates the reason why kernels act diagonally on the two three-point structures in Eq. \eqref{eq:3pt_eigenfunctions}, the reduction of the eigenvalues to chiral double integrals, their evaluation, and the relation between the three-point structures and the eigenfunctions of the Casimir in both sectors.

 \paragraph{Three-point structures and their parities.}
 Under $\text{SL}(2,\mathbb{R})$ transformations, $x\to(a' x+b')/(c' x+d')$ with $a'd'-b'c'=1$, differences transform as $x_{ij}\to x_{ij}/[(c' x_i+d')(c' x_j+d')]$. The orientation of three points on a lightcone is therefore invariant,
 \beq
  \vare(i,j,0)=\sgn{x_{ij}x_{j0}x_{0i}}\;\to\;\sgn{\frac{x_{ij}x_{j0}x_{0i}}{(c' x_i+d')^2(c' x_j+d')^2(c' x_0+d')^2}}=\vare(i,j,0),
 \eeq
 whereas $\sgn{x_{ij}}$ transforms like the sign of a fermion two-point function. Thus, multiplying a covariant structure by $\vare$ gives another covariant structure with the same weights. On the lightcone $x^a$ the structure $f^{a,\ee}$ of Eq. \eqref{eq:3pt_eigenfunctions} carries $\sgn{x_{12}}$ and $f^{a,\oo}$ carries $\sgn{x_{12}}\vare(1,2,0)$; on the lightcone $x^{\bar{a}}$, $f^{a,\ee}$ carries no sign and $f^{a,\oo}$ carries $\vare(1,2,0)$. Since $\vare(1,2,0)$ is odd under $x_1\leftrightarrow x_2$, the exchange $\scP_\pm:x^\pm_1\leftrightarrow x^\pm_2$ of one lightcone coordinate acts as
 \beq
 \label{eq:3-pt_parity}
  \scP_a f^{a,\ee}=-f^{a,\ee},\quad \scP_{\bar{a}}f^{a,\ee}= f^{a,\ee}, \quad \scP_a f^{a,\oo}= f^{a,\oo},\quad \scP_{\bar{a}}f^{a,\oo}=-f^{a,\oo}.
 \eeq

 \paragraph{The kernels acting on the two sectors.}
 We consider a chiral kernel $K$ acting on a function $g$ of two points on one lightcone,
 \beq
  (Kg)(x_1,x_2)=\int dx_3\,dx_4\;K(x_1,x_2,x_3,x_4)\,g(x_3,x_4),
 \eeq
 and let $\scP$ exchange the two arguments, $(\scP g)(x_1,x_2)=g(x_2,x_1)$. Acting first with $\scP$ and then with $K$,
 \beq
  (K \scP g)(x_1,x_2)=\int dx_3\,dx_4\;K(x_1,x_2,x_3,x_4)\,g(x_4,x_3) = \int dx_3\,dx_4\;K(x_1,x_2,x_4,x_3)\,g(x_3,x_4),
 \eeq
 where in the second step we exchanged the integration variables $x_3\leftrightarrow x_4$. Acting first with $K$ and then with $\scP$ exchanges the external points,
 \beq
  (\scP Kg)(x_1,x_2)=(Kg)(x_2,x_1)=\int dx_3\,dx_4\;K(x_2,x_1,x_3,x_4)\,g(x_3,x_4).
 \eeq
 Comparing the two integrands, $\scP K=\pm K \scP$ holds for every $g$ exactly when $K(x_2,x_1,x_3,x_4)=\pm K(x_1,x_2,x_4,x_3)$. Exchanging $x_3\leftrightarrow x_4$ on both sides, this is equivalent to
 \beq
  K(x_2,x_1,x_4,x_3)=\pm K(x_1,x_2,x_3,x_4).
 \eeq
 Thus, the operator $\scP$ commutes or anticommutes with $K$ when $K$ is even or odd, respectively, under the simultaneous exchange $x_1\leftrightarrow x_2$, $x_3\leftrightarrow x_4$.
 
 It is straightforward to check from Eq. \eqref{eq:chiral_kernels} that the diagonal and off-diagonal chiral kernels are symmetric and antisymmetric, respectively, under the simultaneous exchange $1 \leftrightarrow 2$ and $3 \leftrightarrow 4$. Therefore, the diagonal chiral kernels $K^{\pm}_{aa}$ commute with $\scP_\pm$ and the off-diagonal ones $K^{\pm}_{a\bar{a}}$ anticommute,
 \beq
 \label{eq:P_commutes_K}
  \scP_\pm K^{\pm}_{ab} = (-1)^{1-\delta_{ab}} K^{\pm}_{ab} \scP_\pm .
 \eeq

 The kernels $K_{ab}$ map the three-point functions $f^b$ to the functions $f^a$, and by Eq. \eqref{eq:3-pt_parity} $f^{a,\sigma}$ and $f^{b,\sigma}$ have the same parities for $a=b$ and opposite parities for $a \neq b$, in the same sector. Since $K_{ab}$ commutes with the conformal generators (Eq. \eqref{eq:kernel_commutes_conf}), $K_{ab} f^{b,\sigma}$, $\sigma=\ee,\oo$, must be a conformal three-point function with the weights of $f^a$, which is a combination of $f^{a,\ee}$ and $f^{a,\oo}$. By Eq. \eqref{eq:P_commutes_K} only the one with the same $\sigma$ survives and we arrive at
 \beq
 \label{eq:eigenvalue_integral_master_eq}
  \int d^2 \xx_3 \ d^2 \xx_4 K_{ab}(\xx_1,\xx_2,\xx_3,\xx_4) f^{b,\sigma}_{\hpl,\hmi}(\xx_3,\xx_4,\xx_0) = \mathk_{ab}^{\sigma}(\hpl,\hmi) f^{a,\sigma}_{\hpl,\hmi}(\xx_1,\xx_2,\xx_0) .
 \eeq
 In 2D the operator that has to be added to the Casimir to make the spectrum nondegenerate is the single-lightcone exchange $\scP_\pm$, and in cross ratios this is exactly the single reflection $R_\pm$ of the eigenfunctions presented in Sec. \ref{subsec:casimir_eigenfunctions}.

 \paragraph{The eigenvalues as products of two chiral double integrals.} Since the eigenvalues $\mathk_{ab}^{\sigma}(\hpl,\hmi)$ do not depend on the spacetime points, we use conformal invariance and evaluate the integrals in Eq. \eqref{eq:eigenvalue_integral_master_eq} with $x_0^\pm \to \infty$, $x_1^\pm =1$, $x_2^\pm =0$ and rename $x_3^\pm \to x^\pm$, $x_4^\pm \to x'^\pm$. With this choice the three-point functions reduce to Eq. \eqref{eq:3pt_limit}. The common factor $|x_0^+|^{-2\hpl}|x_0^-|^{-2\hmi}$ cancels, and Eq. \eqref{eq:eigenvalue_integral_master_eq} becomes Eq. \eqref{eq:eigenvalue_int}.

 For $K_{++}$ in the even sector, the lightcone $x^+$ carries the signs of the kernel and of $f^{+,\ee}$, and the lightcone $x^-$ carries no sign,
\beal
\label{eq:app_kpp_even}
 \mathk^\ee_{++}(\hpl,\hmi)=\frac{1}{\alpha_{++}}&\int_{-\infty}^{\infty}dx^+\int_{-\infty}^{\infty}dx'^+\,\frac{\sgn{1-x^+}\sgn{-x'^+}\sgn{x^+-x'^+}}{|1-x^+|^{\dpl}|x'^+|^{\dpl}|x^+-x'^+|^{2-\dpl-\hpl}}\\
 \times&\int_{-\infty}^{\infty}dx^-\int_{-\infty}^{\infty}dx'^-\,\frac{1}{|1-x^-|^{\dmi}|x'^-|^{\dmi}|x^--x'^-|^{2-\dmi-\hmi}}\\
 =&\;(q-1)\,\alpha_0^2\,k^\ee_{\dpl}(\hpl)\,\tilde{k}^\ee_{\dmi}(\hmi),
\eeal
and in the odd sector $\sgn{x-x'}$ moves from the $x^+$ integral to the $x^-$ integral, which gives
\beal
\label{eq:app_kpp_odd}
 \mathk^\oo_{++}(\hpl,\hmi)=\frac{1}{\alpha_{++}}&\int_{-\infty}^{\infty}dx^+\int_{-\infty}^{\infty}dx'^+\,\frac{\sgn{1-x^+}\sgn{-x'^+}}{|1-x^+|^{\dpl}|x'^+|^{\dpl}|x^+-x'^+|^{2-\dpl-\hpl}}\\
 \times&\int_{-\infty}^{\infty}dx^-\int_{-\infty}^{\infty}dx'^-\,\frac{\sgn{x^- -x'^-}}{|1-x^-|^{\dmi}|x'^-|^{\dmi}|x^--x'^-|^{2-\dmi-\hmi}}\\
 =&\;(q-1)\,\alpha_0^2\,k^\oo_{\dpl}(\hpl)\,\tilde{k}^\oo_{\dmi}(\hmi).
\eeal
The kernel $K_{+-}$ acts on $f^{-,\ee}$ and $f^{-,\oo}$ yielding the same chiral integrals as $K_{++}$ in the even and odd sector, respectively. Hence, using $\frac{1}{\alpha_{+-}} = q \, \alpha_0^2$ we can write
\beq
 \label{eq:app_kpm}
 \mathk^\ee_{+-}(\hpl,\hmi) = q \, \alpha_0^2 \,k^\ee_{\dpl}(\hpl)\,\tilde{k}^\ee_{\dmi}(\hmi), \qquad \mathk^\oo_{+-}(\hpl,\hmi) = q \, \alpha_0^2 \,k^\oo_{\dpl}(\hpl)\,\tilde{k}^\oo_{\dmi}(\hmi).
\eeq
For the eigenvalues of $K_{-+}$ and $K_{--}$, the sign factors and $\dpm$ get exchanged between the two chiral double integrals, and thus $\hpl$ appears with $\tilde{k}_{\dmi}$ and $\hmi$ appears with $k_{\dpl}$. The eigenvalues can be represented as
 \beq
  \label{eq:app_kmp_kmm}
  \mathk^\sigma_{-+}(\hpl,\hmi) = q \, \alpha_0^2 \,k^\sigma_{\dpl}(\hmi)\,\tilde{k}^\sigma_{\dmi}(\hpl), \qquad \mathk^\sigma_{--}(\hpl,\hmi) = (q-1) \, \alpha_0^2 \,k^\sigma_{\dpl}(\hmi)\,\tilde{k}^\sigma_{\dmi}(\hpl).
 \eeq
 For every kernel $K_{ab}$, the integral on the lightcone $x^a$ is $k^\sigma_{\dpl}(h_a)$ and the integral on the lightcone $x^{\bar a}$ is $\tilde{k}^\sigma_{\dmi}(h_{\bar a})$, independently of $b$, which gives Eq. \eqref{eq:eigenvalues},
 \beq
  \mathk_{ab}^{\sigma}(\hpl,\hmi) = \frac{1}{\alpha_{ab}} k^{\sigma}_{\dpl}(h_a) \tilde{k}^{\sigma}_{\dmi}(h_{\bar{a}}).
 \eeq
 
 \paragraph{Evaluation of the chiral double integrals.}
 We use the Fourier identities in Eqs. \eqref{eq:Fourier_identity1} and \eqref{eq:Fourier_identity2} to determine the functions $k_\Delta^\sigma (h)$ and $\tilde{k}_\Delta^\sigma (h)$. We first evaluate $k^\ee_\Delta(h)$ of Eq. \eqref{eq:eigen_double_int_even}. By Eq. \eqref{eq:Fourier_identity1} with $r=2-\Delta-h$,
\beq
\label{eq:app_step1}
 \frac{\sgn{x-x'}}{|x-x'|^{2-\Delta-h}}=\int\frac{dk}{2\pi}\,e^{-\ii k(x-x')}\,c(2-\Delta-h)\,|k|^{1-\Delta-h}\,\sgn{k}.
\eeq
 In the $x$ integral we change the variable $x \to 1-x$ and perform the integrals over $x$ and $x'$,
\beq
\label{eq:app_step2}
 \int dx\,e^{-\ii kx}\,\frac{\sgn{1-x}}{|1-x|^{\Delta}}=e^{-\ii k}\,c(\Delta)\,|k|^{\Delta-1}\sgn{k},\quad
 \int dx'\,e^{\ii kx'}\,\frac{\sgn{-x'}}{|x'|^{\Delta}}=-c(\Delta)\,|k|^{\Delta-1}\sgn{k}.
\eeq
 Finally, using the inverse Fourier transform we arrive at
\beq
\label{eq:app_k_even}
 k^\ee_\Delta(h)=-c(\Delta)^2\,c(2-\Delta-h)\int\frac{dk}{2\pi}\,e^{-\ii k}\,|k|^{\Delta-h-1}\sgn{k}=-c(\Delta)^2\,\frac{c(2-\Delta-h)}{c(\Delta-h)} .
\eeq
 For $\tilde{k}^\ee_\Delta(h)$ every factor is unsigned. The function $d$ replaces $c$ in each step, there is no $\sgn{k}$ and no minus sign. Thus, in the even sector we obtain
 \beq
 k^\ee_\Delta(h) = -\frac{c(2-\Delta-h)}{c(\Delta-h)}\L c(\Delta)\R^2, \qquad \tilde{k}^\ee_\Delta(h) = \frac{d(2-\Delta-h)}{d(\Delta-h)}\L d(\Delta)\R^2.
 \eeq
 In the odd sector, $\sgn{x-x'}$ is absent in $k^\oo_\Delta$ but is present in $\tilde{k}^\oo_\Delta$. Hence, the chiral double integrals reduce to
 \beq
 k^\oo_\Delta(h) = -\frac{d(2-\Delta-h)}{d(\Delta-h)}\L c(\Delta)\R^2, \qquad \tilde{k}^\oo_\Delta(h) = \frac{c(2-\Delta-h)}{c(\Delta-h)}\L d(\Delta)\R^2.
 \eeq
 These are the results presented in Eq. \eqref{eq:factors_eigen}.

 \paragraph{Properties.}
 The chiral factors $k^\sigma_\Delta(h)$ and $\tilde{k}^\sigma_\Delta(h)$ are invariant under $h \to 1-h$, so are the eigenvalues under single or simultaneous substitutions $\hpm \to 1-\hpm$,
 \beq
  \mathk_{ab}^{\sigma}(\hpl,\hmi) = \mathk_{ab}^{\sigma}(1-\hpl,\hmi) = \mathk_{ab}^{\sigma}(\hpl,1-\hmi) = \mathk_{ab}^{\sigma}(1-\hpl,1-\hmi).
 \eeq
 Under an exchange of the conformal weights, $\hpl \leftrightarrow \hmi$,
 \beq
  \mathk_{ab}^{\sigma}(\hmi,\hpl) = \mathk_{\bar a\bar b}^{\sigma}(\hpl,\hmi).
 \eeq
 A substitution $j\to-j$ exchanges the weights $\dpl$ and $\dmi$ in the eigenvalues. Hence, the eigenvalues of the two sectors are related by relabelling of the chiralities,
 \beq
  \mathk^{\ee}_{ab}|_{-j}=\mathk^{\oo}_{\bar a\bar b}|_{j}.
 \eeq

\section{Zero-rung ladder: projection on the eigenfunctions}\label{app:zero-rung_ladder}

 In this appendix we evaluate the projections of the zero-rung ladder onto the eigenfunctions of its own reflection sector, which enter the sum over ladders in Eq. \eqref{eq:ladder_sum0}. We take $b=+$; the case $b=-$ follows by exchanging $\chp$ and $\chm$. Reflecting $\chp$ in Eq. \eqref{eq:zero_rung_ladder}, the zero-rung ladder in sector $\sigma$ factorises,
 \beq
 \label{eq:zero_rung_sigma}
  \mathf^{(0),\sigma}_+(\chp,\chm)=\frac{1}{2} \L-\sgn{\chp}|\chp|^{\dpl}-\tau\,\sgn{\frac{\chp}{\chp-1}}\left|\frac{\chp}{\chp-1}\right|^{\dpl}\R \L|\chm|^{\dmi}+\tau\left|\frac{\chm}{\chm-1}\right|^{\dmi}\R ,
 \eeq
 with $\tau=+1$ for the even and $\tau=-1$ for the odd sector. The map $\chi\to\chi/(\chi-1)$ takes the region $(0,2)$ onto $\mathbb{R} \setminus[0,2]$ and leaves the integration measure invariant, and the chiral parts of the eigenfunctions transform as $\varphi^\sigma_h \l \chi/(\chi-1) \r = \tau \; \varphi^\sigma_h(\chi)$. Therefore, the substitution $\chi=\bar{\chi}/(\bar{\chi}-1)$ in the second term of each chiral factor gives,
 \beal
  \int_0^2\frac{d\chi}{\chi^2}\,\varphi^\sigma_h(\chi)\L-\sgn{\chi}|\chi|^{\dpl}-\tau\,\sgn{\frac{\chi}{\chi-1}}\left|\frac{\chi}{\chi-1}\right|^{\dpl}\R &=\int_{-\infty}^{\infty}\frac{d\chi}{\chi^2}\,\varphi^\sigma_h(\chi)\l-\sgn{\chi}\r|\chi|^{\dpl},\\
  \int_0^2\frac{d\chi}{\chi^2}\,\varphi^\sigma_h(\chi)\L|\chi|^{\dmi}+\tau\left|\frac{\chi}{\chi-1}\right|^{\dmi}\R &=\int_{-\infty}^{\infty}\frac{d\chi}{\chi^2}\,\varphi^\sigma_h(\chi)\,|\chi|^{\dmi}.
 \eeal
 Thus, each chiral factor of the inner product \eqref{eq:inner_product_sector} is a single integral over the whole $\chpm$ line,
 \beq
  \la \mathf^{(0),\sigma}_\pm,\Psi^\sigma \ra = \frac{1}{2} \,P^{\sigma}_{(\hpm)}\,\tilde{P}^{\sigma}_{(h_{\mp})},
 \eeq
 where
 \beq
  P^{\sigma}_{(h)} = \int_{-\infty}^{\infty} \frac{d\chi}{\chi^2} \l -\sgn{\chi}\r|\chi|^{\dpl} \varphi^\sigma_h(\chi), \qquad \tilde{P}^{\sigma}_{(h)} = \int_{-\infty}^{\infty} \frac{d\chi}{\chi^2}\,|\chi|^{\dmi}\varphi^\sigma_h(\chi).
 \eeq
 To evaluate $P^{\sigma}_{(h)}$ and $\tilde{P}^{\sigma}_{(h)}$ we use the integral representations of $\varphi^{\sigma}_h$ in Eq. \eqref{eq:eigen_integral_rep}. Rescaling $\chi \to y\chi$ maps the integration line onto itself and the integrand factorises, which leads to
 \beal
 \label{eq:factorised_P_and_tildeP}
  P^{\ee}_{(h)} &= -\frac{1}{2} \int_{-\infty}^{\infty} d\chi\,\frac{\sgn{\chi}}{|\chi|^{2-\dpl-h}|1-\chi|^{h}}\int_{-\infty}^{\infty} dy\,\frac{\sgn{y}}{|y|^{1+h-\dpl}|1-y|^{1-h}},\\
  \tilde{P}^{\ee}_{(h)} &=\frac{1}{2} \int_{-\infty}^{\infty} d\chi\,\frac{1}{|\chi|^{2-\dmi-h}|1-\chi|^{h}}\int_{-\infty}^{\infty} dy\,\frac{1}{|y|^{1+h-\dmi}|1-y|^{1-h}},\\
  P^{\oo}_{(h)} &=\frac{1}{2} \int_{-\infty}^{\infty} d\chi\,\frac{\sgn{1-\chi}}{|\chi|^{2-\dpl-h}|1-\chi|^{h}}\int_{-\infty}^{\infty} dy\,\frac{\sgn{1-y}}{|y|^{1+h-\dpl}|1-y|^{1-h}},\\
  \tilde{P}^{\oo}_{(h)} &= -\frac{1}{2} \int_{-\infty}^{\infty} d\chi\,\frac{\sgn{\chi}\sgn{1-\chi}}{|\chi|^{2-\dmi-h}|1-\chi|^{h}}\int_{-\infty}^{\infty} dy\,\frac{\sgn{y}\sgn{1-y}}{|y|^{1+h-\dmi}|1-y|^{1-h}}.
 \eeal
 Each one dimensional integral in Eq. \eqref{eq:factorised_P_and_tildeP} can be evaluated by inserting the Fourier identities in Eq. \eqref{eq:Fourier_identity1} and \eqref{eq:Fourier_identity2}. The results can be written compactly as
 \beq
  \int_{-\infty}^{\infty}d\chi\,\frac{\sgn{\chi}^{\epsilon}\sgn{1-\chi}^{\eta}}{|\chi|^{r}|1-\chi|^{s}}=\frac{e_\epsilon(r)\,e_\eta(s)}{e_{\epsilon+\eta}(r+s-1)}, \qquad \epsilon,\eta\in\{0,1\},
 \eeq
 where $e_1 = c$ and $e_0=e_2=d$ with the functions $c$ and $d$ presented in Eq. \eqref{eq:Fourier_identity1} and \eqref{eq:Fourier_identity2}, respectively. In Eq. \eqref{eq:factorised_P_and_tildeP} the $\chi$ integrals have $(r,s)=(2-\Delta-h,h)$, and the $y$ integrals have $(r,s)=(1+h-\Delta,1-h)$, where $\Delta \to \dpl$ in $P^{\sigma}_{(h)}$ and $\Delta \to \dmi$ in $\tilde{P}^{\sigma}_{(h)}$. Thus, the product of the two integrals with sign factors $\epsilon,\eta$ becomes
 \beq
 \label{eq:int_products}
  \frac{e_\epsilon(2-\Delta-h)\,e_\eta(h)\,e_\epsilon(1+h-\Delta)\,e_\eta(1-h)}{e_{\epsilon+\eta}(1-\Delta)^2} = s_\epsilon s_\eta\,e_{\epsilon+\eta}(\Delta)^2\,\frac{e_\epsilon(2-\Delta-h)}{e_\epsilon(\Delta-h)}
 \eeq
 with $s_0=1,\ s_1=-1$, where we have used the identities
 \beq
  c(a)\,c(1-a)=-2\pi,\qquad d(a)\,d(1-a)=2\pi,
 \eeq
 which follow from $c(a) = 2\ii \, \Gamma(1-a)\cos\frac{\pi a}2$ and $d(a)=2 \, \Gamma(1-a)\sin\frac{\pi a}2$. The integrals in Eq. \eqref{eq:factorised_P_and_tildeP} can also be computed by splitting the integration region into three parts, $(-\infty,0) \cup (0,1) \cup (1,\infty)$, and reducing each part into elementary beta integrals. This yields the same result as presented in Eq. \eqref{eq:int_products}. Now, it is straightforward to verify from Eq. \eqref{eq:factors_eigen} that the right hand side of Eq. \eqref{eq:int_products} is exactly $k^{\ee}_{\Delta}(h)$, $\tilde{k}^{\ee}_{\Delta}(h)$, $k^{\oo}_{\Delta}(h)$ and $\tilde{k}^{\oo}_{\Delta}(h)$ for $(\epsilon,\eta) = (1,0),(0,0),(0,1)$ and $(1,1)$, respectively. Hence, we obtain
 \beal
 \label{eq:P_tildeP_results}
  P^{\ee}_{(h)} &= -\frac{1}{2} k^{\ee}_{\dpl}(h), \qquad \tilde{P}^{\ee}_{(h)} = \frac{1}{2} \tilde{k}^{\ee}_{\dmi}(h), \\
  P^{\oo}_{(h)} &= \frac{1}{2} k^{\oo}_{\dpl}(h), \qquad \quad \tilde{P}^{\oo}_{(h)} = -\frac{1}{2} \tilde{k}^{\oo}_{\dmi}(h),
 \eeal
 which lead to the inner products,
 \beal
 \label{eq:app_inner_prod_result}
  \la \mathf^{(0),\sigma}_{+},\Psi^\sigma \ra &=-\frac{1}{8} \, k^\sigma_{\dpl}(\hpl)\,\tilde{k}^\sigma_{\dmi}(\hmi),\\
  \la \mathf^{(0),\sigma}_{-},\Psi^\sigma \ra &=-\frac{1}{8} \, k^\sigma_{\dpl}(\hmi)\,\tilde{k}^\sigma_{\dmi}(\hpl) .
 \eeal
 The integral representations of $\varphi_h^{\sigma}$ in Eq. \eqref{eq:eigen_integral_rep} do not converge for the discrete states, where Eq. \eqref{eq:P_tildeP_results} holds by analytic continuation of $h$.

 The inner products in Eq. \eqref{eq:app_inner_prod_result} are the two-dimensional analogues of the corresponding result in the 1D SYK model \cite{Maldacena_2016}, where the projection of the zero-rung ladder onto the eigenfunctions of the Casimir is proportional to the kernel eigenvalue. The mechanism is same in both cases. The integral over $0<\chi<2$ is half of an integral over the whole real line, and the integral representation of the eigenfunction turns the inner products into the same convolutions that define the eigenvalues.

\bibliographystyle{unsrt}
\bibliography{testbib}

\end{document}